\documentclass[twocolumn]{aastex63}

\received{}
\revised{}
\accepted{\today}

\submitjournal{ApJ}

\shorttitle{UV upturn evolution}
\shortauthors{Ali et al.}
\graphicspath{{./}}

\begin{document}

\title{Evolution of the Ultraviolet Upturn at $0.3<z<1$: exploring helium rich stellar populations}

\correspondingauthor{Sadman S. Ali}
\email{sali@naoj.org}

\author[0000-0003-3883-6500]{Sadman S. Ali}
\affiliation{
Subaru Telescope,  National Astronomical
Observatory of Japan, 650 North Aohoku Place, Hilo, HI, 96720, USA}

\author[0000-0003-1455-7339]{Roberto De Propris}
\affiliation{Finnish Centre for Astronomy with ESO, University of Turku, Vesilinnantie 5, Turku, Finland}

\author[0000-0001-6812-4542]{Chul Chung}
\affiliation{Center for Galaxy Evolution Research, Department of Astronomy, Yonsei University, Seoul 03722, Republic of Korea}

\author{Steven Phillipps}
\affiliation{H. H. Wills Physics Laboratory,
University of Bristol, Tyndall Avenue, Bristol, BS1 TL8, United Kingdom}

\author{Malcolm N. Bremer}
\affiliation{H. H. Wills Physics Laboratory,
University of Bristol, Tyndall Avenue, Bristol, BS1 TL8, United Kingdom}

\begin{abstract}

We measure the near-UV (rest-frame $\sim 2400$\AA) to optical
color for early-type galaxies in 12 clusters 
at $0.3 < z < 1.0$. We show that this is a suitable
proxy for the more common far-ultraviolet bandpass
used to measure the ultraviolet upturn and 
find that the upturn is detected to $z=0.6$ in these
data, in agreement with previous work. We find
evidence that the strength of the upturn starts to
wane beyond this redshift and largely disappears at
$z=1$. Our data is most consistent with models where
early-type galaxies contain minority stellar populations with
non-cosmological helium abundances, up to around 46\%,
formed at $z \geq 3$, resembling multiple stellar population
globular clusters in our Galaxy. This suggests that
elliptical galaxies and globular clusters share
similar chemical evolution and star formation 
histories. The vast majority of the stellar mass
in these galaxies also must have been in place at $z > 3$.

\end{abstract}


\section{Introduction} \label{sec:intro}

It has long been known that potentially all early-type galaxies (ETGs)
contain excess flux at short (vacuum ultraviolet)
wavelengths when compared to expectations from their
old, metal-rich stellar populations extrapolated to
this wavelength regime (e.g., \citealt{Code1979,
Bertola1982,Burstein1988} -- see also
\citealt{OConnell1999,yi2010} for reviews and discussions). Typical observed $FUV-V$ colors for local ETGs are
2--3 magnitudes bluer than expected from spectrum
synthesis models (e.g., \citealt{conroy2009}) that,
with standard initial conditions (solar metallicity,
formation redshift of 4, star-formation e-folding 
timescale of 0.3 Gyr\footnote{The star-formation rate decays by a factor of $e$ during this timescale.}) reproduce the observed optical and infrared colors of ETGs and their evolution even at redshift $\sim2$. 

There is now a broad consensus that
hot (blue) horizontal branch (HB) stars are
responsible for producing the excess UV flux
\citep{Greggio1990,Dorman1993,Dorman1995}. These stars
have been directly identified as the sources of the 
UV light in the nearby bulges of M31 and M32
\citep{Brown1998,tbrown2000b}. \cite{Rosenfield2012} find that the vast majority of the UV light in the bulge of M31 can only be produced by hot HB stars (unresolved in their imaging). Since the bulge of M31 is 
both old \citep{Saglia2010} and metal-rich
\citep{Sarajedini2005}, it provides a good
counterpart to the stellar populations
of more distant ETGs. Contributions from sources other than blue 
HB stars are not fully consistent with observational evidence from local galaxies: Post-AGB (PAGB) stars \citep{Lee1999,Werle2020} are absent
from the UV color-magnitude diagram of 
M32 \citep{tbrown2000b,tbrown2004} and provide only a relatively small ($\sim 20\%$ -- \citealt{Rosenfield2012}) fraction of the UV flux in the M31 bulge: those found are actually
descendants of the blue HB population 
producing the bulk of the UV flux \citep{brown1997,Rosenfield2012}. $FUV$
spectra of nearby ETGs from {\it Astro HUT}
\citep{Ferguson1993,brown1997} lack the broad absorption features that would be
expected by a population of intermediate
luminosity white dwarfs \citep{Werle2020}.
Young stars \citep{Vazdekis2016,Rusinol2019,Werle2020}
cannot easily account for the UV excess light as well. No star hotter than B1V is observed
in M31's bulge \citep{OConnell1992,Rosenfield2012}. Images of several ETGs from {\it Astro UIT} \citep{Stecher1997} exhibit none of the clumpiness usually associated with star formation elsewhere. $FUV$ Spectra of six nearby ETGs from \cite{brown1997} and spectral energy distributions covering
1500--3000\AA\ for several dozen ETGs in
Coma and Abell 1689 \citep{ali2018a,ali2018b} are consistent with a single blackbody and unlike the flux distribution produced by young stellar populations for normal initial mass functions. Finally, estimates of the
star formation rate in ETGs from \cite{Hakobyan2012} and \cite{Sedgwick2021}
using Type II supernovae imply much lower
star formation rates ($\sim 0.01$ to $0.1$  $M_{\odot}$ yr$^{-1}$) than needed to account 
for the observed UV flux: furthermore, star formation tends to occur among lower mass ETGs, whereas the UV upturn is stronger in more massive galaxies.

While neither of the above contributions to the UV upturn flux can be conclusively ruled out from current data (many deriving from nearly 30 years old observations from space telescopes flown on the Space Shuttle), blue HB stars appear to
be the most likely contributor to the bulk
of the UV flux, at least for nearby galaxies. A fundamental assumption of this (and other papers) is that the redshift evolution of the UV upturn is due to one of these sources at all epochs (which of course is a reflection of the comparatively
poor degree of spatial and spectral evolution available, especially for high redshift galaxies, that are extremely faint in the UV), and that such sources should also have local counterparts in the Milky
Way (e.g., blue He-rich stars in globular clusters, sdB binaries in the field, etc.)

However, standard stellar
evolution models do not produce blue HB stars for 
ages and metallicities typical of ETGs (e.g.,
\citealt{Catelan2009}). Blue HB stars are naturally produced by low
metallicity stellar populations (\citealt{Park1997} -- e.g., as in metal-poor globular clusters in our Galaxy). However, if 
such stars existed in sufficient numbers to account
for the observed far-UV flux in ETGs, the optical 
colors and spectra of these galaxies would look dramatically different.

High metallicity stars (of standard composition),
on the other hand, can never evolve to the blue HB within
cosmological timescales, unless they lose sufficient
envelope mass during their first ascent of the red
giant branch prior to the helium flash \citep{yi1997},
to expose the hotter inner core, but this is ruled out
by observations showing no evidence for extra mass 
loss in local open and globular clusters
\citep{Miglio2012,Zijlstra2015,Williams2018} over 
a range of 3 dex in metal abundance\footnote{e.g., \cite{Percival2011} postulate this mechanism to force the existence of blue HB stars in metal-rich isochrones}. 

Mass loss during
the evolution of close binaries \citep{Han2007,Hernandez2014},
losing their envelopes by angular momentum transfer 
as one star evolves off the main sequence and expands,
has also been proposed as a possible mechanism, but
such stars are rare in the globular clusters of our
Galaxy \citep{MoniBidin2009,Kamann2020} and in the bulge
\citep{Badenes2018}. As already noted by \cite{Smith2012}, if such binaries existed, they would have to be systematically more frequent and tighter as a function of galaxy mass and metallicity to account for the observed trends of the UV upturn color with galaxy mass and metallicity \citep{Burstein1988,ali2018a}. They would also have to be tighter and more frequent as a function of galactocentric radius to explain the radial color gradients in the UV upturn color and their correlation with radial gradients in $Mg_2$ strength and overall metallicity as observed by \cite{carter2011} and \cite{jeong2012}. The $FUV-V$ color of cluster ETGs at $z > 0.6$ \citep{ali2018c} and the apparent decrease in
the fraction of blue HB stars at $0.6 < z < 1.0$ as inferred from mid-UV spectral indices in \cite{lecras2016}, are not well
reproduced by this model, where the binary contribution to the UV colors is instead nearly constant after the first Gyr.

However, stars enriched in helium can evolve to the blue HB even at high metallicities (\citealt{lee2005b}; \citealt{chung2011,chung2017}); the extra helium causes faster evolution of stars by increasing their mean molecular weight, which in turn results in smaller mass at the He-burning stage at a given age. Such He-rich stars are observed in the Milky Way globular clusters, where they produce anomalously blue HBs (\citealt{piotto2007,Gratton2012}) and are directly associated with the multiple stellar populations observed in such systems (see review by \citealt{bastian2018}). \cite{peacock2017} found evidence for hot HB stars in M87's metal rich globular clusters while \cite{Goudfrooij2018} suggests that the disintegration of large numbers of such clusters may supply the UV upturn stars in galaxies.

Each of the above models produce different scenarios
for the evolution of the UV upturn color with redshift
\citep{yi2010}. \cite{Brown1998b,Brown2000,Brown2003} obtained a UV color by differencing two STIS filters for a few bright galaxies in clusters at $z=0.33, 0.37$ and $0.55$; these data are consistent with mild or no evolution of the UV upturn over this redshift range. \cite{Ree2007} used bright ellipticals from GALEX out to $z=0.2$ and again finds little evidence for any significant change in the UV upturn color. \cite{Donahue2010} and \cite{Loubser2011} study a small sample of brightest cluster galaxies (BCG) at moderate redshifts ($z < 0.2$) and also find no or modest evolution. \cite{boissier2018} use a large sample of BCGs in the background of the Virgo cluster and detect upturn in these objects out to $z=0.35$ (showing no clear signs of evolution in their $FUV-NUV$ color), with an excess of upturn sources found at $z\sim0.25$. In a follow-up study of GAMA galaxies, \cite{Dantas2018} suggest an overabundance of upturn carriers at $z\sim0.25$, but with no clear statistical evidence to trends over z=0.25. The evolution
regarding the amount of galaxies carrying the UV upturn beyond this redshift remains to be probed. These data are in mild disagreement with the metal-poor and metal-rich HB scenarios (with no extra He) but do not provide strong constraints on either of the models discussed above. Most of these studies, however, have limited redshift range and/or target only a small number of very bright galaxies. We \citep{ali2018a,ali2018b,ali2018c} have used archival
data from the Hubble Space Telescope (HST) to measure
the evolution of the UV upturn out to $z=0.7$ and for
galaxies reaching to or below the $M^*$ point in the
luminosity function (i.e., the more normal population, as opposed to the brighter cluster galaxies
used in most previous studies as cited). Our data
show that there is no significant evolution in the
color and scatter of the UV upturn out to $z=0.55$,
but there is clear evidence that at $z=0.7$ the
$FUV-V$ color has become significantly redder.
Although only a lower (blue) limit to the mean UV
upturn color of red sequence ETGs at $z=0.7$ could be
derived, this is $\sim 1.2$ mag. redder (at the 3$\sigma$
level) than at lower redshifts \citep{ali2018c}. This observation is not consistent with all other models we have discussed earlier for the origin of the UV
upturn, except best explained by a minority population of He-rich stars, while other models cannot reproduce this pattern of evolution, and constrains their helium abundance to be $Y > 0.42$, nearly twice the cosmological value from Big Bang nucleosynthesis. Similar evolutionary trends are also observed for group/field ETGs at $z\sim 0.05$ \citep{phillipps2020} and to $z=0.6$ \citep{Atlee2009} and $z=1$ \citep{lecras2016}.

A testable prediction of this model is that (a)
$FUV-V$ evolves rapidly to the red above a redshift
specified by the epoch of galaxy formation and the
helium abundance, and (b) the scatter in this color
decreases as the upturn fades to match the small
scatter in the optical/infrared colors of ETGs (i.e.,
the extra component due to the anomalous blue HB
disappears). However, this generally requires
deep observations in the vacuum UV that are hard to
obtain. While not ideal, near-UV colors (around 2300\AA)
are also a suitable proxy for the UV upturn, as the UV
sources contribute more than ¾ of the total flux in
this region of the spectrum \citep{ali2019,phillipps2020}. The UV
upturn `leaks' out to nearly 3000\AA\ and the older
stellar populations of normal composition contribute
little to this regime (e.g., see
\citealt{ali2018a,ali2018b}), although the 
sensitivity to evolution is lower at increasing
wavelength.

In this paper we extend the cluster-wide analysis of
the UV upturn to red sequence galaxies of 12 clusters,
reaching out to $z\sim1$ - the
highest redshift as allowed by currently available
data, at which it is possible to reach down to and
beyond the $M^*$ point\footnote{The luminosity function of galaxies is well described by the Schechter function $\Phi(M)dM=\Phi^* 10^{-0.4(M-M^*)(\alpha+1)} \exp (10^{-0.4(M-M^*))}$ where $M^*$ is the characteristic magnitude of bright galaxies and $\alpha$ the slope of the faint end.}. At this redshift we explore 
an epoch in which the hot horizontal branch stars are
no longer expected to exist, as the low mass stars in
ETGs simply have not had enough time 
to occupy the late stages (core helium burning) of
stellar evolution. Typical evolutionary timescales for stars to reach the HB are $>6.5$ Gyrs. Given $z=0.6$ as the point at which they appear to ‘switch on the UV upturn’, one would get ages of around 12 Gyrs, i.e. $z\sim3-4$; similar to globular clusters. The results for cluster galaxies
between $0.3<z<1$ are compared to one another and to
predictions of theoretical models of UV upturn. We also measure
the scatter in the UV colors of cluster galaxies at
increasing redshift and compare them to theoretical
expectations. The aforementioned analyses will allow
us to place stringent constraints on any helium
enhancement present in these galaxies, their formation
epochs and probe the evolution of the upturn
phenomenon at large.

We describe in section 2 our dataset and present
results in section 3. In section 4 we discuss our
results in the light of helium-rich models and
theories of galaxy formation and evolution. We use 
the conventional cosmological parameters for this
analysis H$_0 = 67$ km s$^{-1}$ Mpc$^{-1}$, $\Omega_m=0.3$, $\Omega_{\Lambda}=0.7$  (\citealt{Planck2020}). All magnitudes
quoted are in the AB system.

\section{Data and Photometry}

\begin{deluxetable*}{ccccccc}
\tablenum{1}
\tablecaption{Summary of Observational Data\label{table1}}
\tablewidth{0pt}
\tablehead{
\colhead{Cluster} & \colhead{Redshift} & \colhead{Filters} & \colhead{Rest-Frame Central} \vspace{-0.2cm} & \colhead{Total Exposure} & \colhead{Proposal ID} & \colhead{PI} \\ \colhead{} & \colhead{} & \colhead{} & \colhead{Wavelength (\AA)} & \colhead{Time (ks)} & \colhead{} & \colhead{} \\
}
\startdata
Abell 2744 & 0.31 & F336W & 2569 & 27.7 & 13389 & Siana \\
           &      & F435W & 3326 & 45.7 & 13495 & Lotz \tablenotemark{1} \\
           &      & F606W & 4633 & 23.6 & 13495 & Lotz \tablenotemark{1}\\
           &      & F814W & 6223 & 104.3 & 13495 & Lotz \tablenotemark{1}\\
Abell S1063 & 0.35 & F336W & 2493 & 22.6 & 14209 & Siana \\
            &      & F435W & 3227 & 46.4 & 14037 & Lotz \tablenotemark{1}\\
            &      &  F606W & 4496 & 25.8 & 14037 & Lotz \tablenotemark{1}\\
            &      &  F814W & 6039 & 120.2 & 14037 & Lotz \tablenotemark{1}\\
Abell 370   & 0.38 &  F336W & 2446 & 22.2 & 14209 & Siana \\
            &      &  F435W & 3164 & 51.4 & 14038 & Lotz \tablenotemark{1}\\
            &      &  F606W & 4407 & 25.3 & 14038 & Lotz \tablenotemark{1}\\
            &      &  F814W & 5920 & 130.6 & 14038 & Lotz \tablenotemark{1}\\
MACS J0416$-$2403 & 0.40 & F336W & 2407 & 22.2 & 14209 & Siana \\
                  &      & F435W & 3116 & 54.5 & 13496 & Lotz \tablenotemark{1}\\
                  &      & F606W & 4341 & 33.5 & 13496 & Lotz \tablenotemark{1}\\
                  &      & F814W & 5831 & 129.9 & 13496 & Lotz \tablenotemark{1}\\
MACS J0717.5+3745 & 0.54 & F390W & 2516 & 4.9 & 12103 & Postman \tablenotemark{2}\\
                  &      & F555W & 3580 & 8.9 & 9722 & Ebeling \\
                  &      & F606W & 3910 & 27.0 & 13498 & Lotz \tablenotemark{1}\\
                  &      & F814W & 5252 & 114.6 & 13498 & Lotz \tablenotemark{1}\\
MACS J1149.5+2223 & 0.54 & F390W & 2516 & 4.9 & 12068 & Postman \tablenotemark{2}\\
                  &      & F555W & 3580 & 9.0 & 9722 & Ebeling \\
                  &      & F606W & 3910 & 24.8 & 13504 & Lotz \tablenotemark{1}\\
                  &      & F814W & 5252 & 104.2 & 13504 & Lotz \tablenotemark{1}\\
MACS J2129.4$-$0741 & 0.59 & F390W & 2453 & 4.6 & 10493 & Gal-Yam \\
                    &      & F555W & 3491 & 8.9 & 9722 & Ebeling \\
                    &      & F814W & 5119 & 13.4 & 10493 & Gal-Yam \\
                    &      & F125W & 7862 & 2.4 & 12100 & Postman \tablenotemark{2}\\
                    &      & F160W & 10063 & 5.0 & 12100 & Postman \tablenotemark{2}\\
SDSS 1004+4112      & 0.68 & F435W & 2589 & 13.4 & 10509 & Kochanek \\
                    &      & F555W & 3304 & 8.0 & 10509 & Kochanek \\
                    &      & F814W & 4845 & 5.4 & 10509 & Kochanek \\
MACS J0744.8+3927   & 0.69 & F435W & 2574 & 4.0 & 12067 & Postman \tablenotemark{2}\\
                    &      & F555W & 3284 & 17.8 & 12067 & Postman \tablenotemark{2}\\
                    &      & F814W & 4817 & 25.8 & 10493 & Gal-Yam \\
                    &      & F125W & 7396 & 2.5 & 12067 & Postman \tablenotemark{2}\\
                    &      & F160W & 9467 & 3.1 & 12067 & Postman \tablenotemark{2}\\
\enddata
\tablenotetext{1}{\cite{lotz2017}}
\tablenotetext{2}{\cite{postman2012}}
\end{deluxetable*}

\begin{deluxetable*}{ccccccc}
\tablenum{1}
\tablecaption{Summary of Observational Data (Continued) }
\tablewidth{0pt}
\tablehead{
\colhead{Cluster} & \colhead{Redshift} & \colhead{Filters} & \colhead{Rest-Frame Central} \vspace{-0.2cm} & \colhead{Total Exposure} & \colhead{Proposal ID} & \colhead{PI} \\ \colhead{} & \colhead{} & \colhead{} & \colhead{Wavelength (\AA)} & \colhead{Time (ks)} & \colhead{} & \colhead{} \\
}
\startdata
RCS 2327-0204 & 0.70 & F435W & 2559 & 4.2 & 10846 & Gladders \\
              &      & F814W & 4788 & 5.3 & 10846 & Gladders  \\
                  &      & F125W & 7353 & 4.6 & 13177 & Bradac \tablenotemark{3}\\
                  &      & F160W & 9412 & 4.6 & 13177 & Bradac  \tablenotemark{3}\\
CL 1226.9+3332    & 0.89 & F475W & 2513 & 4.4 & 12791 & Postman \tablenotemark{2}\\
                  &      & F606W & 3206 & 32.0 & 9033 & Ebeling \\
                  &      & F125W  & 6614 & 2.5 & 12791 & Postman \tablenotemark{2}\\
                  &      & F160W & 7407 & 2.3 & 12791 & Postman \tablenotemark{2}\\
SPT-CL 2011$-$5228 & 0.96 & F475W & 2423 & 5.4 & 14630 & Collett \\
                  &       & F606W & 3092 & 5.4 & 14630 & Collett \\
                  &       & F125W & 6378 & 1.4 & 14630 & Collett \\
                  &       & F140W & 7143 & 1.4 & 14630 & Collett \\
\enddata
\tablenotetext{2}{\cite{postman2012}}
\tablenotetext{3}{\cite{bradac2014}}
\end{deluxetable*}

In order to perform a cluster-wide analysis of the upturn in the red sequence population (and not just of the brighter luminous red galaxies - LRGs; e.g. \citealt{lecras2016}, or brightest cluster galaxies - BCGs; e.g. \citealt{Brown1998b,brown2000a,Brown2003}) at high redshifts, one requires rest-frame UV data of sufficient depth below $\sim$2500\AA\ down to the Lyman limit, the wavelength regime most sensitive to the upturn. At $z < 1$ these bandpasses still reside in the vacuum ultraviolet and need deep space-based observations. In this paper we measure the evolution
of the UV upturn flux at $\sim 2400$ \AA, where the
main contribution to the observed flux is still 
dominated by the UV upturn sources (from the full
UV spectral energy distributions we have derived in
our previous work -- \citealt{ali2018a,ali2018b,ali2019,phillipps2020}). 
\cite{lecras2016} carry out a similar measurement 
using a series of spectral indices at rest-frame
wavelengths $\sim 2000 - 3000$ \AA, i.e., within the 
near ultraviolet regime we probe here.

We have identified 12 clusters in the Hubble Legacy
Archive with data in filters $F336W$, $F390W$, $F435W$
and $F475W$ probing a rest-frame bandpass centered at
$\sim 2400$\AA\ out to $z \sim 1$. The clusters 
are identified in Table~\ref{table1}, where we 
detail all filters used for each cluster, their rest-frame central wavelengths, the respective programs and the relative exposure times. The clusters cover the redshift range from $z=0.31$ (Abell 2744) to $z=0.96$ (CL2011). All images were retrieved as drizzled and fully processed data on which photometry could be performed. In the cases where multiple observations exist in each band, those frames were combined together using $IRAF$'s $imcombine$ function to produce the deepest possible images.

\begin{figure*}
\centering
{\includegraphics[width=0.325\textwidth]{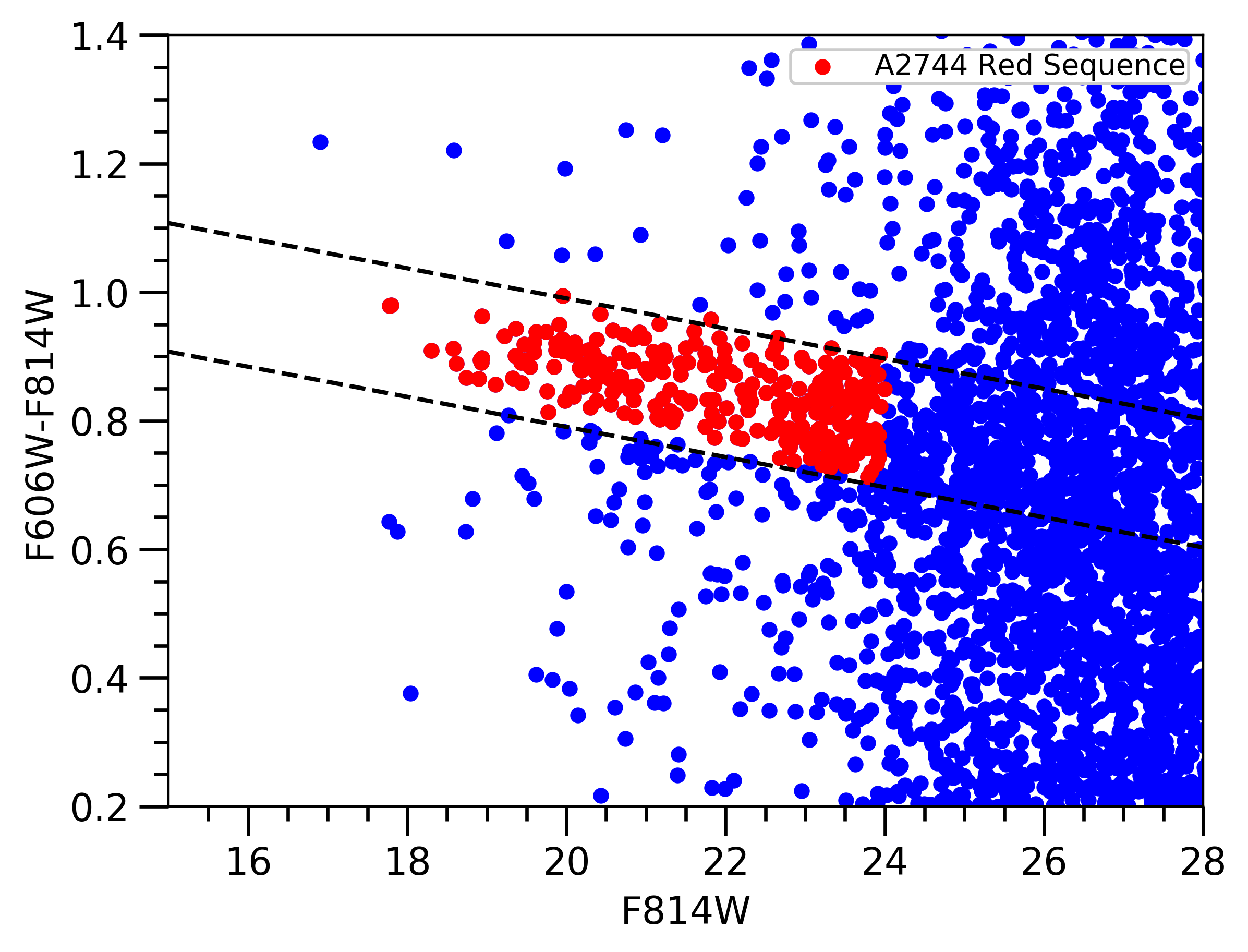}}
{\includegraphics[width=0.325\textwidth]{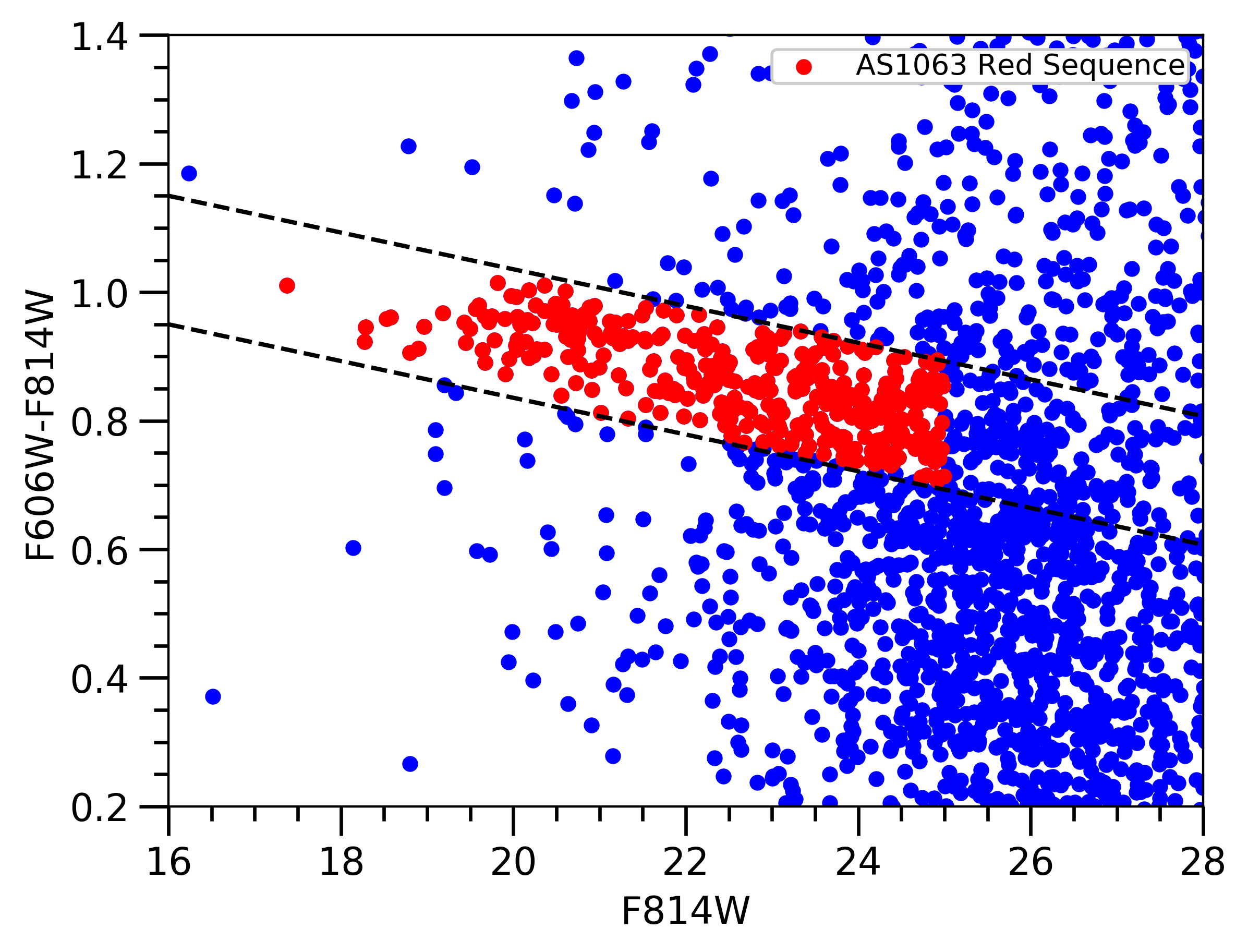}}
{\includegraphics[width=0.325\textwidth]{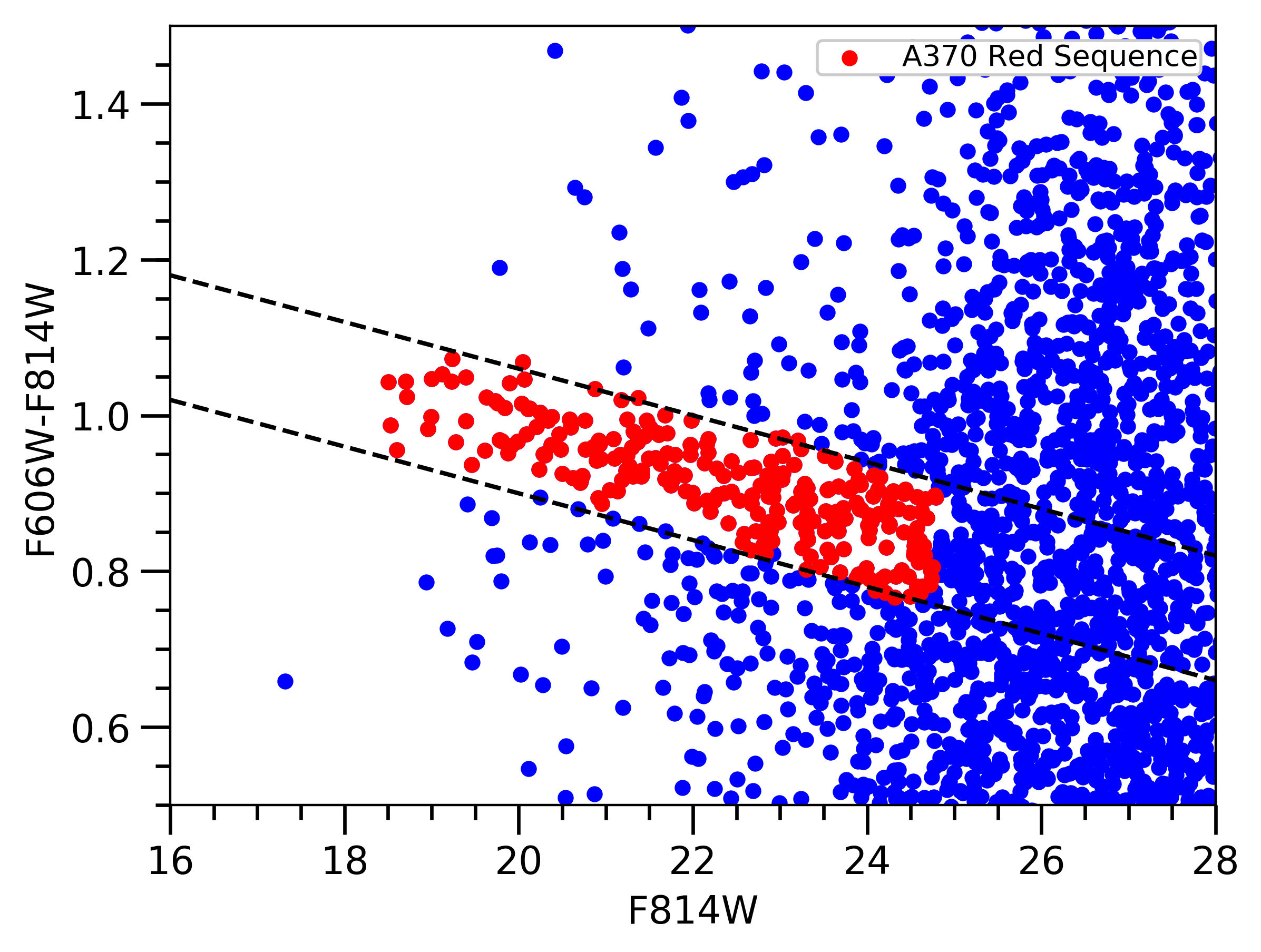}}
{\includegraphics[width=0.325\textwidth]{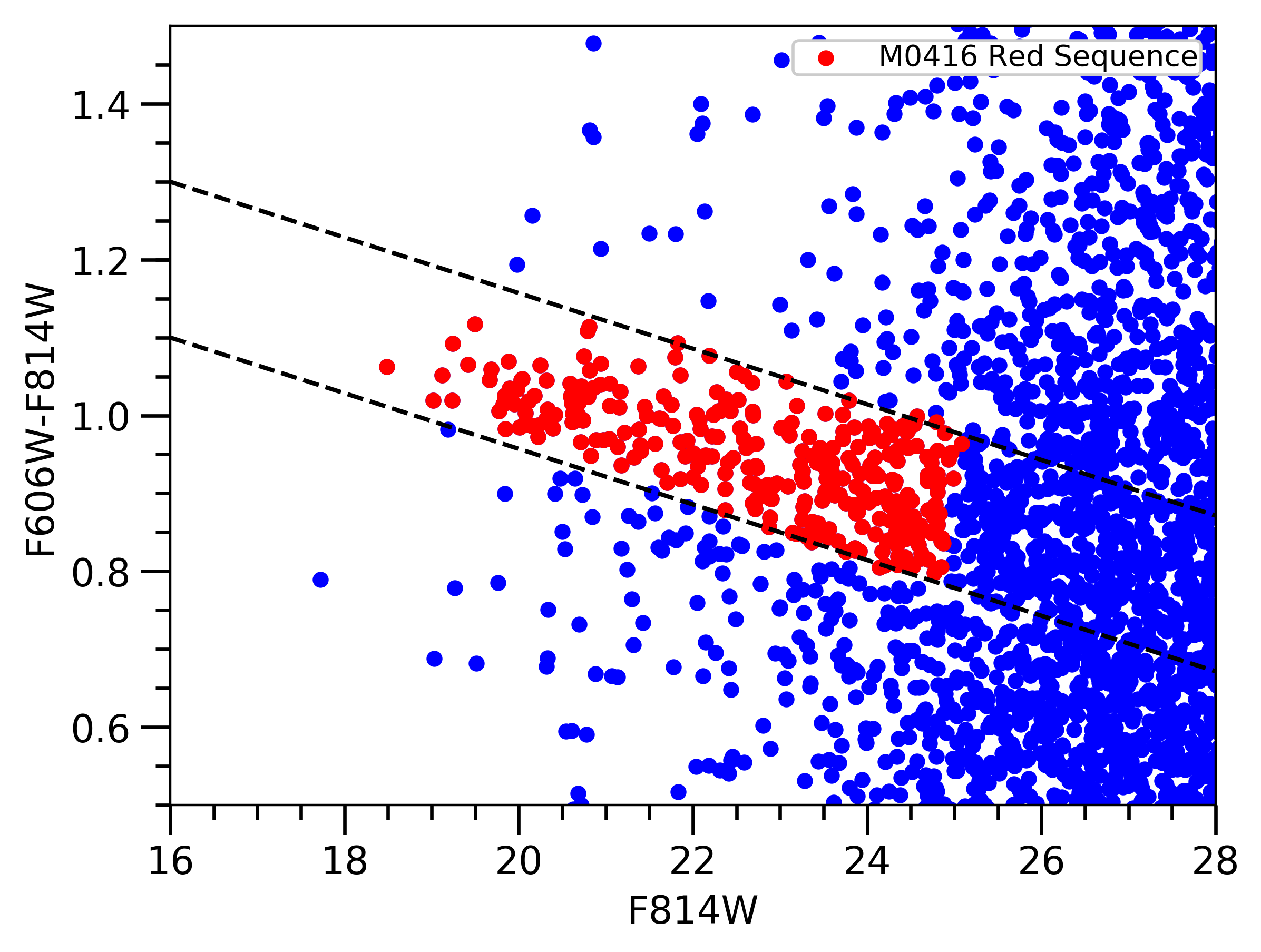}}
{\includegraphics[width=0.325\textwidth]{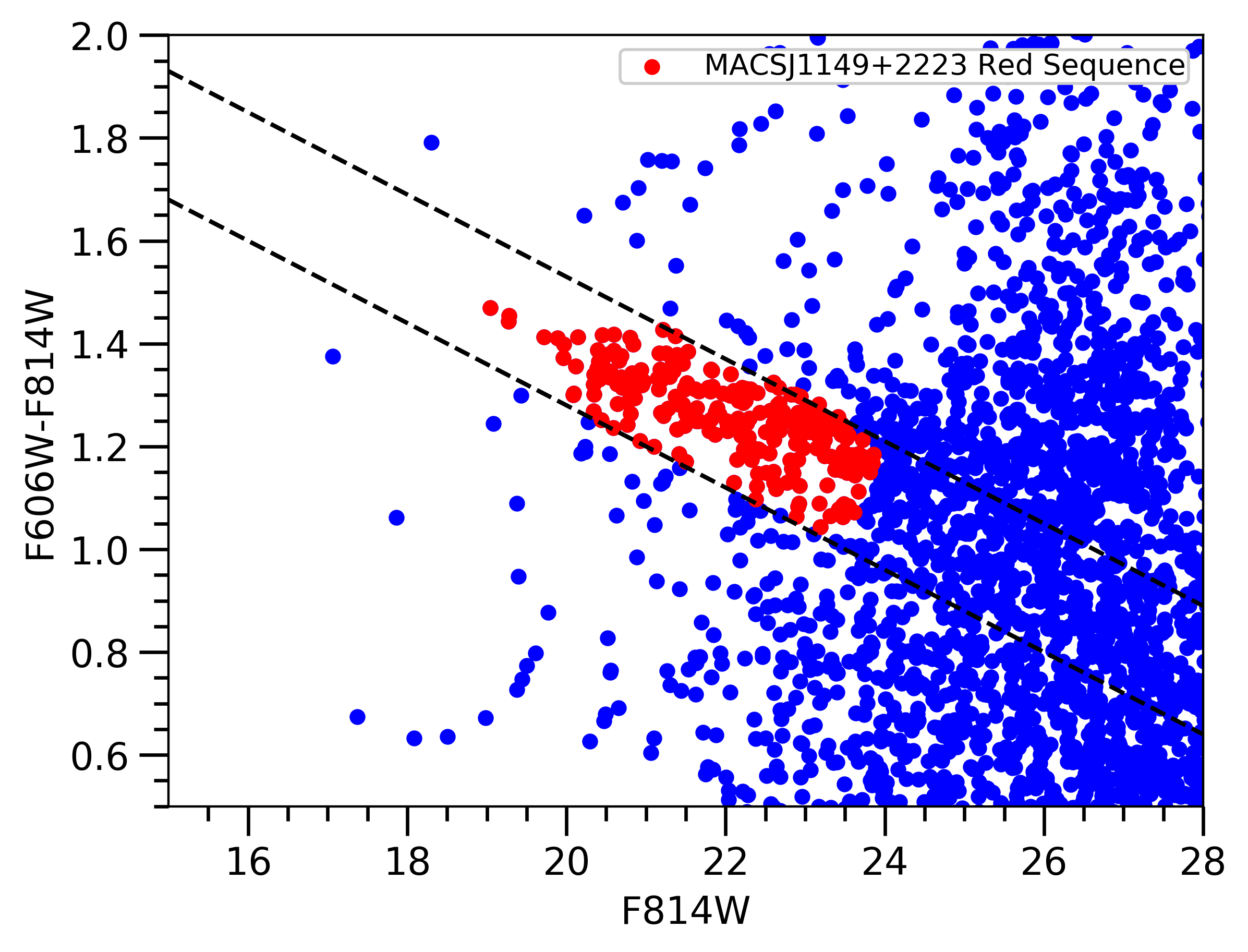}}
{\includegraphics[width=0.325\textwidth]{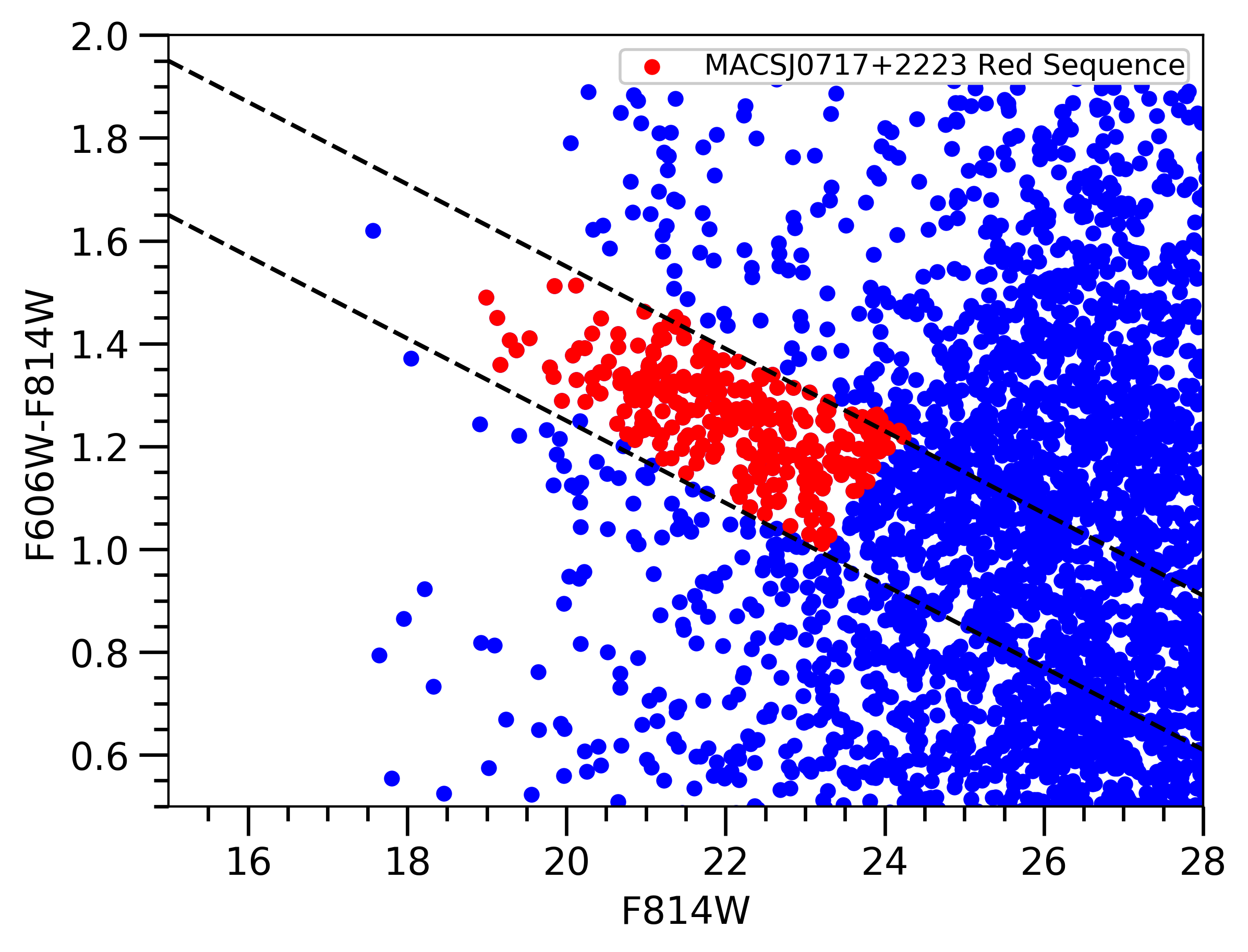}}
{\includegraphics[width=0.325\textwidth]{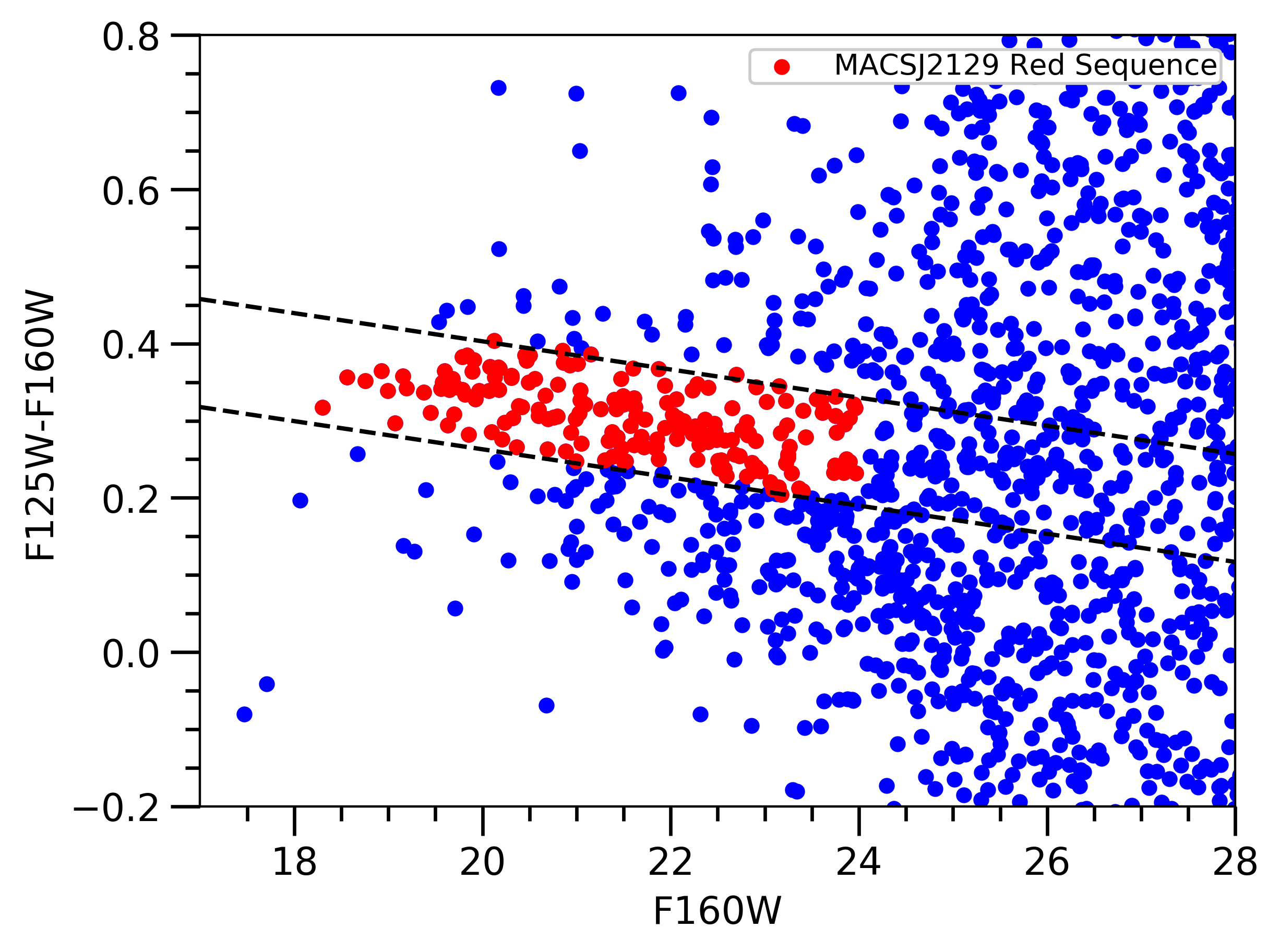}}
{\includegraphics[width=0.325\textwidth]{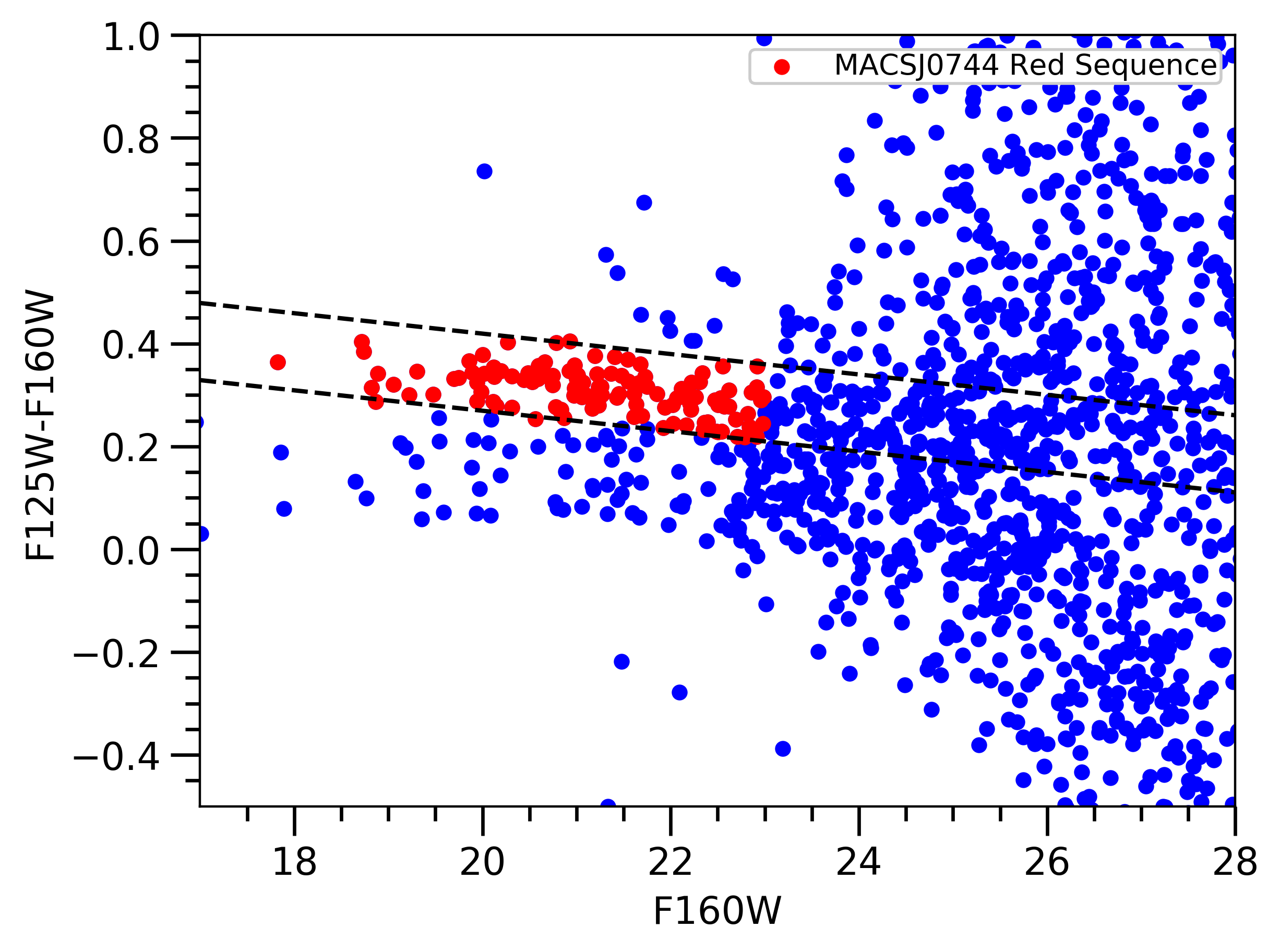}}
{\includegraphics[width=0.325\textwidth]{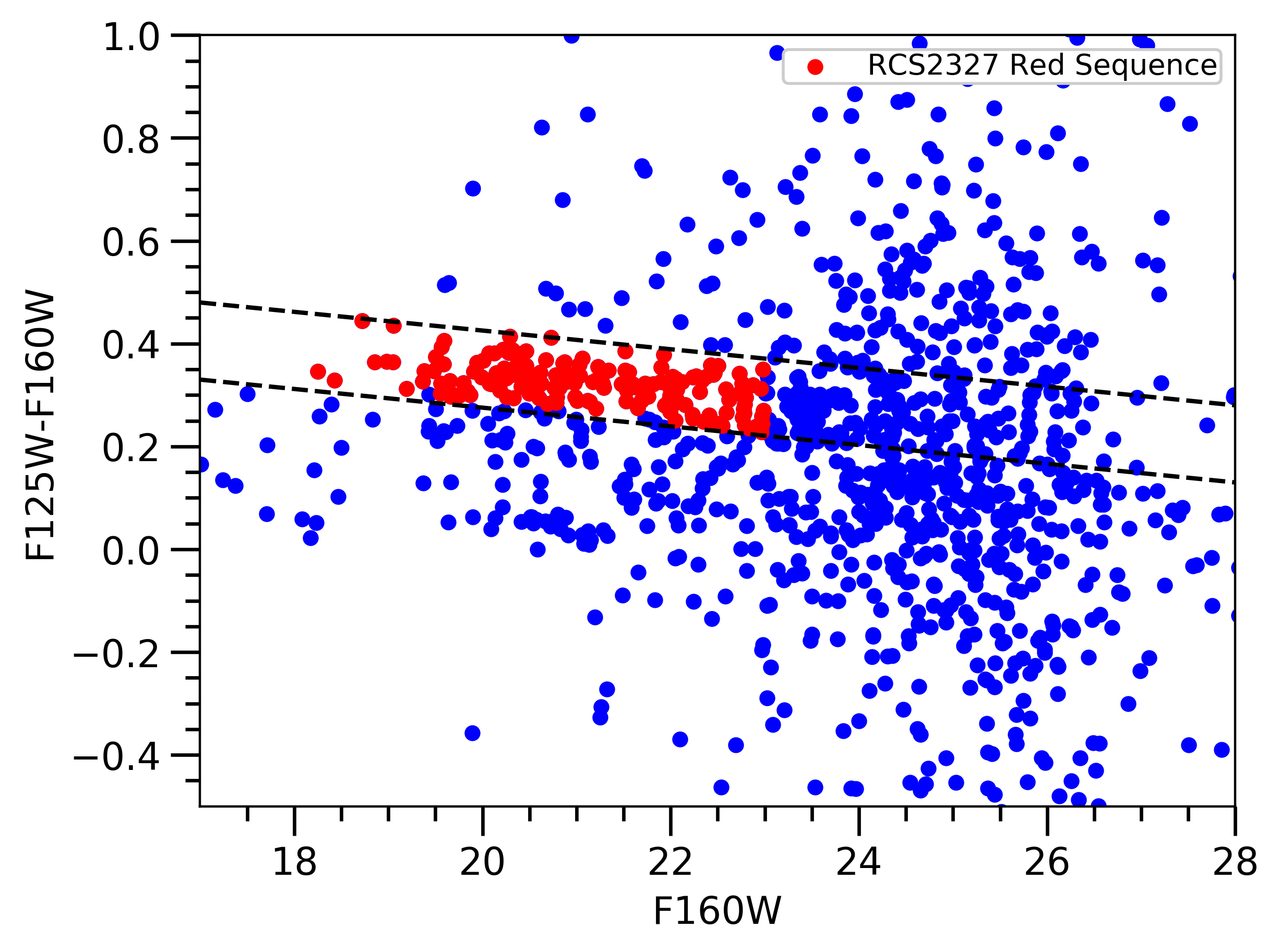}}
{\includegraphics[width=0.325\textwidth]{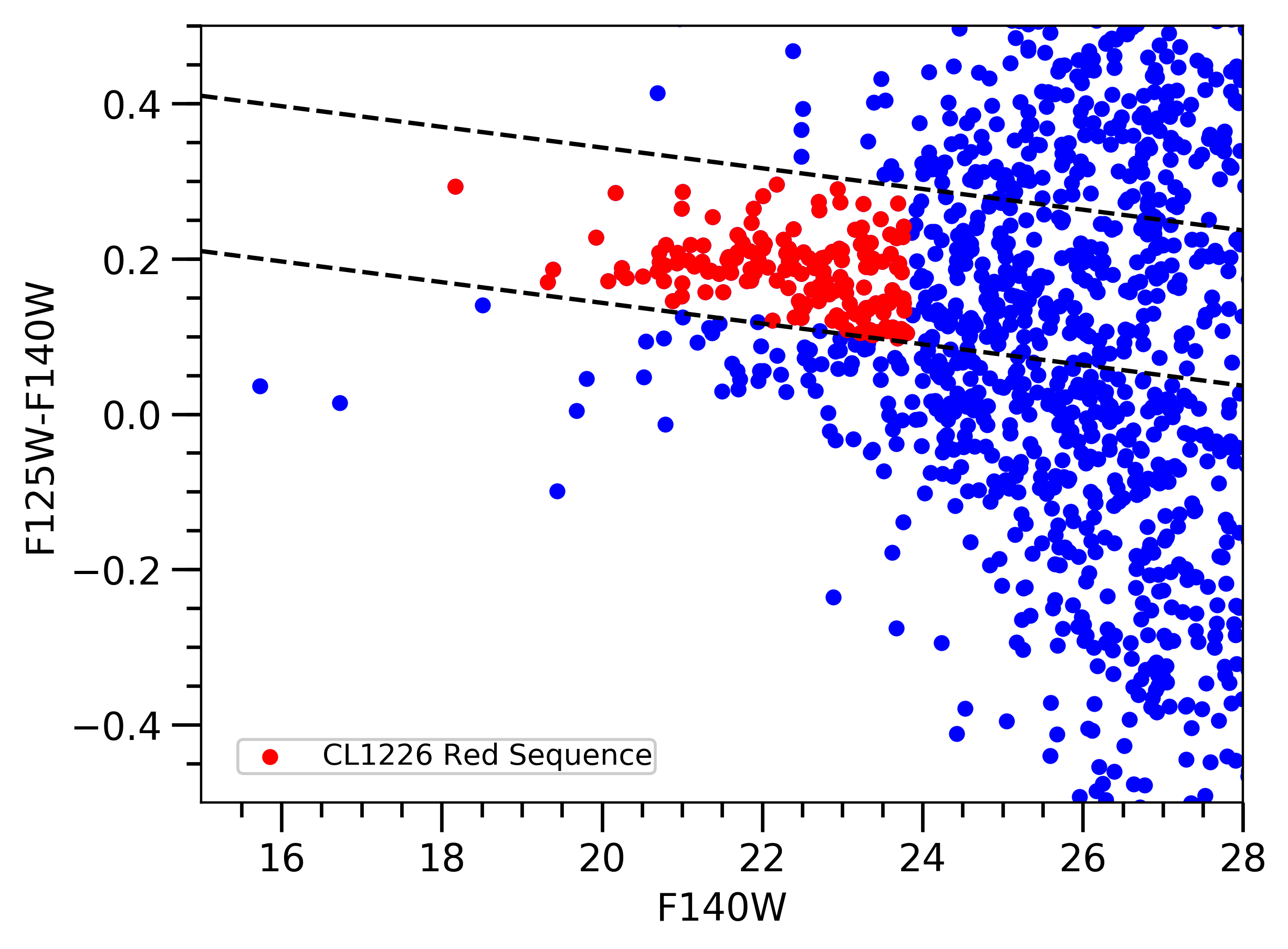}}
{\includegraphics[width=0.325\textwidth]{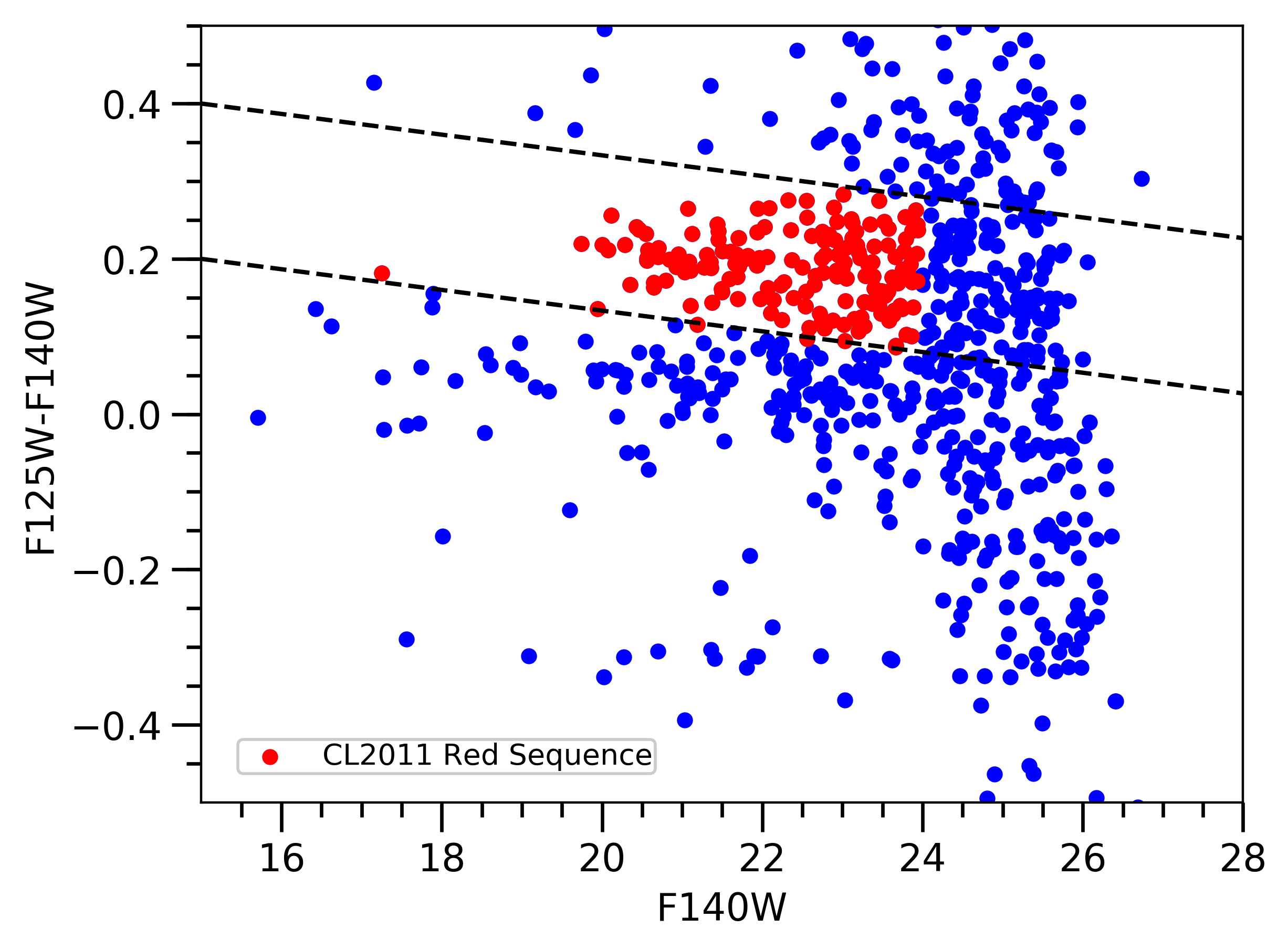}}
\caption{Optical color-magnitude diagrams for 11 of the clusters analysed in this paper. The red sequence galaxies are denoted with the red filled circles within the dashed lines and have photometric uncertainties of $<0.05$ magnitudes in their optical colors. The dashed lines show the selection region for quiescent cluster ETGs (as described in the text, within $\pm 0.1$ mag. of the mean ridge line for the red sequence). SDSS1004 did not have any equivalent optical (rest-frame) data to construct a corresponding CMD.}
\label{fig:1}
\end{figure*}
 
 This provides us with a set of clusters at multiple
 redshift intervals, thus allowing for a study of the
 evolution of the upturn with lookback time;
 particularly important as most theoretical models predict a decline in the
 strength of the upturn at higher redshifts (except \cite{Han2007} where the UV upturn color is nearly constant with redshift to $z \sim 6$).
 Pinpointing the specific epoch at which this
 decline occurs can allow us to constrain the key
 model parameters, such as the He-enhancement $Y$
 and the age of formation of the stellar population.

\begin{figure*}
\centering
{\includegraphics[width=0.325\textwidth]{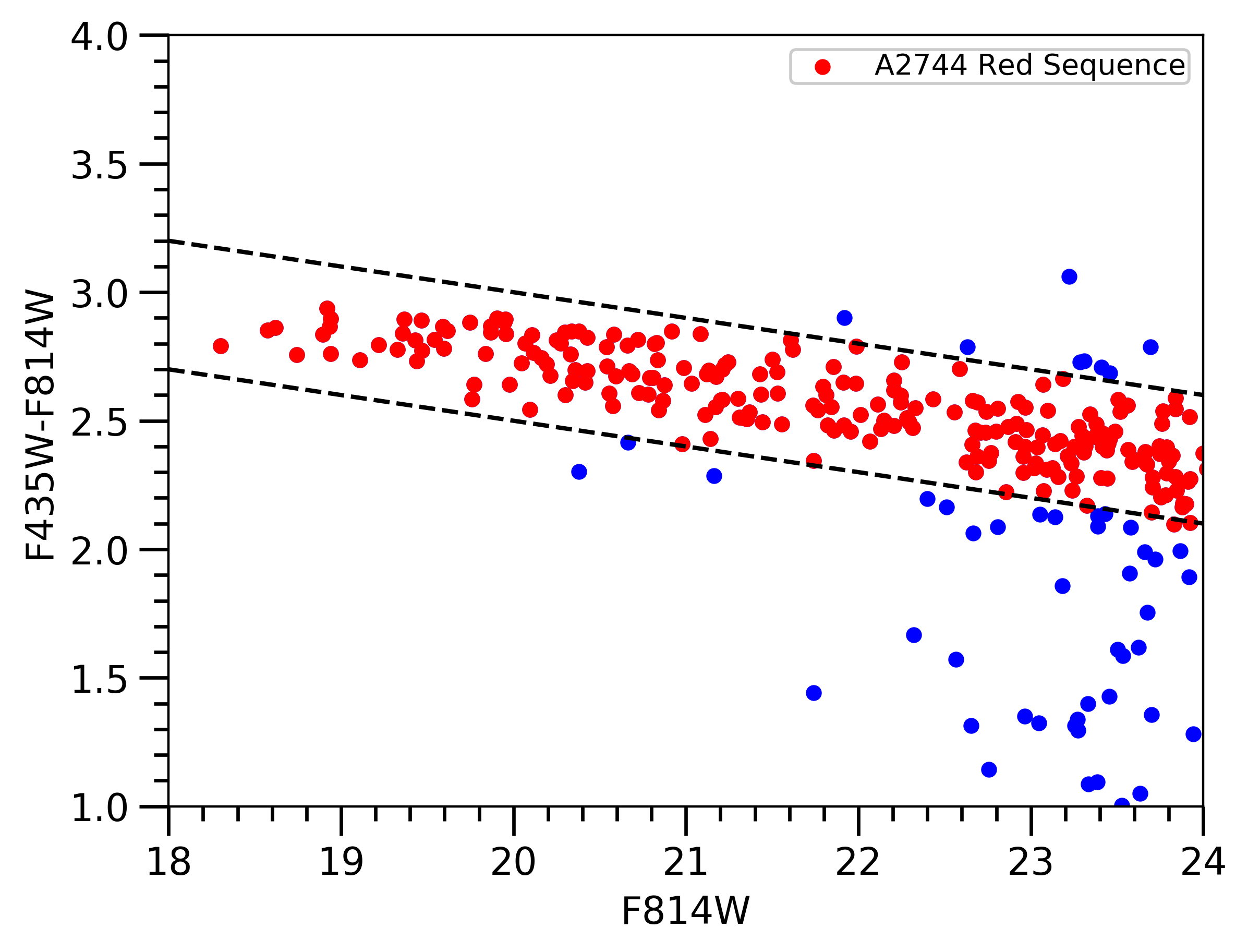}}
{\includegraphics[width=0.325\textwidth]{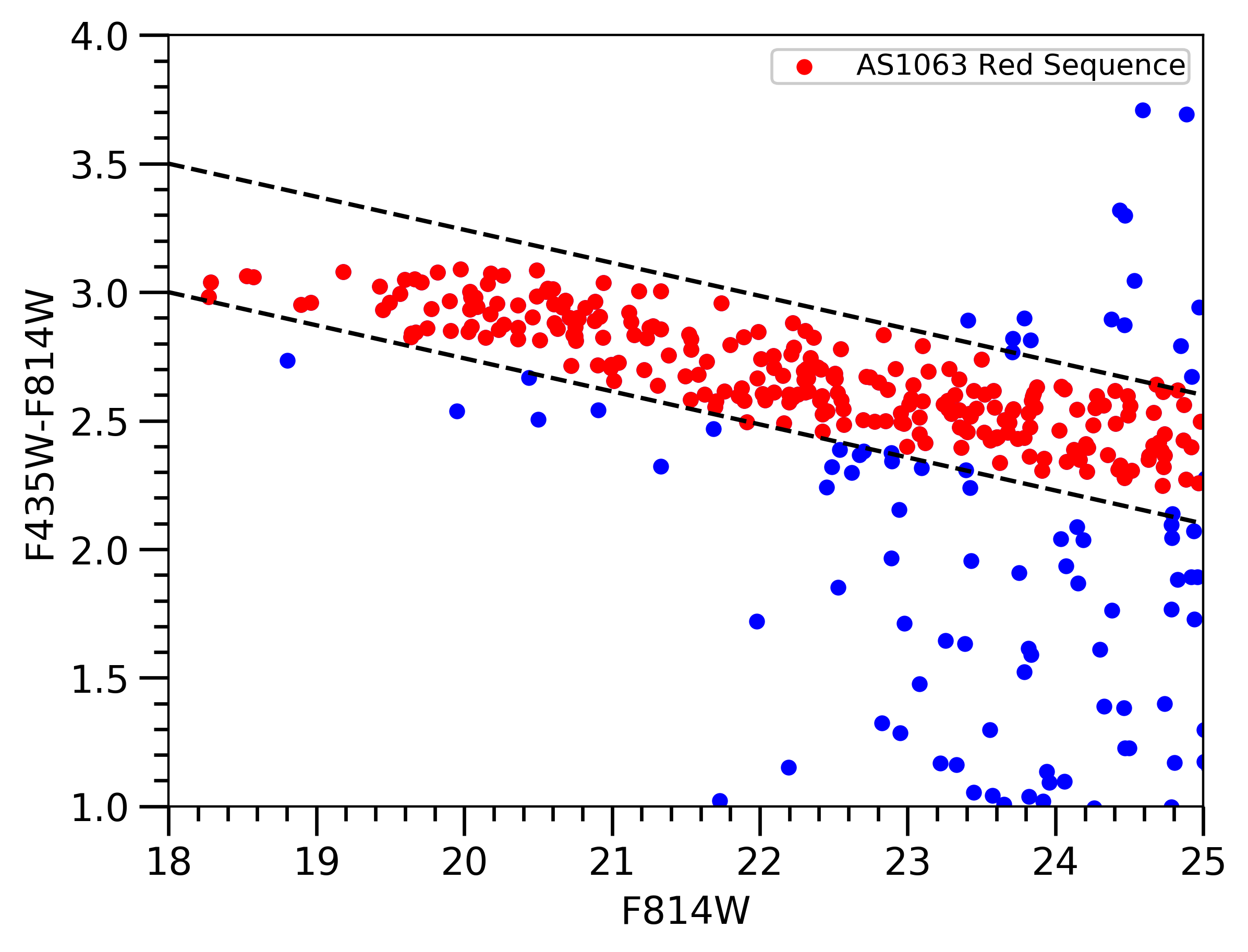}}
{\includegraphics[width=0.325\textwidth]{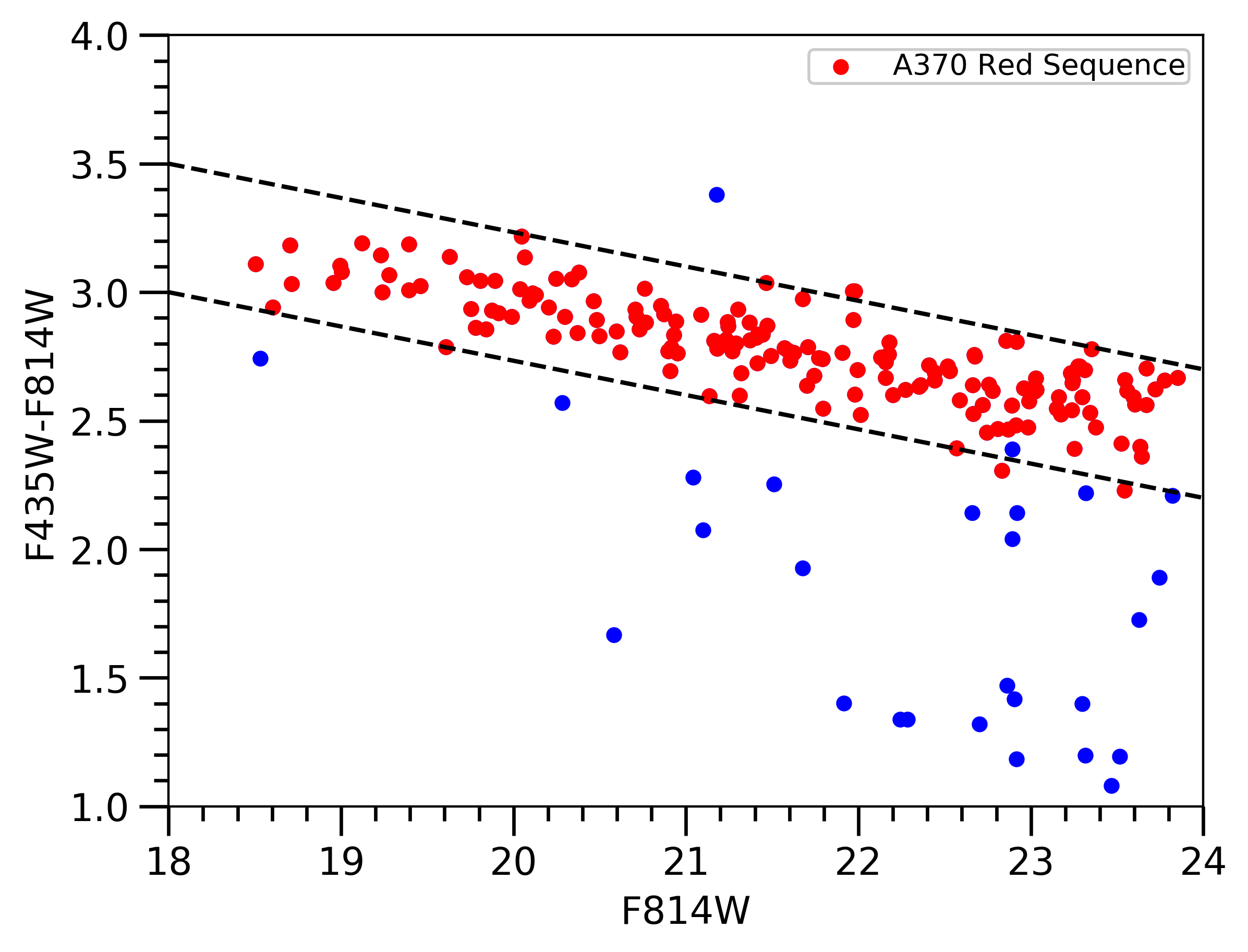}}
{\includegraphics[width=0.325\textwidth]{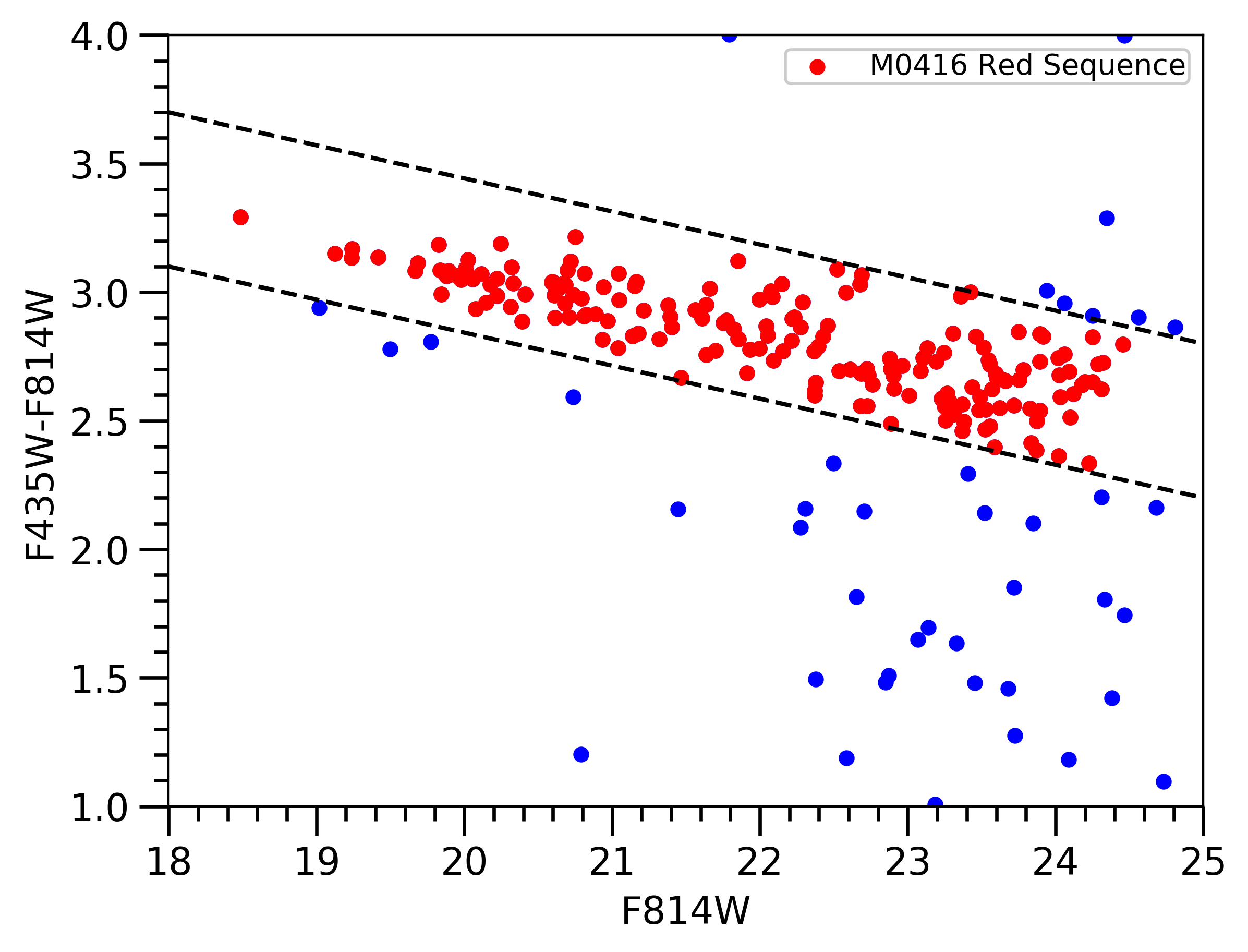}}
{\includegraphics[width=0.325\textwidth]{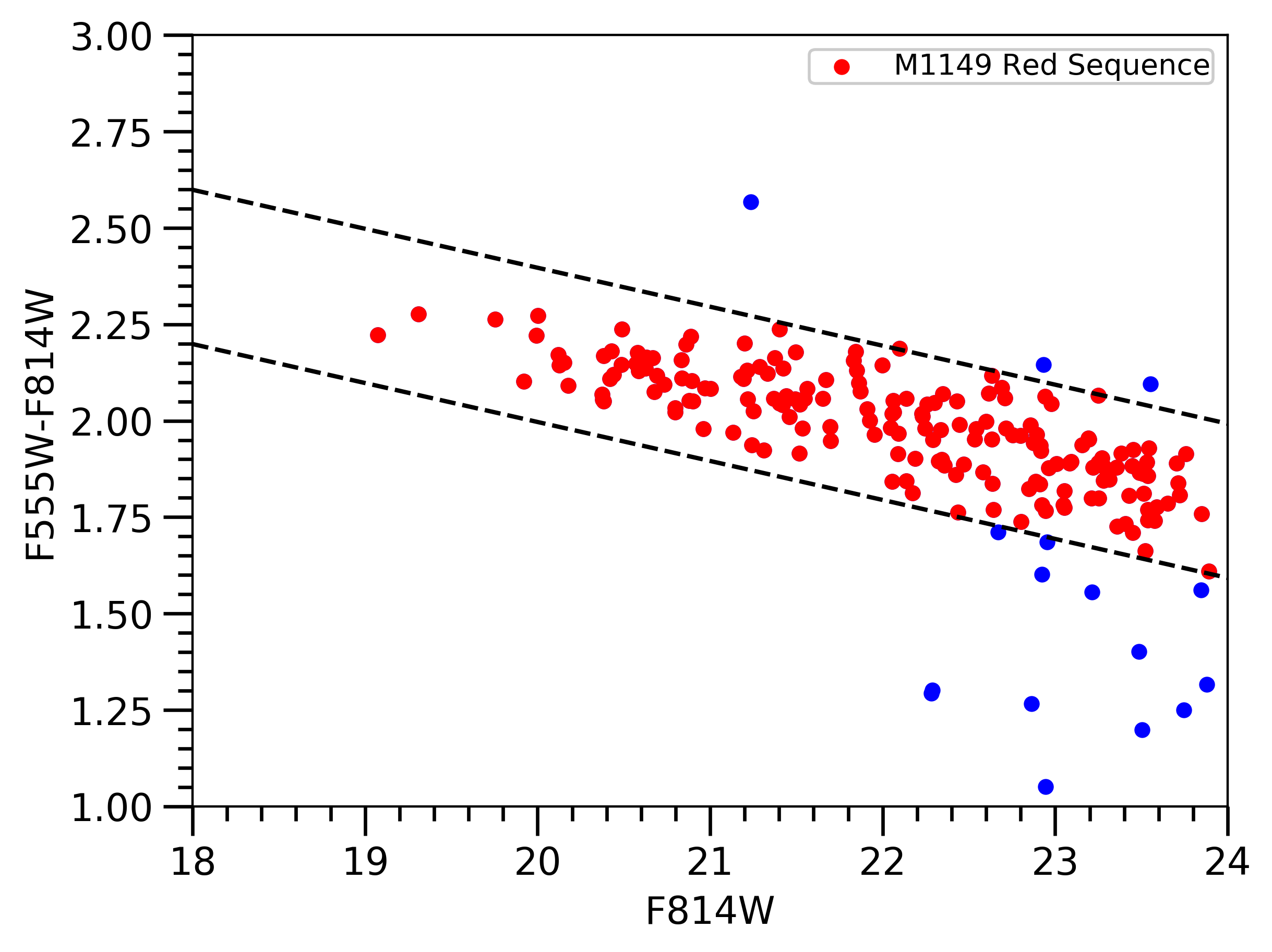}}
{\includegraphics[width=0.325\textwidth]{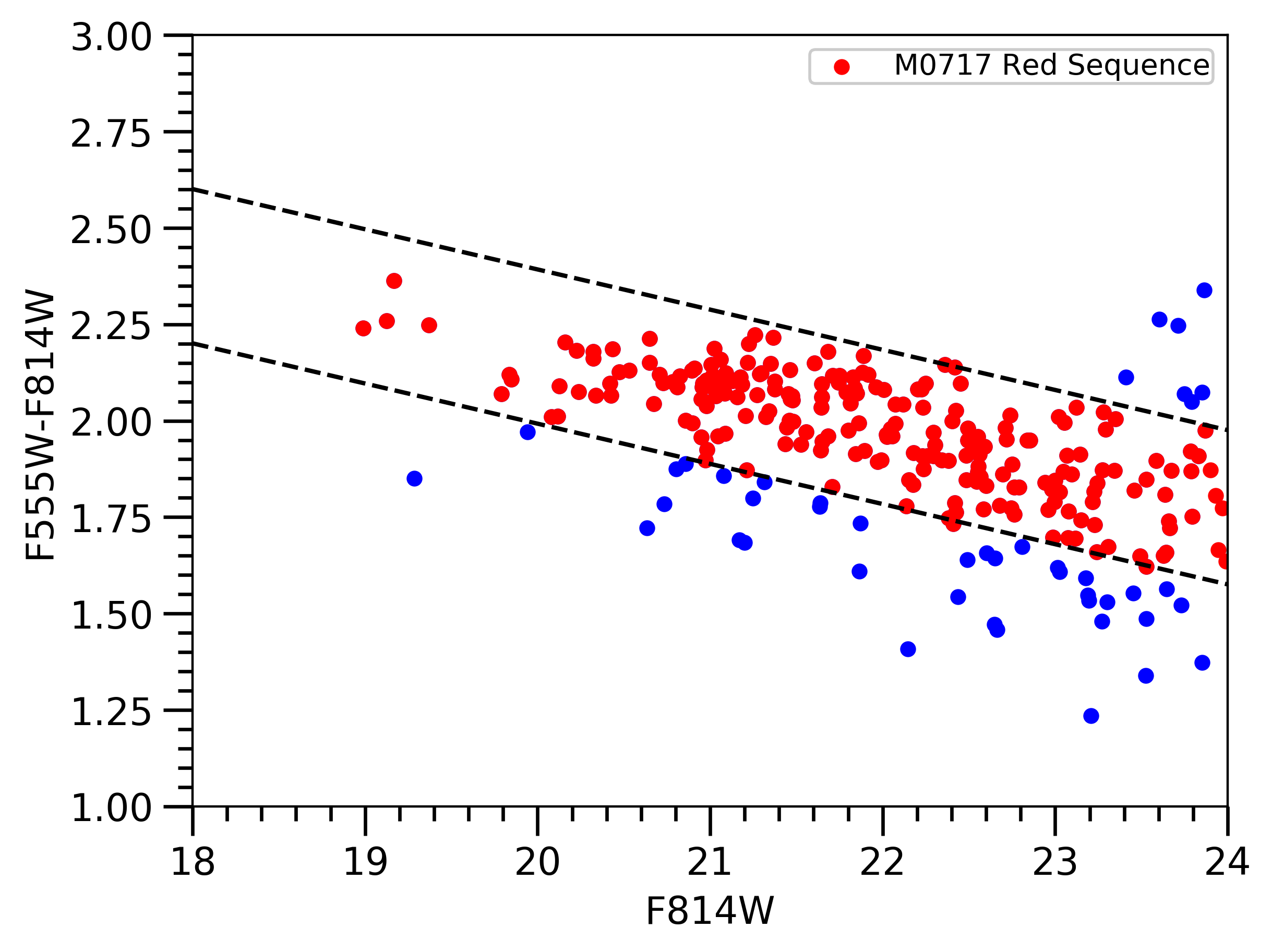}}
{\includegraphics[width=0.325\textwidth]{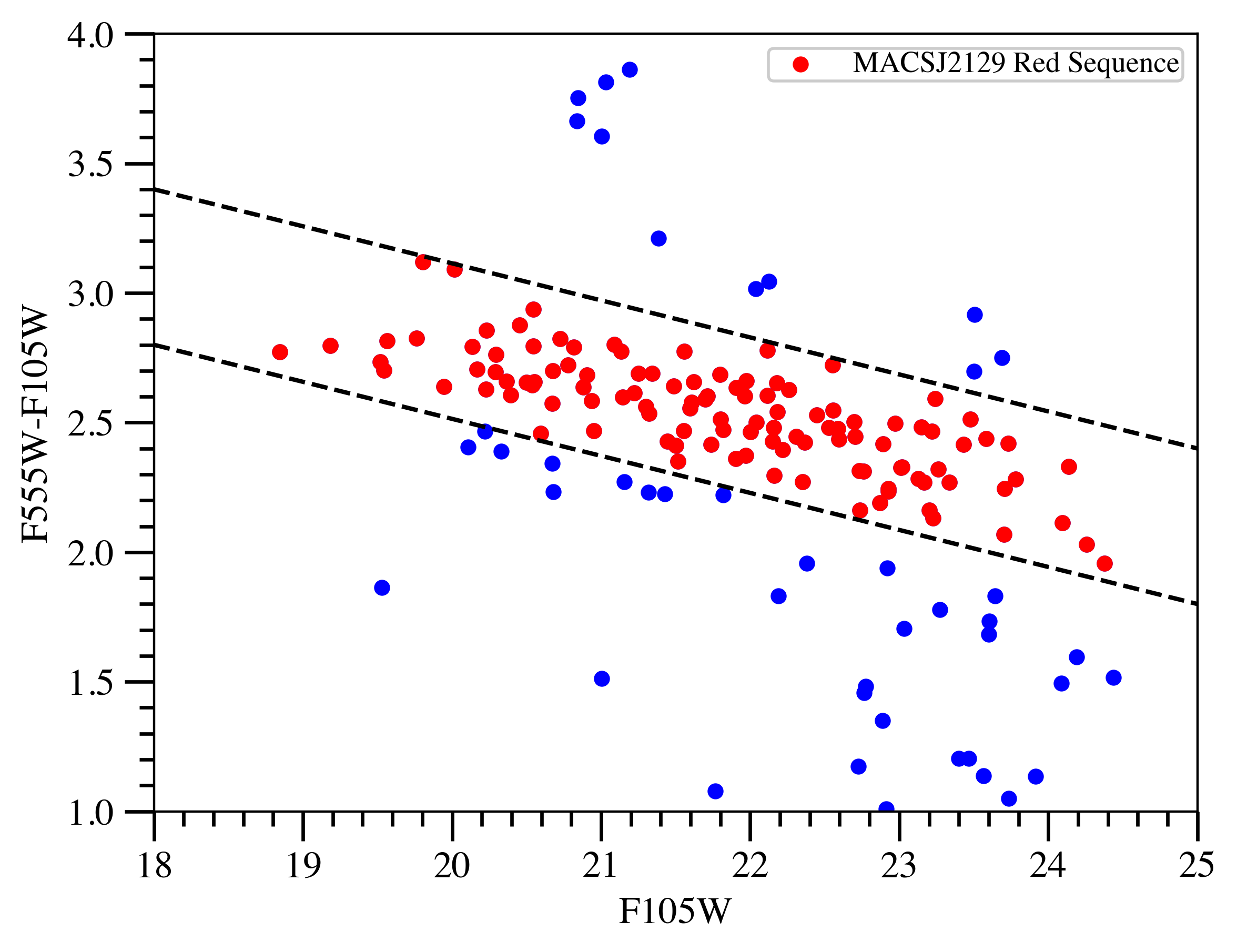}}
{\includegraphics[width=0.325\textwidth]{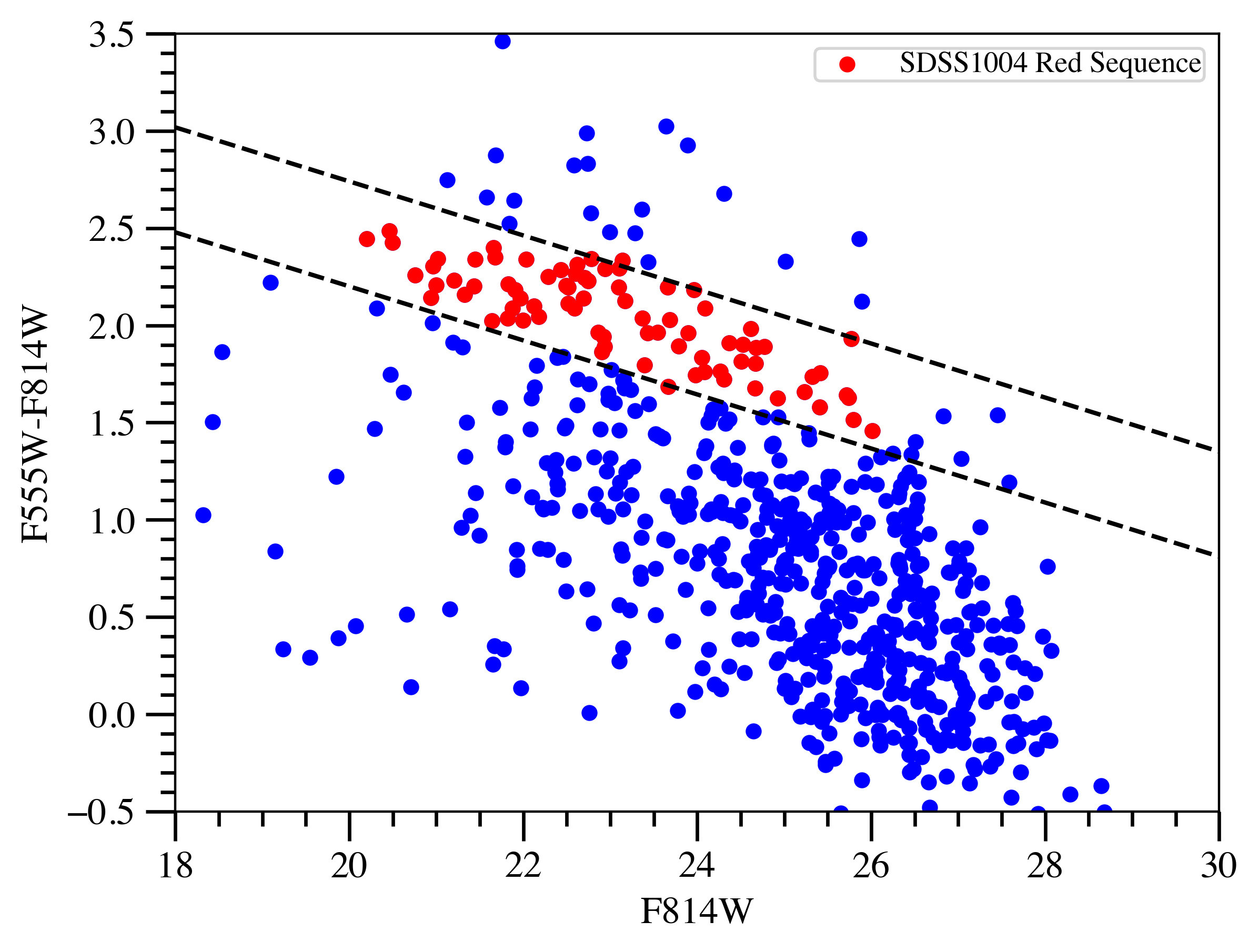}}
{\includegraphics[width=0.325\textwidth]{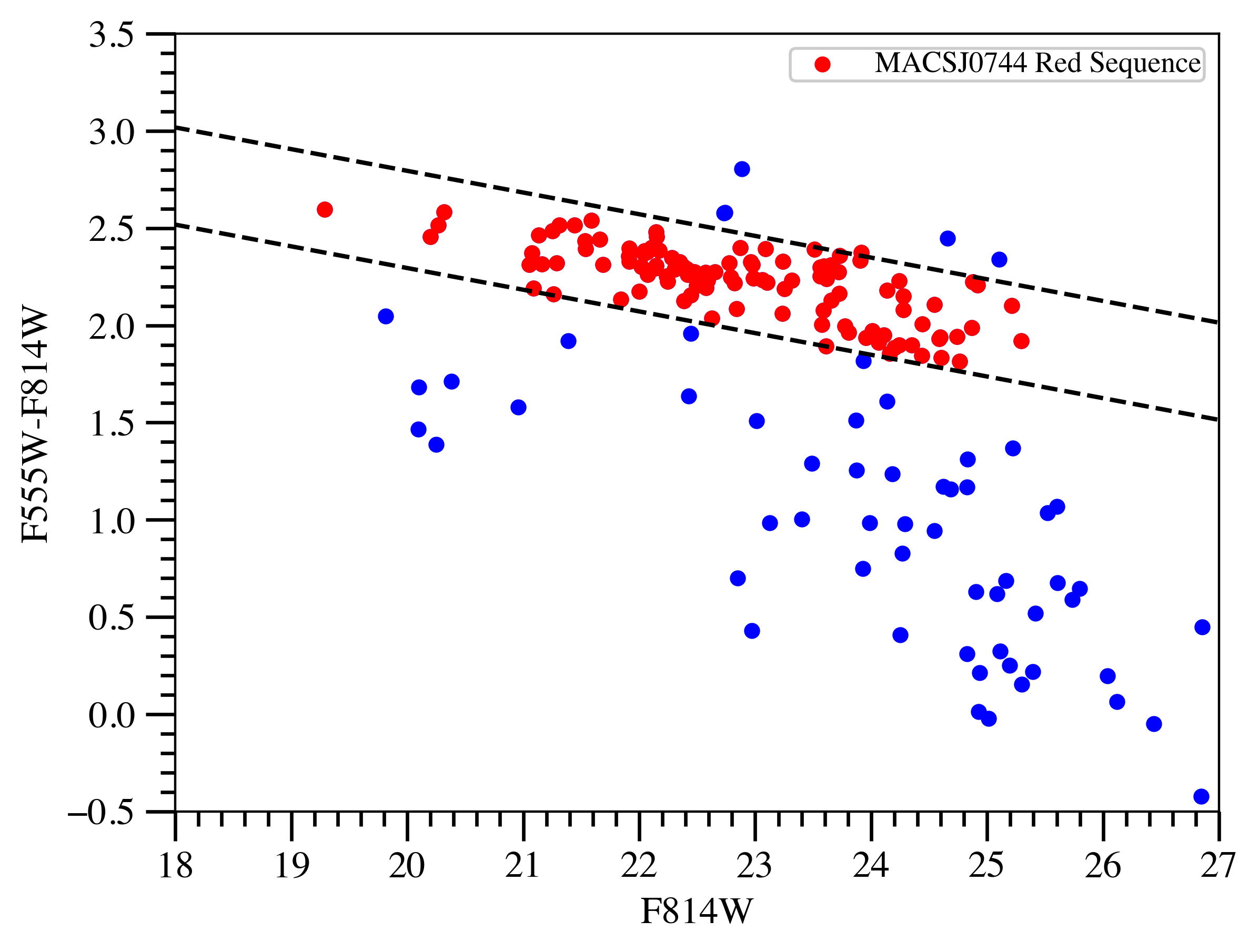}}
{\includegraphics[width=0.325\textwidth]{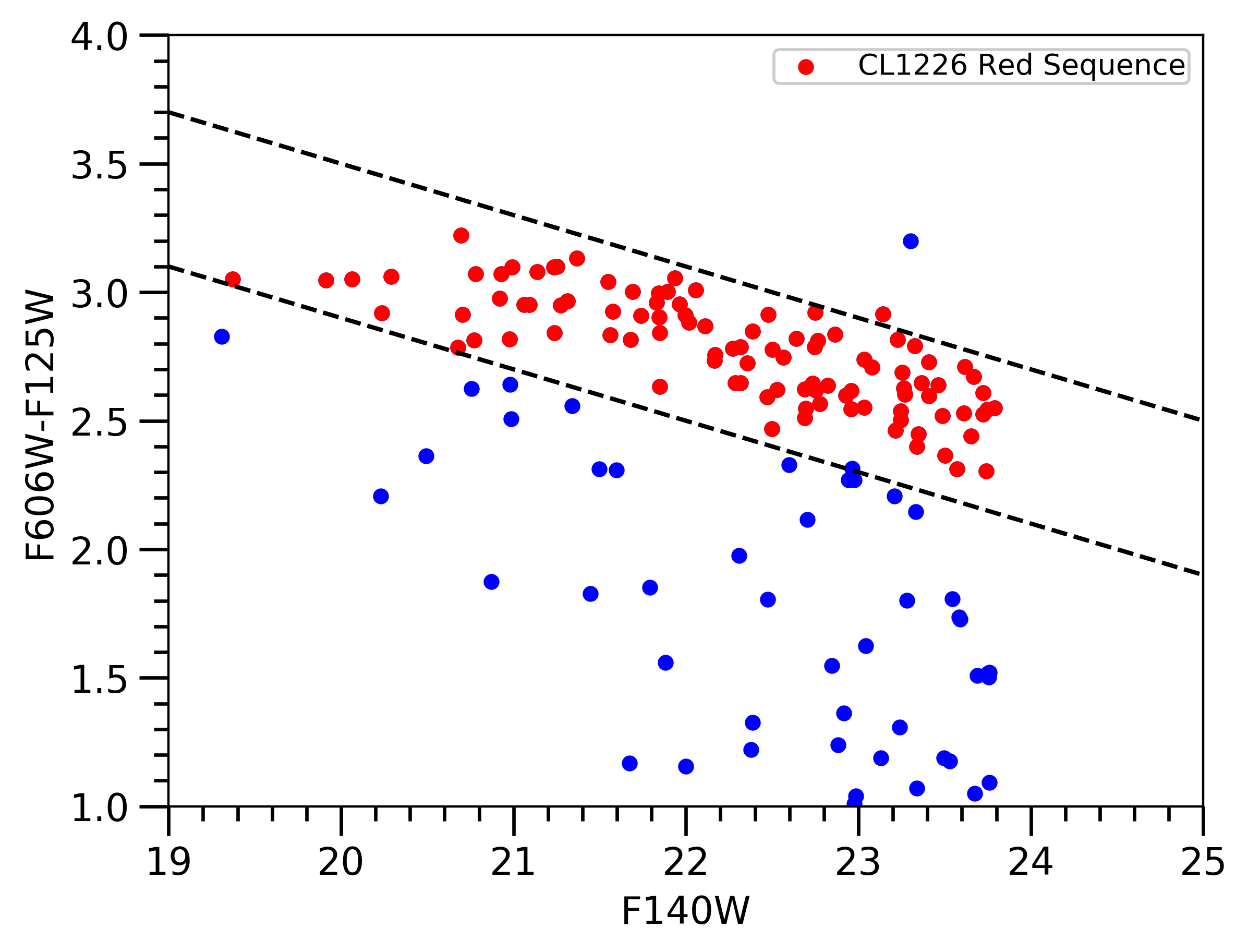}}
{\includegraphics[width=0.325\textwidth]{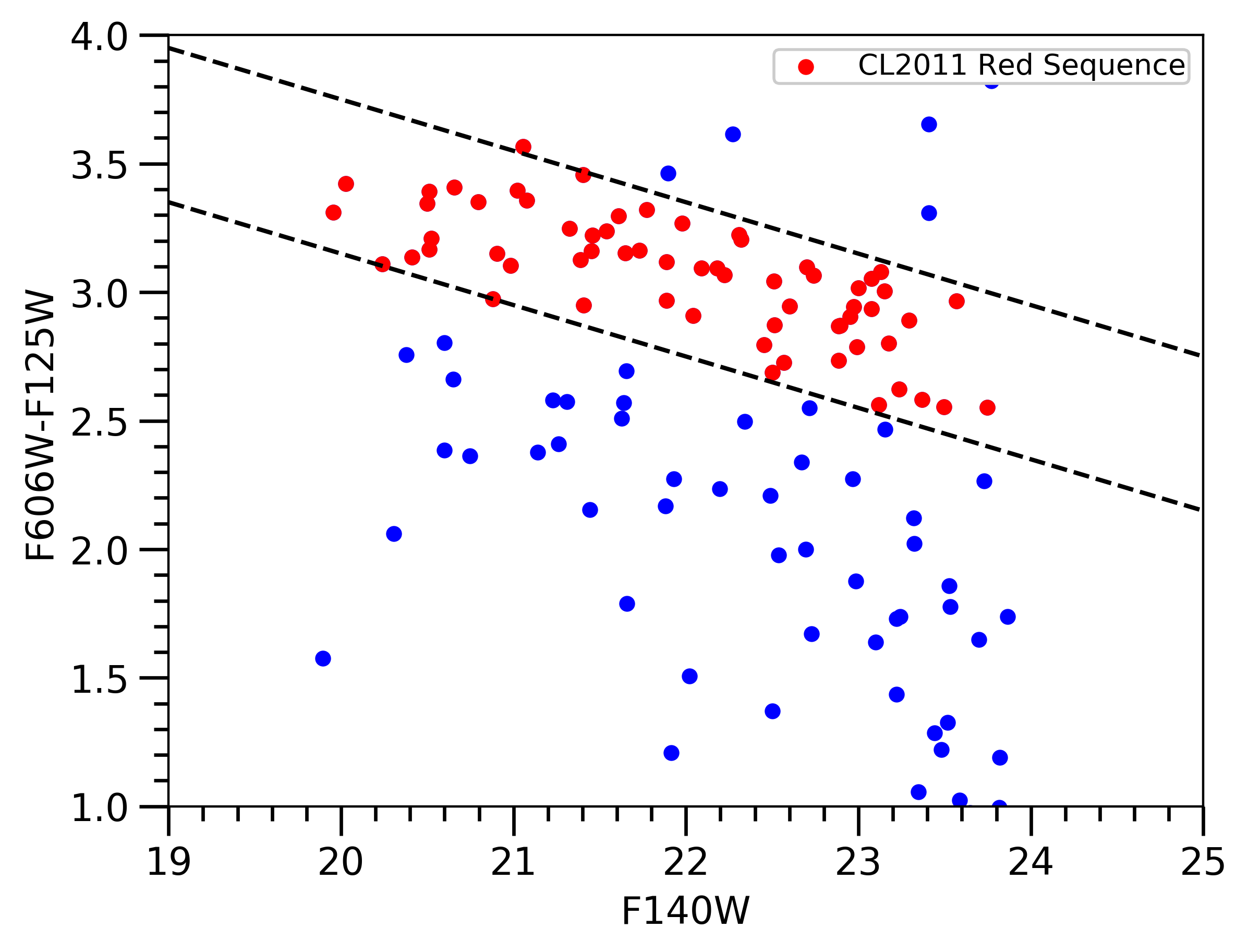}}
\caption{U-band (rest-frame) color-magnitude diagrams for clusters analysed in this paper. The red sequence is denoted with the red filled circles within the dashed lines and have photometric uncertainties of $<0.05$ magnitudes in their optical colors. RCS2327 did not have any $u$-band (rest-frame) data available to construct a corresponding CMD.}
\label{fig:2}
\end{figure*}

As in our previous work we select a sample of passive
cluster ETGs using red sequence galaxies (spectroscopic samples -- e.g., \citealt{rozo2015,depropris2016} -- confirm that red sequence galaxies have $>95\%$ likelihood of being quiescent cluster members, usually having morphologies consistent with their being early-type systems). The initial red sequence in all clusters is selected from an optical color - $F606W-F814W$ for Abell 2744/Abell S1063/Abell 370/MACSJ0416 MACSJ1149/MACSJ0717, $F555W-F814W$ for SDSS1004, $F125W-F160W$ for RCS2327/MACSJ0744 and $F125W-F140W$ for CL1226/CL2011. Optical photometry for all clusters was performed using Source Extractor (\citealt{bertin1996}) to measure the \cite{Kron1980} style total magnitude and an aperture magnitude within a fixed 10 kpc diameter for each galaxy in the image. The optical color-magnitude diagrams (CMDs) and the identification of red sequence galaxies for all clusters are shown in Fig. \ref{fig:1} (apart from SDSS1004, which did not have data in multiple optical bands). All clusters exhibit a tight red sequence in their CMD with a scatter of roughly $\pm$0.1 mags as is typical of ETGs. This red sequence is observed up to the point where the likelihood of contamination from non-cluster members and dusty foreground galaxies increases significantly, as seen in Fig. \ref{fig:1}. The two thick lines around the red sequence identify the likely early-type cluster members we analyze in this paper.

We then further constrained our sample to exclude
objects with even low levels of possible residual star
formation, by deriving red sequences in (rest-frame)
$u$-band CMDs for the optically selected red sequence
galaxies, and excluding objects bluer than the derived
red sequence in these $u$-band colors. The use of
multiple colors, particularly the $u$-band allows
for the selection of a truly passive sample of ETGs
(\citealt{ali2019,phillipps2020}). The standard $u$-band is very sensitive to even low levels of star formation, while contamination from the putative UV upturn sources is relatively small, allowing us to discriminate against red sequence galaxies with possible on-going low-level star formation. The $u-r$ vs. $r$
CMDs for all clusters are shown in Fig. \ref{fig:2}
(apart from RCS2327, which had no rest-frame $u$-band
data available). We finally check by eye all galaxies
that survived both color cuts and classify them, to reject any that 
may have non-early type morphologies, in order to
maintain a likely purely passive sample of galaxies. Several of these clusters are common to our earlier work \citep{depropris2016} where we have measured the Sersic indices of red sequence galaxies, showing that they are consistent with those of local clusters ETGs (at least to $z=0.8$). This three-fold selection method allows us to discriminate truly quiescent early-type cluster members from all
other interlopers, thus presenting us with a sample of
galaxies for which the UV SED will almost entirely be
dominated by the upturn, if present.

The rest-frame $\sim2400-optical$ color in all
clusters can then be used to probe the near-UV 
output of the selected galaxies and gauge its
evolution. While this color is clearly not as
efficient for analysing the upturn as the classical
$FUV-optical$ colors used in our previous studies
\citep{ali2018a,ali2018b,ali2018c} and in other works,
it still retains reasonable sensitivity to flux from
sources of the upturn, as will be explored later. The
UV photometry was carried out using $IRAF$'s aperture
photometry package, centred at the RA and DEC of each
galaxy's optical position, within a metric 10 kpc
diameter aperture after properly aligning the UV and
optical images. This ensures the same physical
aperture sizes for the measurement of $UV-optical$
colors across all clusters, irrespective of redshift.
A 5$\sigma$ signal-to-noise photometric cut was then made and each
object checked by eye to ensure their detection. In
all clusters we detect galaxies in the UV about
$2\sim3$ magnitudes fainter than $M^*$ in the luminosity
function, as seen in Fig. \ref{fig:3}. However, for
CL1226/CL2011, due to the large distance and
relatively limited exposure times, we directly detect
only the brightest galaxies in these clusters (above
$M^*$). To overcome this limitation, for both clusters
we stacked all galaxies per magnitude bin in the UV from 
their optical positions between
$21<M_{F125W}(r)<24$ (i.e. 3 bins) where very few
galaxies are directly detected. For each magnitude
bin the stacking procedure yielded at least a
$5\sigma$ detection, thus allowing us to push beyond
the direct detection limit and calculate the average
$UV-optical$ colors of the galaxies beyond $M^*$ in
these higher redshift clusters and compare them to
their lower redshift counterparts. 

Furthermore, we also consider the potential of contamination of our sample by dust-reddened green valley or blue cloud galaxies. In a previous study, \cite{phillipps2020} sought to define truly passive galaxies in the GAMA survey that exhibit a UV upturn by using a multi-band color cut similar to that employed in this paper in order to isolate even low level star-forming galaxies. For the galaxies that survived the initial optical/u-band cuts, a further cut was made in the WISE $W2-W3$, which is a completely independent measure of star-formation (essentially from re-processed UV light) with minimal effect from internal dust extinction on the IR light directly. This color allows for the identification of potential star-forming galaxies that may have been dust-reddened into the red sequence and thought to be passive systems (\citealt{fraser2016}). Almost all galaxies that survived the initial selection but not the IR cut had $NUV-r\lesssim5$. Similar $NUV-r$ cuts have also been used in other studies to identify potential green valley galaxies and residual star-formers from true ETGs (e.g. \citealt{salim2014,crossett2014}). As such we check all clusters in our sample to identify the number of red sequence members that have colors bluer than $NUV-r<5$ (by using a simple SSP as described in the next sub-section to transform GALEX $NUV-r=5$ to the equivalent rest-frame $NUV-optical$ color of each cluster). We find that the number of galaxies that fall into this criterion for all clusters is roughly $\sim5\%$ or less, and even then most of these blue galaxies are not significantly bluer than $NUV-r=5$ (within approximately 0.5 mags). As such, we do not explicitly reject them from our sample, but these galaxies are identified later (in box plots) as statistical outliers in the overall UV color distribution of clusters.

\subsection{Extinction and k-corrections}
\label{sec:kcorr}

\begin{deluxetable}{ccccccc}
\tablenum{2}
\tablecaption{Extinction and k-corrections for all clusters. K-corrections are used for Fig. \ref{fig:yeps} and \ref{fig:chem}.\label{table2}}
\tablewidth{0pt}
\tablehead{\colhead{Cluster} & \colhead{Observed} \vspace{-0.2cm} & \colhead{Rest-frame} & \colhead{Extinction} & \colhead{K-correction}\\  \colhead{} & \colhead{Color} & \colhead{Color} & \colhead{Correction} & \colhead{}}
\startdata
CL2011      & F475-F125 & 2400-r    & 0.168  	&   0\\     
CL1226	    & F475-F125 & 2500-r    & 0.049	&   0.31\\
RCS2327	    & F435-F814 & 2550-g    & 0.162	&   0.93\\
MACSJ0744   & F435-F814 & 2550-g    & 0.12	&   0.96\\
SDSS1004	& F435-F814 & 2550-g    & 0.038	&   0.99\\
MACSJ2129   & F390-F814 & 2450-g    & 0.18	&   0.58\\
MACSJ1149   & F390-F814 & 2500-V    & 0.054	&   0.6\\
MACSJ0717   & F390-F814 & 2500-V    & 0.182	&   0.6\\
MACSJ0416	& F336-F814 & 2400-r    & 0.119	&   -0.12\\
Abell370	& F336-F814 & 2450-r    & 0.095	&   -0.02\\
AbellS1063	& F336-F814 & 2500-r    & 0.035	&   0.1\\
Abell2744	& F336-F814 & 2550-r    & 0.039	&   0.2\\
\enddata
\end{deluxetable}

\begin{figure}
\includegraphics[width=0.49\textwidth]{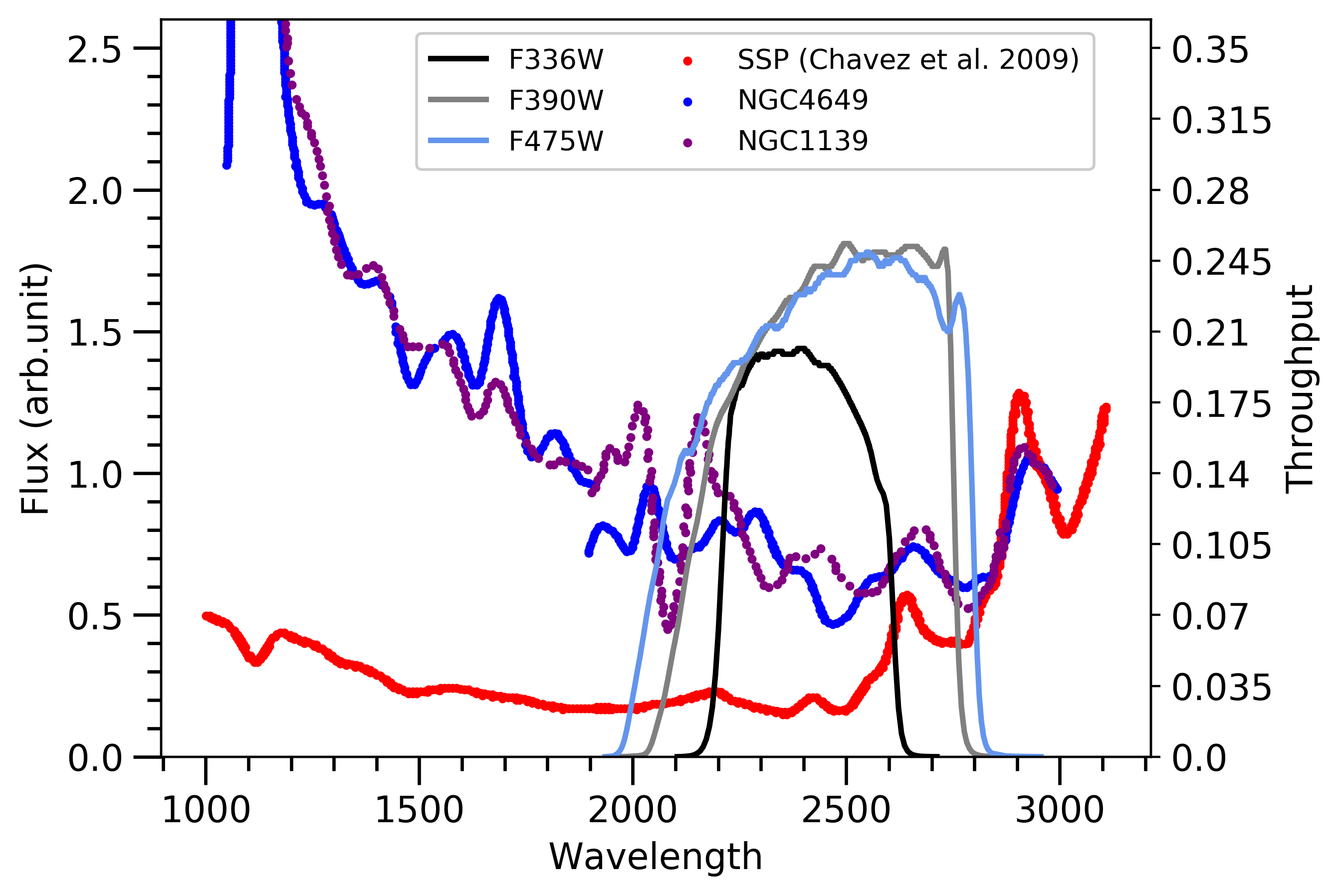}
\caption{IUE far and near-UV SEDs of the giant elliptical galaxies NGC1139 and NGC4649 from \protect\cite{chavez2011}. The purple and blue curves show the observed UV upturn in their spectra while the red curve is a theoretical spectrum from a stellar population template of solar metallicity and age of 10 Gyrs from \protect\cite{chavez2009}, which represents an ETG with no UV upturn component. Plotted on top are the filter response curves of three filters used for the near-UV observations of the clusters studied in this paper, to show their coverage in the upturn dominated region of the spectra in ETGs.}
\label{fig:sed}
\end{figure}

Extinction corrections are taken from the NASA/IPAC database\footnote{https://ned.ipac.caltech.edu}, which uses the reddening maps from \cite{schlafly2011} and extrapolates to the UV following the Milky Way extinction law of \cite{Cardelli1989}. For simplicity, we assume all galaxies within a cluster to have the same line-of-sight extinction and given the distance to each cluster, there should be little variation in extinctions from galaxy to galaxy. The extinction corrections mode to the $UV-optical$ colors for all clusters are given in Table \ref{table2}. We also assume that all red sequence members have little to no internal dust content, as is the case for ETGs in general and as demonstrated through model fitting of Coma red sequence members in \cite{ali2018a}, where $A{_V}$ was found to be $<0.1$ in most cases. In agreement with this assumption, a measurement for nearby ETGs from the SDSS returns a mean internal extinction $A_z$ of 0 \citep{Barber2007}. Observations in several nearby ETGs from {\it Astro UIT} and {\it Astro HUT} also imply very low internal extinction ($E(B-V) < 0.03$ -- \citealt{Ferguson1993}) in the $FUV$ bandpass, while \cite{Rettura2006,Rettura2011} also find internal extinction values $E(B-V) < 0.05$ (and consistent with zero) for the majority of ETGs in clusters up to $z=1.3$.

The clusters we use in this study were carefully selected such that despite their varying redshifts, they have observations in filters (of increasing wavelength at higher cluster redshifts) that probe roughly the same wavelength range in the rest-frame UV (centred at $\sim$2400\AA). This was done in order to avoid the large k-corrections that become necessary when using the same filter to probe the upturn across large redshift ranges. It is particularly important to minimise the magnitude of the k-corrections in the UV, where the spectra of ETGs are not as rigorously studied, and the stochastic nature of the upturn (which causes a large scatter in the UV colors compared to optical) introduces uncertainty to the k-corrections applied to each galaxy without a priori knowledge of their spectra, which we of course do not have.

Despite our careful selection of clusters, it is still necessary to perform some k-corrections to the data. This k-correction is driven by two key factors - 1) the difference in the rest-frame wavelengths covered by the filters; and 2) the difference in shape of the filters used to observe the rest-frame UV. These two elements are illustrated visually in Fig. \ref{fig:sed}, where the different filters used do not exactly align in the wavelength regime and have varying shapes in their response curves. The k-correction term we apply seeks to correct for these two factors. However, given that the UV data for all clusters are centred between $2400\sim2500$\AA, k-corrections only need to account for about a $100$\AA\ shift at maximum in the wavelength regime, so the correction is mostly dominated by the varying shape of the filters (i.e. bandpass correction).

To perform this correction, we use the YEPS spectrophotometric models of \cite{chung2017} which seek to replicate the UV output of ETGs through He-enhancement (any model that reproduces the observed UV SED could instead have been used, yielding similar results). We take two SSPs, both of which have $Z$=\(Z_\odot\), $z_f=4$ and an age of 12 Gyrs, but with $Y_{ini}$\footnote{The helium abundance $Y$ of a stellar population is related to the initial helium abundance $Y_{ini}$ and the metallicity $Z$ through the following equation: $Y = \Delta Y/\Delta Z \times $ $Z$\, +\, $Y_{ini}$, where $\Delta Y/\Delta Z$ is the galactic helium enrichment parameter, assumed to be 2.0.}$=0.23$ (no upturn) and $Y_{ini}=0.43$ (maximum upturn). The SSPs are shifted to the cluster redshift and k-corrections from both SSPs are calculated in the observed colors; then averaged. An average of the two SSPs is used as the vast majority of our observed galaxies have colors that lie between those of the aforementioned SSPs (see Fig. \ref{fig:chem} for the color evolution of the SSPs). Due to the stochastic nature of the upturn and large scatter in the UV colors, using a single k-correction value for all galaxies in a cluster introduces an uncertainty to the results, as not all galaxies will be fit by the same SSP, which is an issue that cannot be mitigated without detailed knowledge of the UV SED of each galaxy. However for any given galaxy, we find that the k-correction term used has an error of approximately $\pm 0.2$ mags, as the k-correction calculated using the $Y_{ini}=0.23$ (no upturn) and $Y_{ini}=0.43$ (maximum upturn) SSPs, which form the two extreme cases for what the galaxies' UV colors can be, only vary by $\sim \pm 0.2$ mags from our adopted values.

The k-corrections as described above are particularly important for Fig. \ref{fig:yeps}, where we directly compare the evolution of the rest-frame $UV-optical$ colors of cluster galaxies across redshift (will be discussed in detail in later sections). We apply the k-correction to all clusters to bring them in line with those of CL2011's observed $F475W-F125W$ at $z=0.9$ (rest-frame $2400-r$) and because the evolution of the models are also plotted in the observed (and redshifted) bands of CL2011. To elaborate further on the error in the k-correction term, we can take for example MACSJ1149, where the observed color is $F390W-F814W$ - roughly $2500-V$ in rest-frame. The k-correction calculated (as described earlier) to bring this color to the rest-frame $2400-r$ of CL2011 is 0.6 mags. Using the $Y_{ini}=0.23$ SSP, the k-correction is 0.74 mags and using the $Y_{ini}=0.43$ SSP yields 0.45 mags, which are both within 0.2 mags of our adopted correction of 0.6.

The k-corrections as applied to all clusters in Fig. \ref{fig:yeps} and \ref{fig:chem} are shown in Table \ref{table2}. The clusters for which the rest-frame color is closer to CL2011 have smaller k-corrections as expected. No age/evolutionary corrections are made as these figures seek to analyse the age evolution of the UV colors in cluster galaxies.

A similar k-correction is also applied in each of the redshift bins of Fig. \ref{fig:3} (detailed in next section). In each sub-plot, all clusters are k-corrected to the redshift of one selected cluster to account for the small redshift difference between clusters in each redshift bin. For example, in the top left sub-plot (which shows the colors of Abell 2744 at $z=0.31$ and Abell S1063 at $z=0.35$), a k-correction is applied to the Abell S1063 galaxies to bring them in line with the rest-frame colors of Abell 2744 (accounting for the redshift difference of 0.04). Similarly for the rest of the sub-plots the clusters to which the others were k-corrected to are MACSJ0416, MACSJ1149, RCS2327 and CL2011. It should be noted that in each case the k-correction is rather small ($<0.2$ mags in most cases) as the redshift difference between the clusters in each bin is 0.05 or less and because the clusters in each bin are observed using the same filters.

\section{Results}

\subsection{Near-UV color-magnitude diagrams}

\begin{figure*}
\centering
\includegraphics[width=0.49\textwidth]{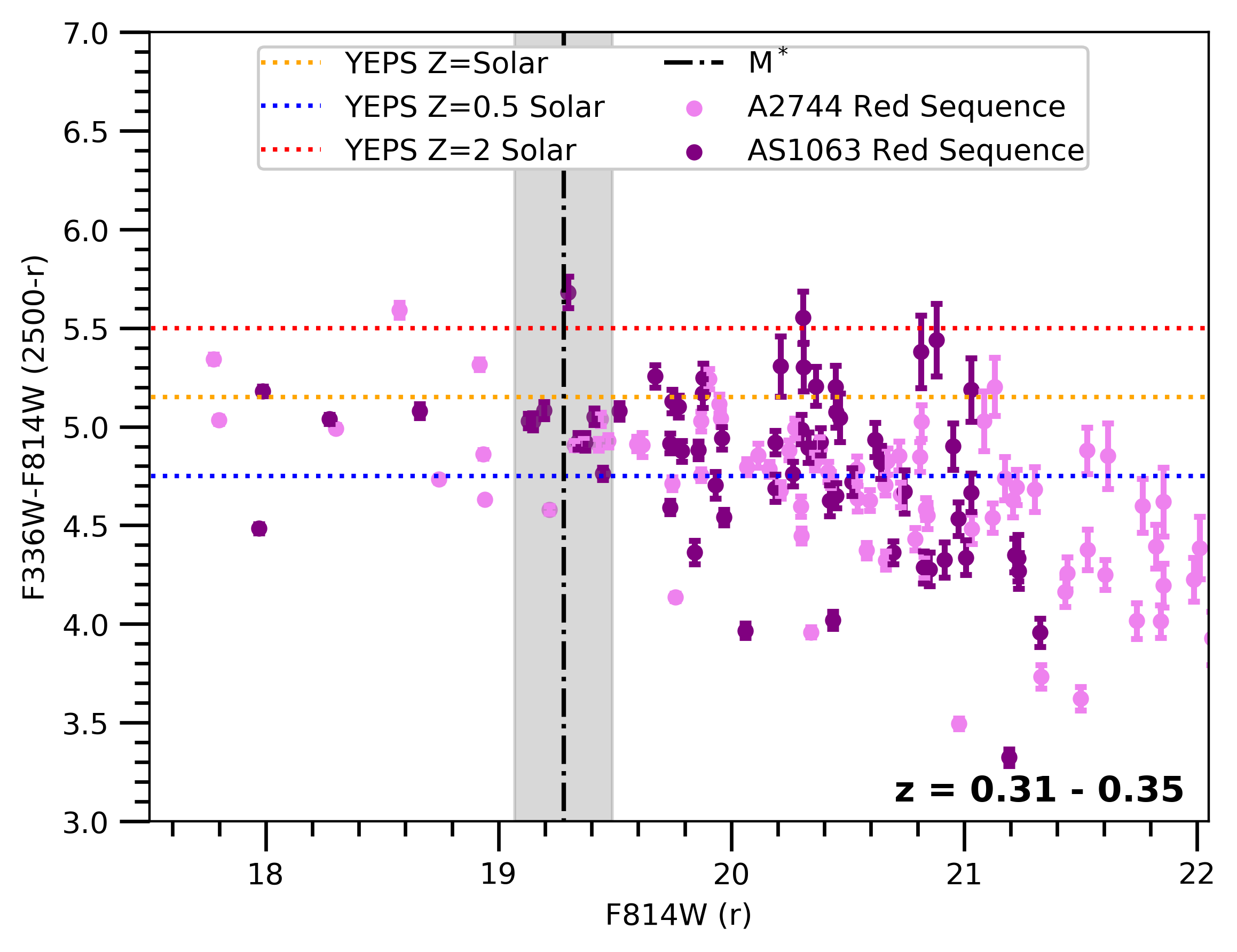}
\includegraphics[width=0.49\textwidth]{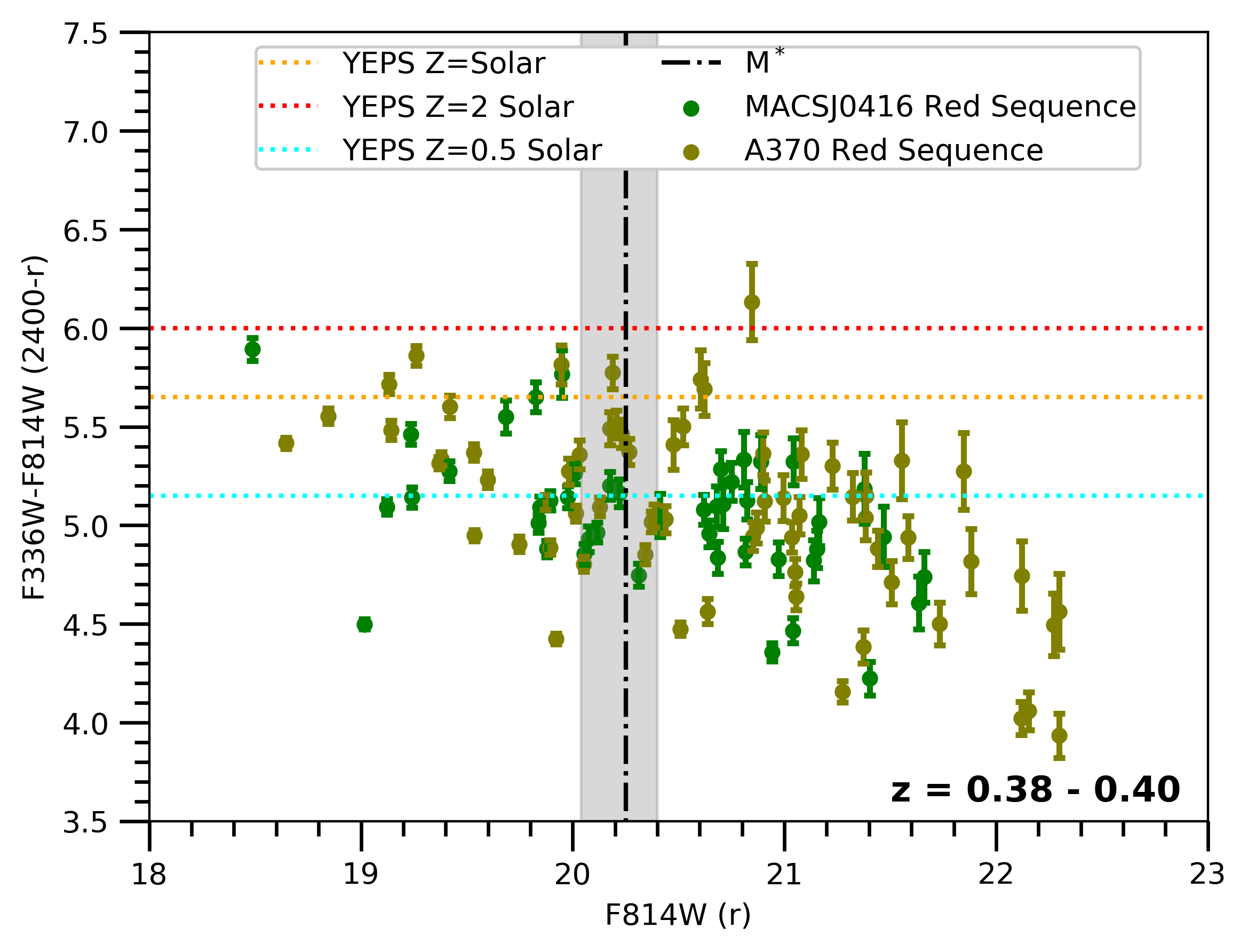}
\includegraphics[width=0.49\textwidth]{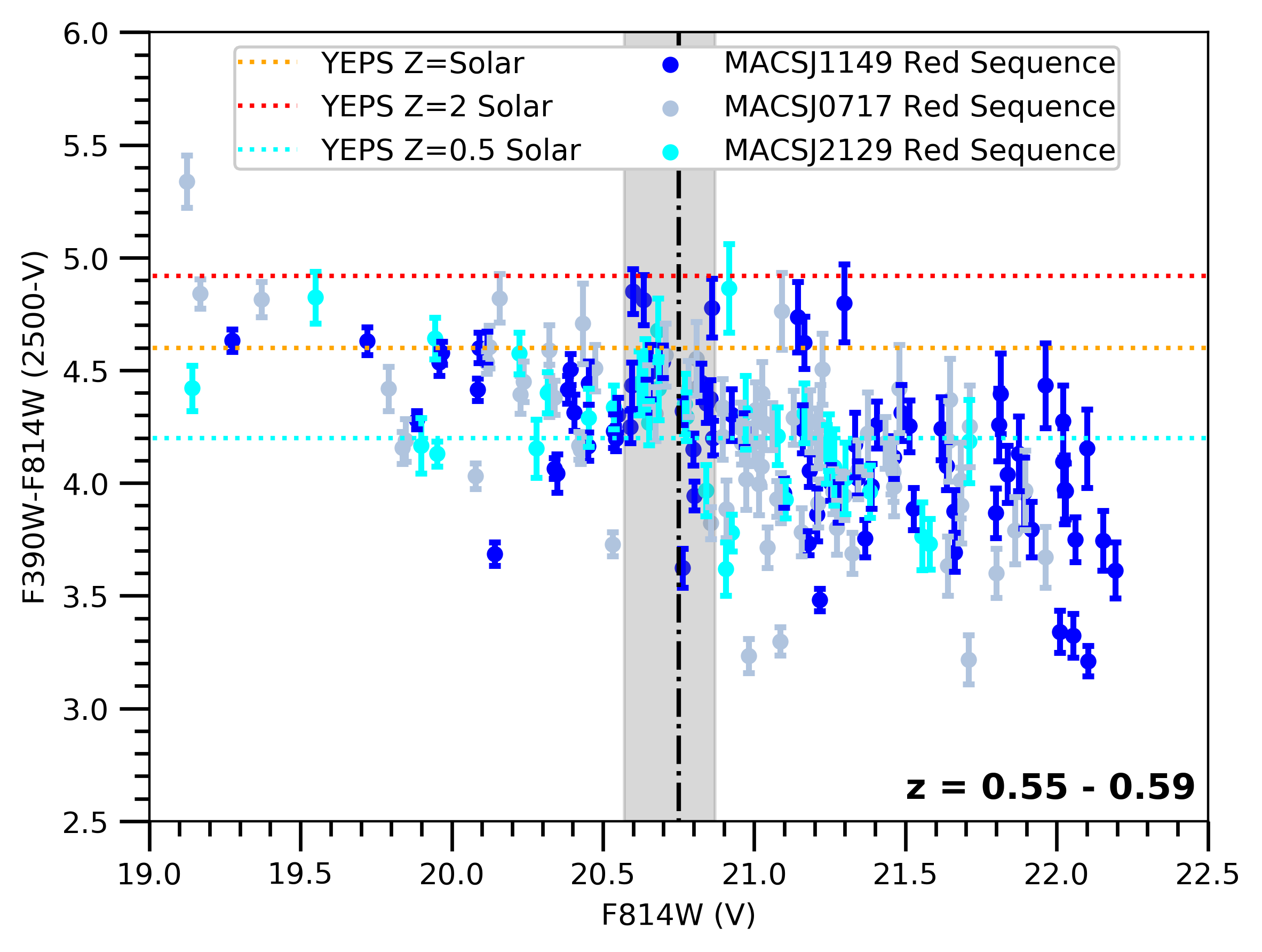}
\includegraphics[width=0.49\textwidth]{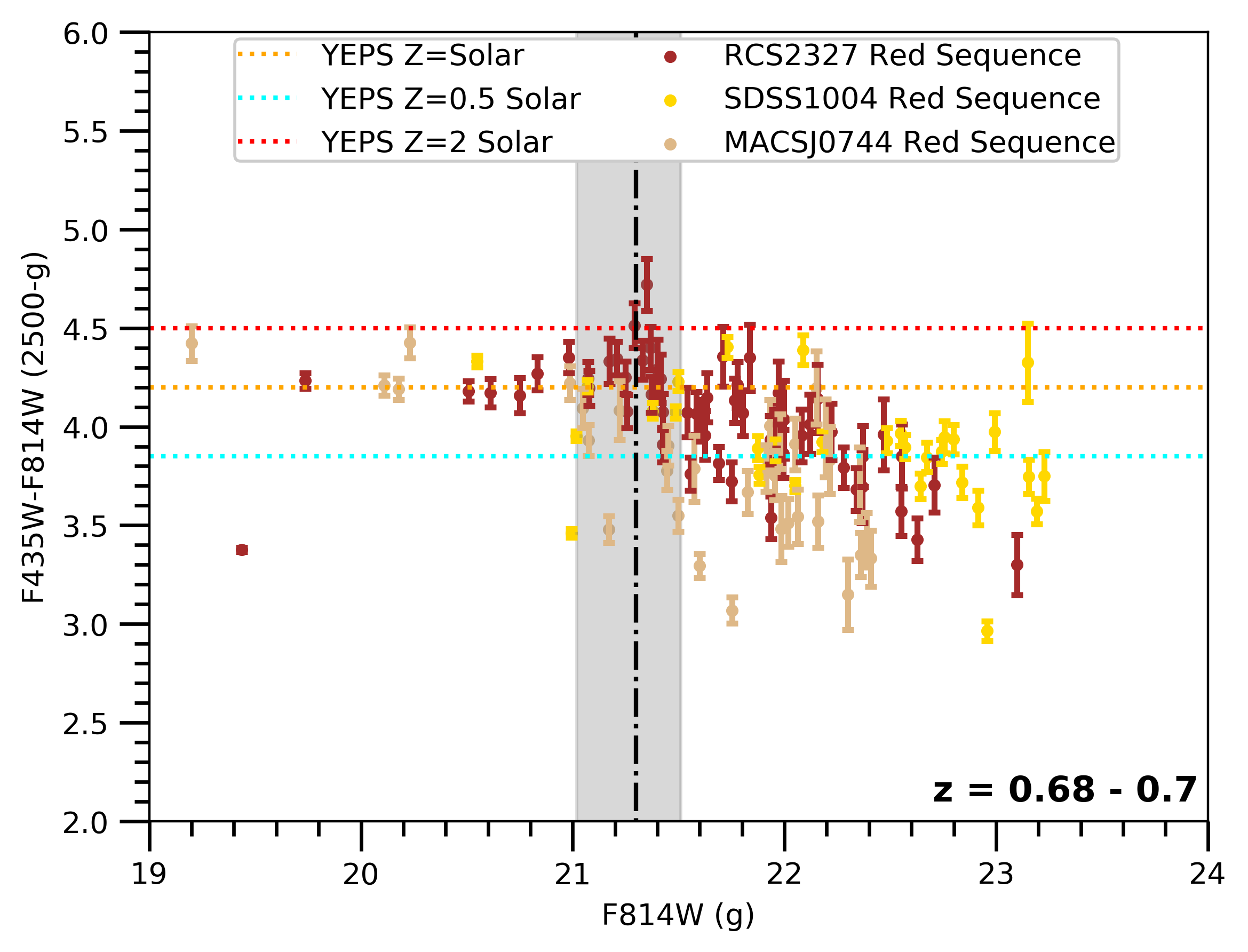}
\includegraphics[width=0.49\textwidth]{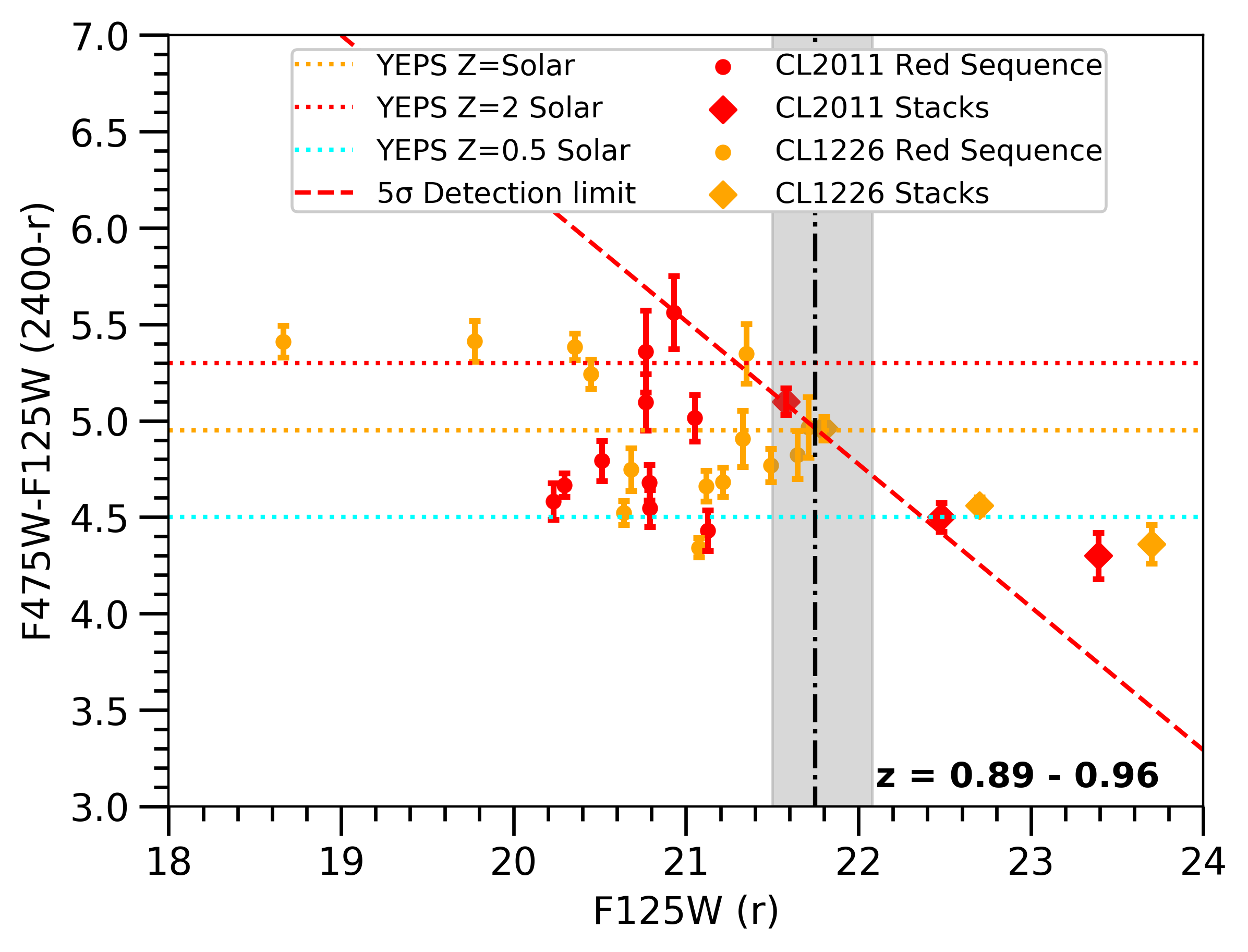}
\caption{Rest-frame $2400\sim2500-optical$ CMDs of clusters at five redshift bins between $0.3<z<1$ as denoted in each figure. The vertical black line represents the respective cluster $M^{*}$ and the shaded region denotes the error in the measurement, taken and adjusted from \protect\cite{connor2017} and \protect\cite{depropris2013}. The horizontal dotted lines in each plot are the colors as given by an infall model from the \protect\cite{chung2017} with cosmological He ($Y_{ini}=0.23$) and $z_f$=4, but varying metallicities at $Z$=\(Z_\odot\), 0.5\(Z_\odot\) \& 2\(Z_\odot\).}
\label{fig:3}
\end{figure*}

We present in Fig. \ref{fig:3} the rest-frame $2400-r$ or $2500-g$ / $2500-V$ colors for selected cluster ETGs (as previously described) in 5 redshift bins: Abell 2744/Abell S1063 at $z=0.31\sim0.35$ (top left), Abell 370/MACSJ0416 at $z=0.38\sim0.40$ (top right), MACSJ1149/MACSJ0717/MACSJ2129 at $z=0.55\sim0.59$ (middle left), SDSS1004/RCS2327/MACSJ0744 at $z=0.68\sim0.7$ (middle right) and CL1226/CL2011 at $z=0.89\sim0.96$ (bottom). For the redshift bins this rest-frame color lies in the observed $F336W-F814W$, $F390W-F814W$, $F435W-F814W$ and $F475W-F125W$ respectively. For each set of clusters, also plotted is the prediction for the same color from the YEPS (Yonsei Evolutionary Population Synthesis) infall model of \protect\cite{chung2017} with $Z$=\(Z_\odot\), 0.5\(Z_\odot\), 2\(Z_\odot\), a fixed redshift of formation - $z_f=4$, cosmological helium ($Y_{ini}=0.23$) and a delta burst star-formation history (i.e. all stars formed instantly at the same age). These models represent the UV colors of galaxies with no upturn component (and standard composition) - the model colors are purely driven by age and metallicity effects, and are comparable to results from other population synthesis models that do not treat the UV upturn and reproduce the typical optical and infrared colors of ETGs even at these redshifts -- e.g., see \cite{mei2006}.

While the upturn has been classically analysed using photometry at shorter wavelengths, particularly the $FUV$, previous studies (\citealt{schombert2016}; \citealt{phillipps2020}) have shown that even the $NUV$ flux of quiescent ETGs is mostly dominated by the upturn. Given that the UV wavelength probed in this study is extremely close to the classical GALEX $NUV$ waveband (centred at 2250\AA), we expect the near-UV bands (centred at $\sim$2400\AA) to also contain significant flux from the upturn sub-population, albeit with a stronger contribution from the short wavelength tail of the blackbody emission of the majority main sequence/red giant branch stellar population. To visualise this, we plot in Fig. \ref{fig:sed} the UV SEDs below 3000\AA\ of the giant elliptical galaxies NGC1139 and NGC4649 (\citealt{chavez2011}) as taken by the IUE. These are galaxies known for a strong UV upturn. Also plotted for comparison is the SED of a 10 Gyr old SSP with solar metallicity from \cite{chavez2009}, which acts as a representation for ETGs with no upturn component (such a model well fits the optical and infrared colors of normal ETGs in the local universe and even their evolution to high redshift). Superimposed on top of the SEDs are the response curves of the filters used to make the UV observations of the clusters analysed in this paper. From the figure we can see that at $>3000$\AA, the SEDs of the observed ETGs and the reference model are nearly identical, while at $<3000$\AA, the upturn starts to contribute and the UV flux rises with decreasing wavelength up to the Lyman limit in the observed galaxies. Conversely the UV SED of the theoretical spectrum is nearly flat with almost no flux, as would be expected from a purely quiescent stellar population. Even in the near-UV region probed in this study, there is $\sim5-6$ times more flux from galaxies with upturn than one without. As such we expect this wavelength region to still be reasonably sensitive to the upturn, while remaining vigilant about the contribution from the standard stellar population in ETGs in our analysis.

In Fig. \ref{fig:3} we see in the first four redshift bins that the colors of a majority of our observed galaxies cannot entirely be reproduced with composite
stellar populations (CSPs) of solar metallicity or higher, as is otherwise expected for standard ETGs of this luminosity. A large number of the brightest galaxies in our sample (around or above $M^*$) have UV colors that can only be reproduced with metallicities below solar, which is implausible given that the brightest cluster ETGs are known to have solar or super-solar abundances from their optical spectra/colors. Even lower mass galaxies that are several magnitudes below $M^*$, and are naturally expected to be bluer because of their lower metallicities \citep{gallazzi2005,gallazzi2014}, are much bluer than the half-solar CSP in this near-UV color. In both instances of high and low mass galaxies, this suggests that metallicity (and age) alone cannot account for the extremely blue UV colors as observed in our galaxies. Instead there needs to exist a `second parameter' that drives the bluer UV colors in these otherwise quiescent galaxies. Interestingly however, in the final highest redshift bin, most galaxies across the luminosity function seem to be reasonably fit with 0.5\(Z_\odot\) $< Z <$ 2\(Z_\odot\) models with no upturn. Even in the fainter galaxies, including stacks (containing the average flux of galaxies with optical luminosities well below $M^*$), the colors are largely consistent with models down to $Z\sim0.5$\(Z_\odot\), as would be reasonably expected from the mass-metallicity relation (\citealt{gallazzi2005,gallazzi2014}). Virtually no galaxy brighter than $M^*$ (and very few overall) in these two $z\sim 0.9$ clusters has $2400-r \lesssim 5$ while most galaxies in the lower redshift clusters are this blue or even bluer. Any such object (if it existed) would be easily detected in these data to well below the $M^*$ luminosity. Similarly, if such blue objects existed they would drive the colors of galaxy stacks in these two higher redshift clusters to the blue, but this is not observed. This suggests that at this redshift the upturn is significantly weakened, in agreement with the findings of \cite{lecras2016}, using near-UV spectral indices. We cannot however exclude that a small upturn component may still be present for the brighter, more He-rich galaxies which formed earlier (as in Coma -- \citealt{ali2018a}) and as recently discovered by \cite{lonoce2020} for one $z=1.4$ galaxy in the COSMOS field (although this is a 2$\sigma$ detection based on a single index while all other indices are consistent with no upturn), but our data in these two $z \sim 1$ clusters are consistent with a model where the UV upturn component clearly observed at low redshift has now largely if not totally disappeared. 

\subsection{Intrinsic scatter}

\begin{figure}
\includegraphics[width=0.49\textwidth]{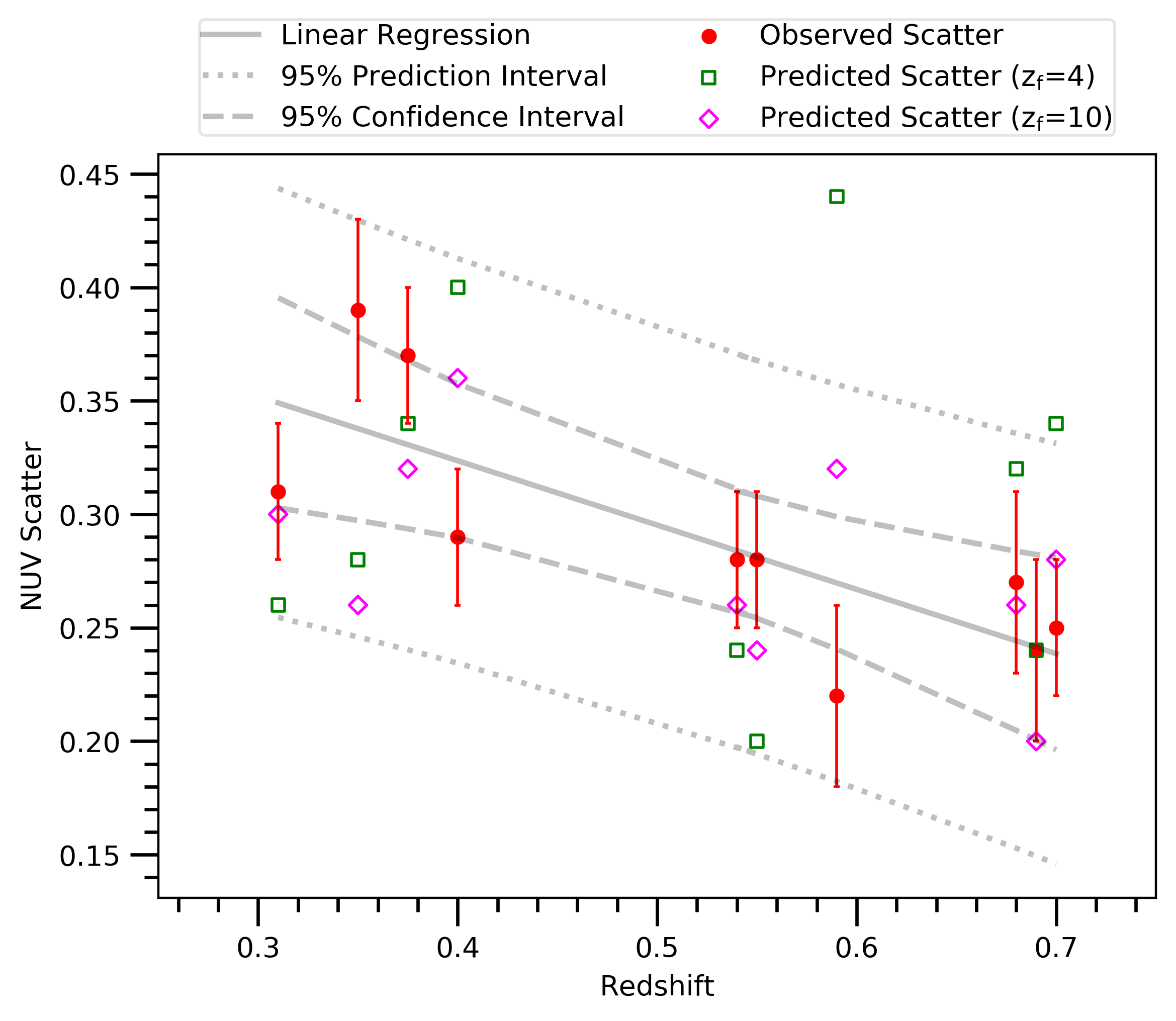}
\caption{Observed scatters about the near-UV color magnitude relations of Fig \ref{fig:3} vs. redshift of cluster. Also plotted are the predicted near-UV scatters for each cluster assuming a pure age spread around the color-magnitude relation from the optical data for $z_f=4$ (green squares) and $z_f=10$ (pink diamonds). Also shown is a linear fit to the data and the 95\% confidence and prediction intervals around the fit.}
\label{fig:scatter}
\end{figure}

Another parameter of the CMDs that can be used to probe the strength of the upturn is the intrinsic scatter in the UV colors. The scatter in the optical colors of cluster ETGs is generally found to be extremely tight and of the order $\sim0.1$ mags in $g-r$, due to their similar ages and metallicities, leading to little variance in their SEDs. However, the same galaxies in the $UV-optical$ colors exhibit several times larger scatter, e.g. $\sim1.5$ mags in $FUV-V$ (e.g. \citealt{ali2018a}). While we expect the scatter to naturally increase at shorter wavelengths, this increase can be predicted using the optical colors and compared with observations to check if the UV scatter can be attributable to simple age/metallicity effects as is the case in the optical regime. We use the Hyper-Fit software (\citealt{robotham2015}) to measure the intrinsic scatter in the optical and UV CMDs (as in Fig. \ref{fig:1} and \ref{fig:3}) for each cluster. The code assumes a Gaussian distribution in the data and no covariance between errors in magnitude and color to fit the best representative linear model to the data, and estimate the intrinsic scatter from the model. In Fig. \ref{fig:scatter} we plot the observed scatter in the near-UV color for all of our clusters (apart from CL1226/CL2011 as we only individually detect a small subset of the brightest galaxies) against redshift. Qualitatively, we can see that the scatter in the colors seem to show a general decrease with increasing redshift - the largest decrease being seen around the $z=0.7$ clusters, albeit with significant differences across the redshift range, possibly indicating cluster-to-cluster variations in the degree of helium enhancement or formation epoch. This variation is manifested statistically when we plot a linear fit to the data and the 95\% confidence/prediction intervals to the fit. The fit shows a general decreasing trend with redshift, however the confidence/predictions intervals are very wide, suggesting a large variation in cluster-to-cluster scatter. Despite this large variation, the scatter still shows a gradual decrease with redshift.

We then use a model from \cite{conroy2009} with parameters as described above to estimate the scatter in optical colors, assuming that the entire intrinsic scatter around the mass-metallicity relation is due to age, and from these derive the expected scatter in the $UV-optical$ colors for various assumed formation redshifts, which are also plotted in Fig. \ref{fig:scatter}. At $z < 0.6$ there is clear excess scatter in the near-UV colors in most clusters, for any assumed formation redshift. However, and in agreement with our earlier work and the analysis shown from simple models discussed later, at $z > 0.6$ the scatter in near-UV colors appear closer to the predicted scatter from the optical colors, or even have lower scatters than predicted for certain formation redshifts. This suggests a weakening of the UV upturn at these redshifts, as observed in \cite{ali2018c} and, by a different approach, \cite{lecras2016}.

One mechanism for this is that the UV upturn is produced by a population with non-cosmological helium abundance, with a significant fraction of stars having
helium abundances larger than 0.23, as observed in 
Galactic globular clusters (e.g., in NGC 2808). The most extreme He-enhanced stars contribute most to the
$FUV$ flux. As these stars disappear (when they have 
younger ages and are too massive to reach the
extreme HB), the remaining He-rich population cannot extend so far into the blue HB and
provides more and more flux to the NUV regime,
appearing gradually over the redshift range $0.6 
< z < 1$ as observed by \cite{lecras2016}.

\section{Discussion}

\subsection{Comparison with YEPS models}
\begin{figure*}
\includegraphics[width=\textwidth]{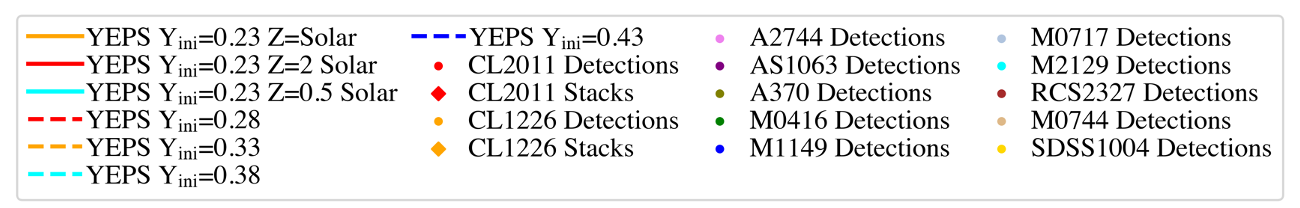}
\includegraphics[width=0.325\textwidth]{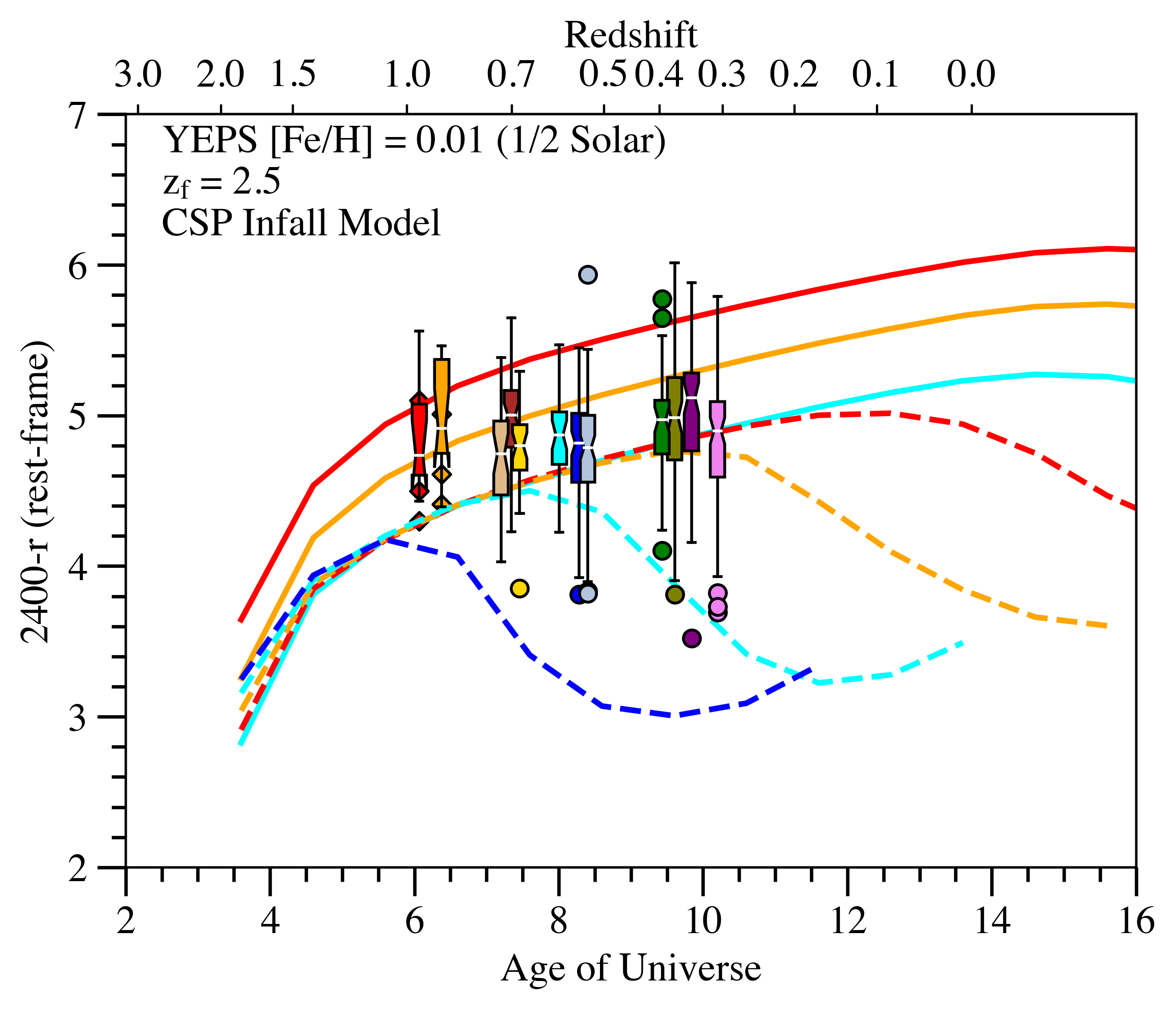}
\includegraphics[width=0.325\textwidth]{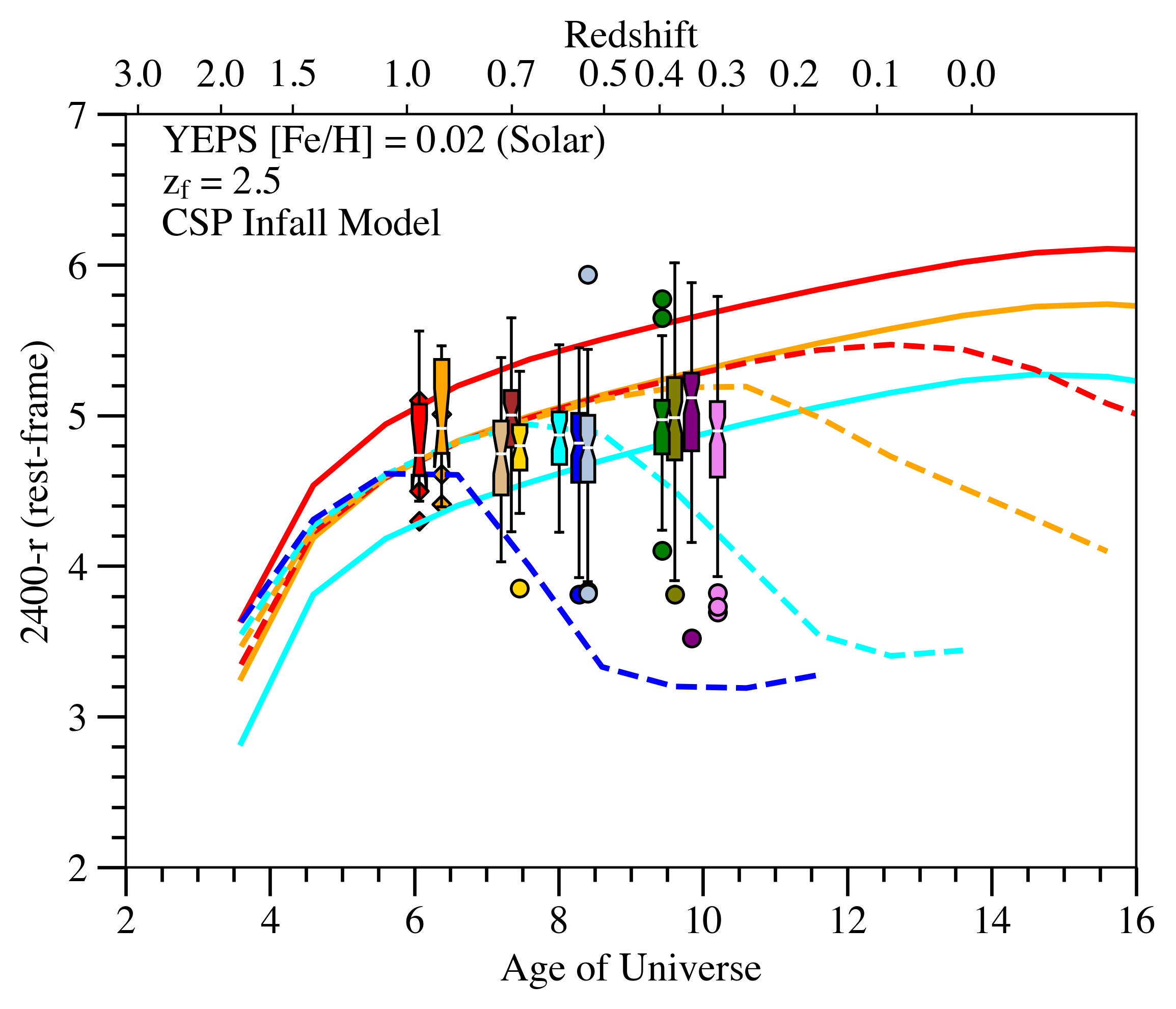}
\includegraphics[width=0.325\textwidth]{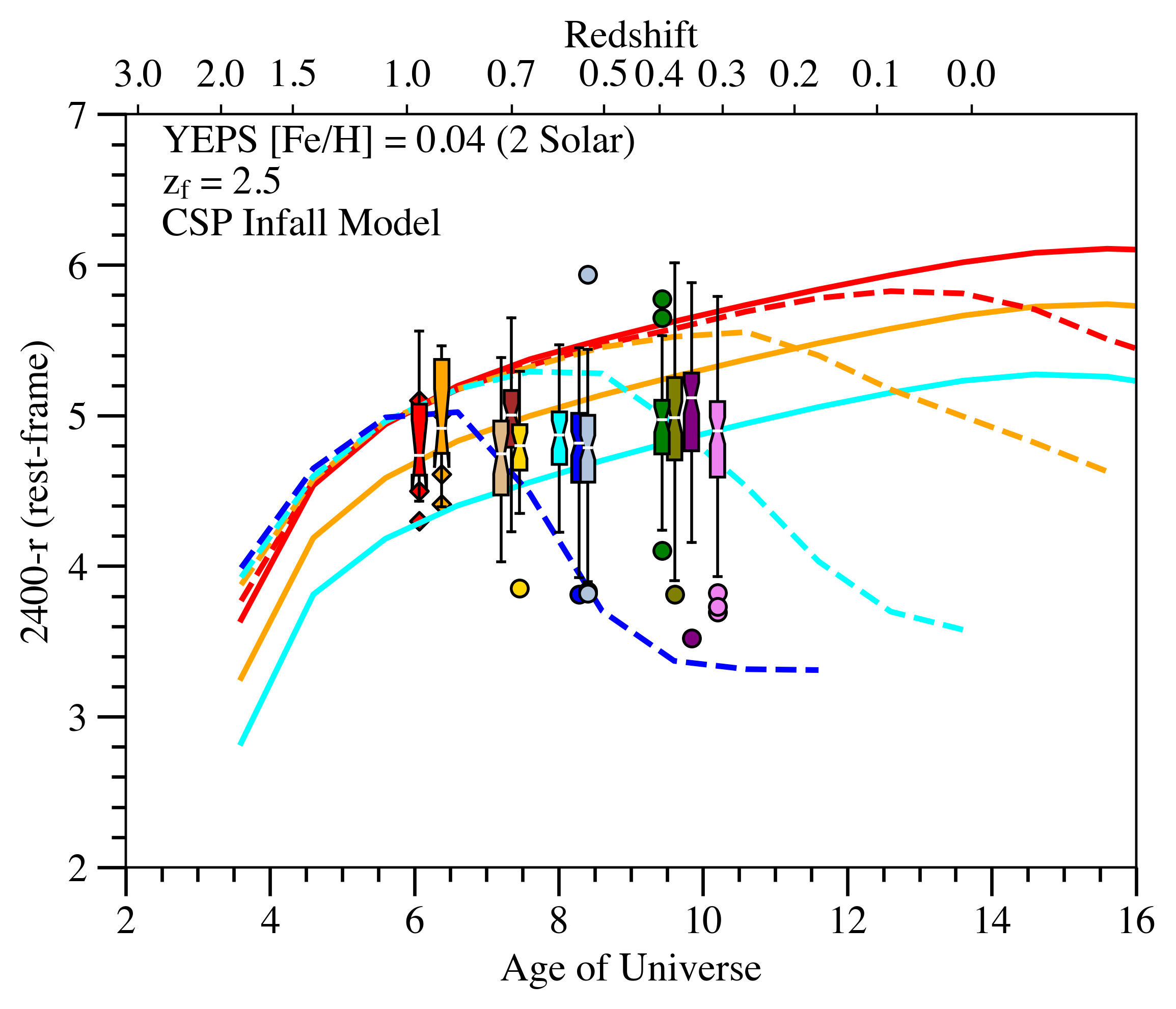}
\includegraphics[width=0.325\textwidth]{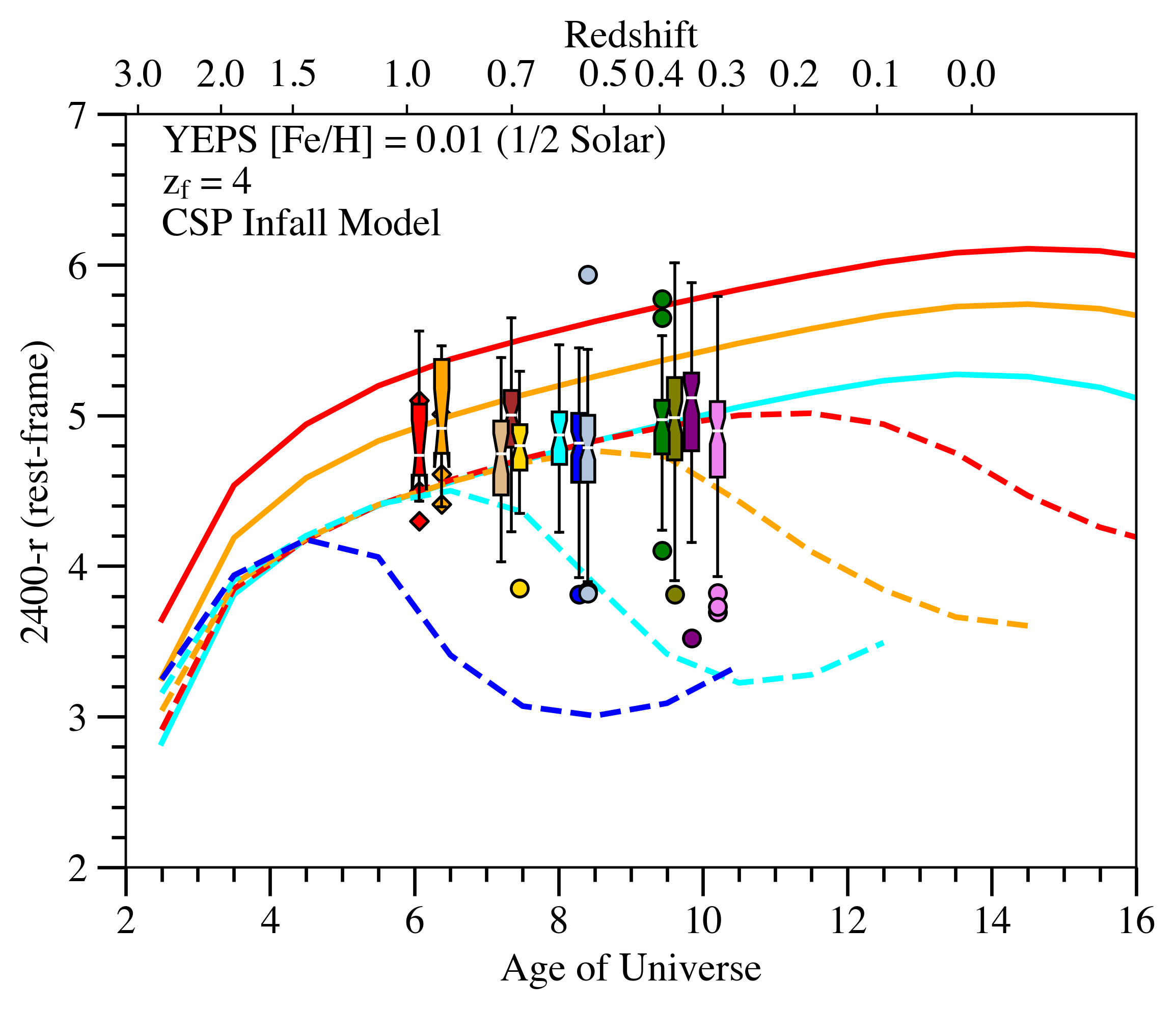}
\includegraphics[width=0.325\textwidth]{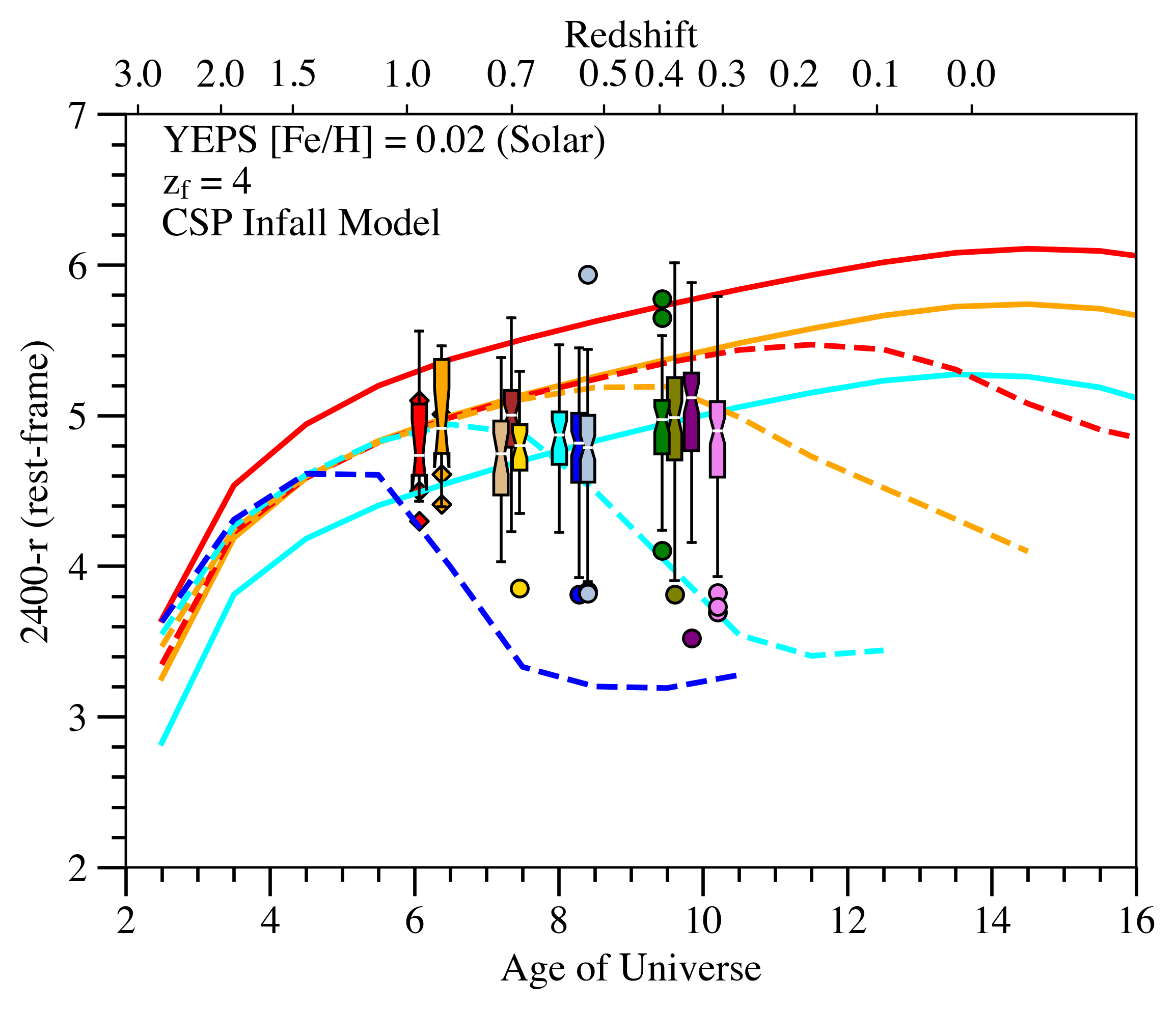}
\includegraphics[width=0.325\textwidth]{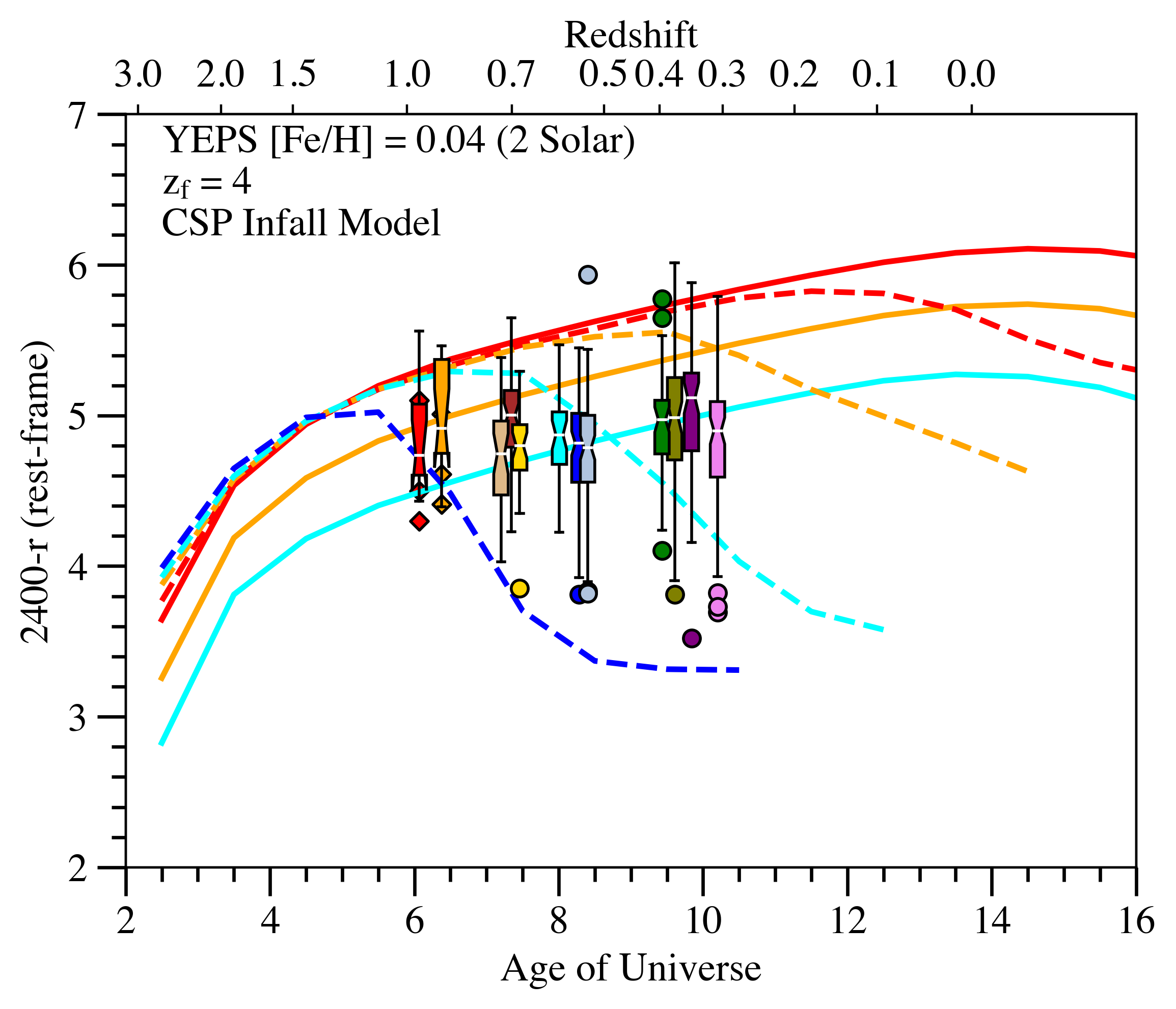}
\includegraphics[width=0.325\textwidth]{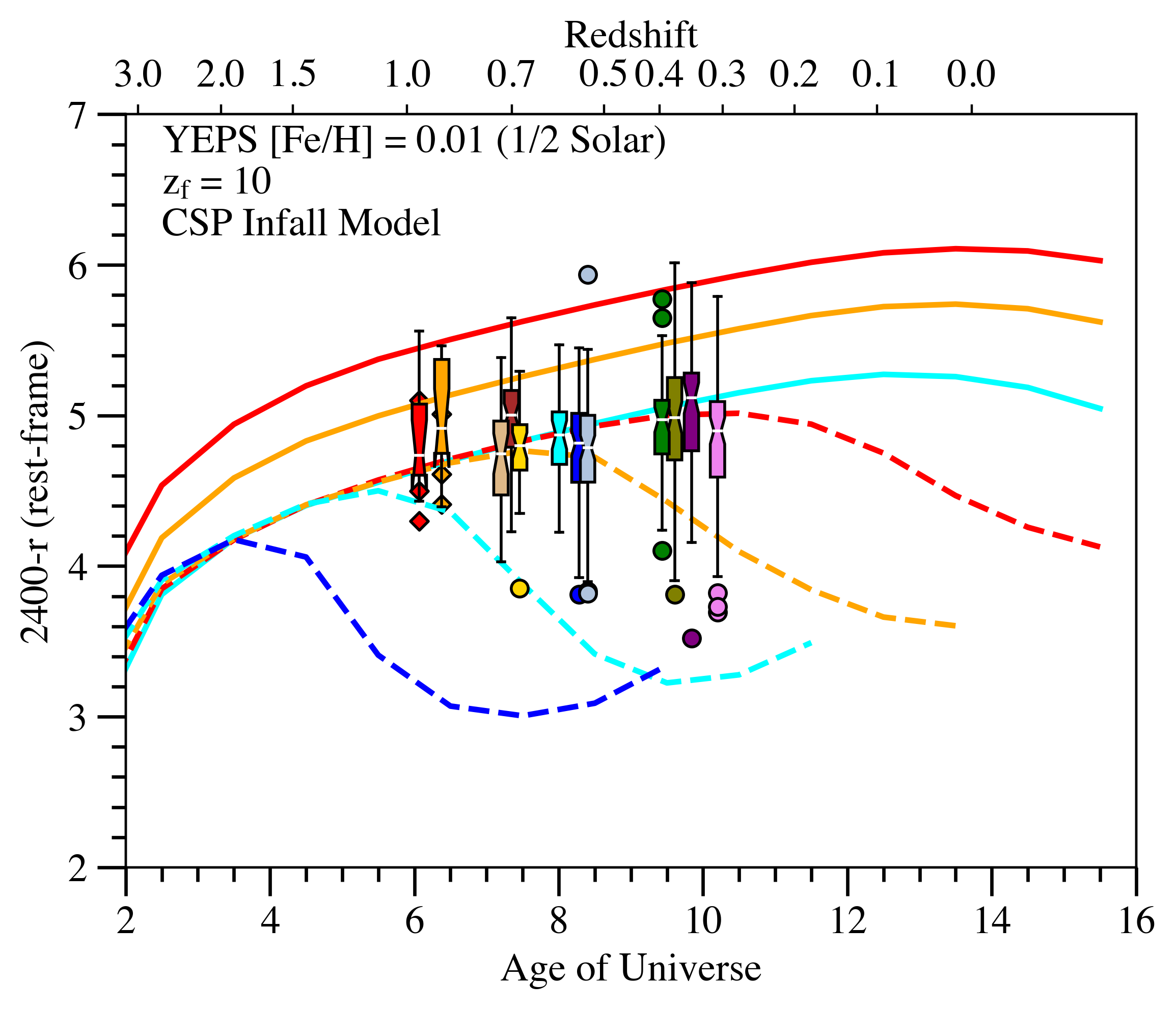}
\includegraphics[width=0.325\textwidth]{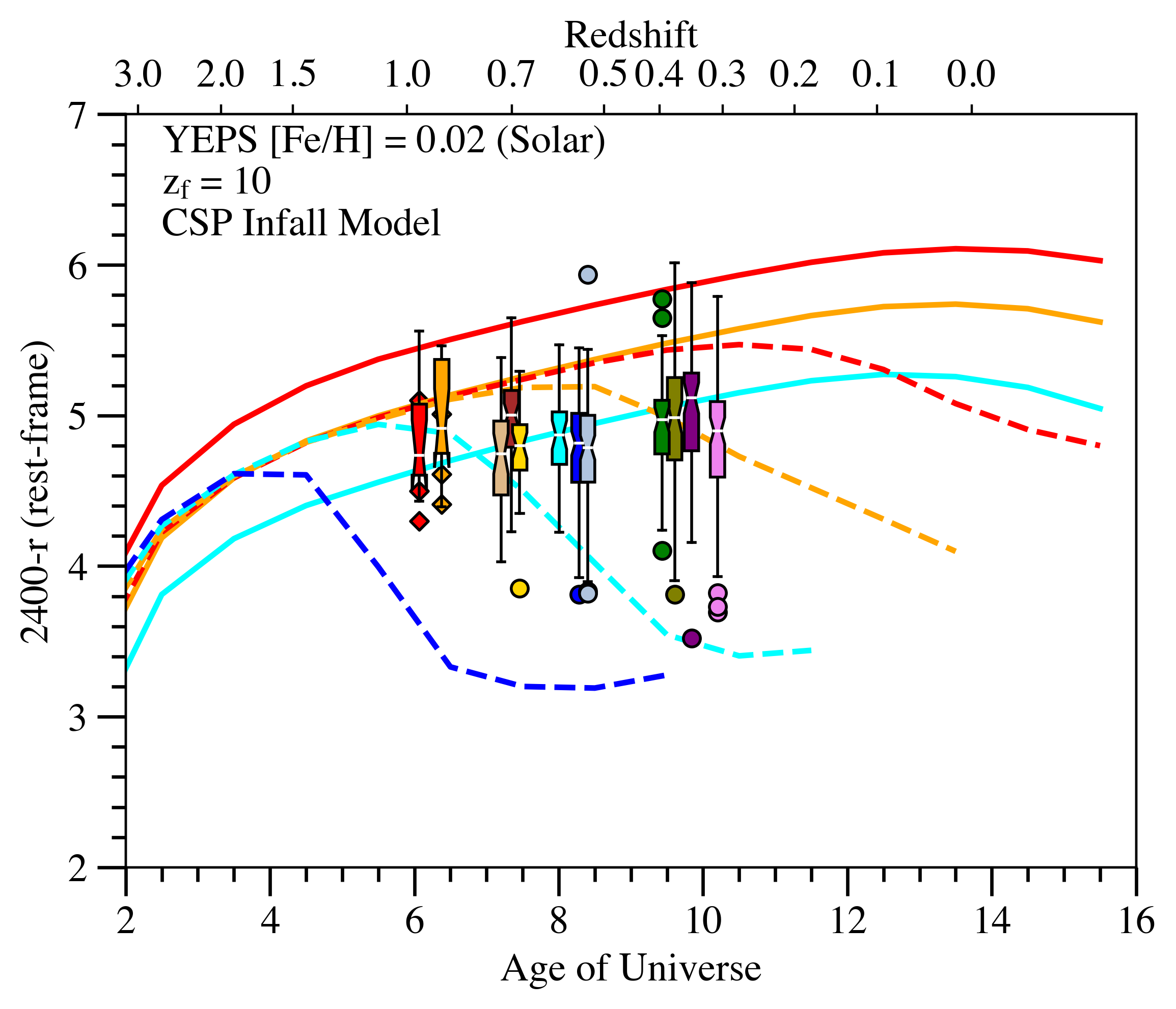}
\includegraphics[width=0.325\textwidth]{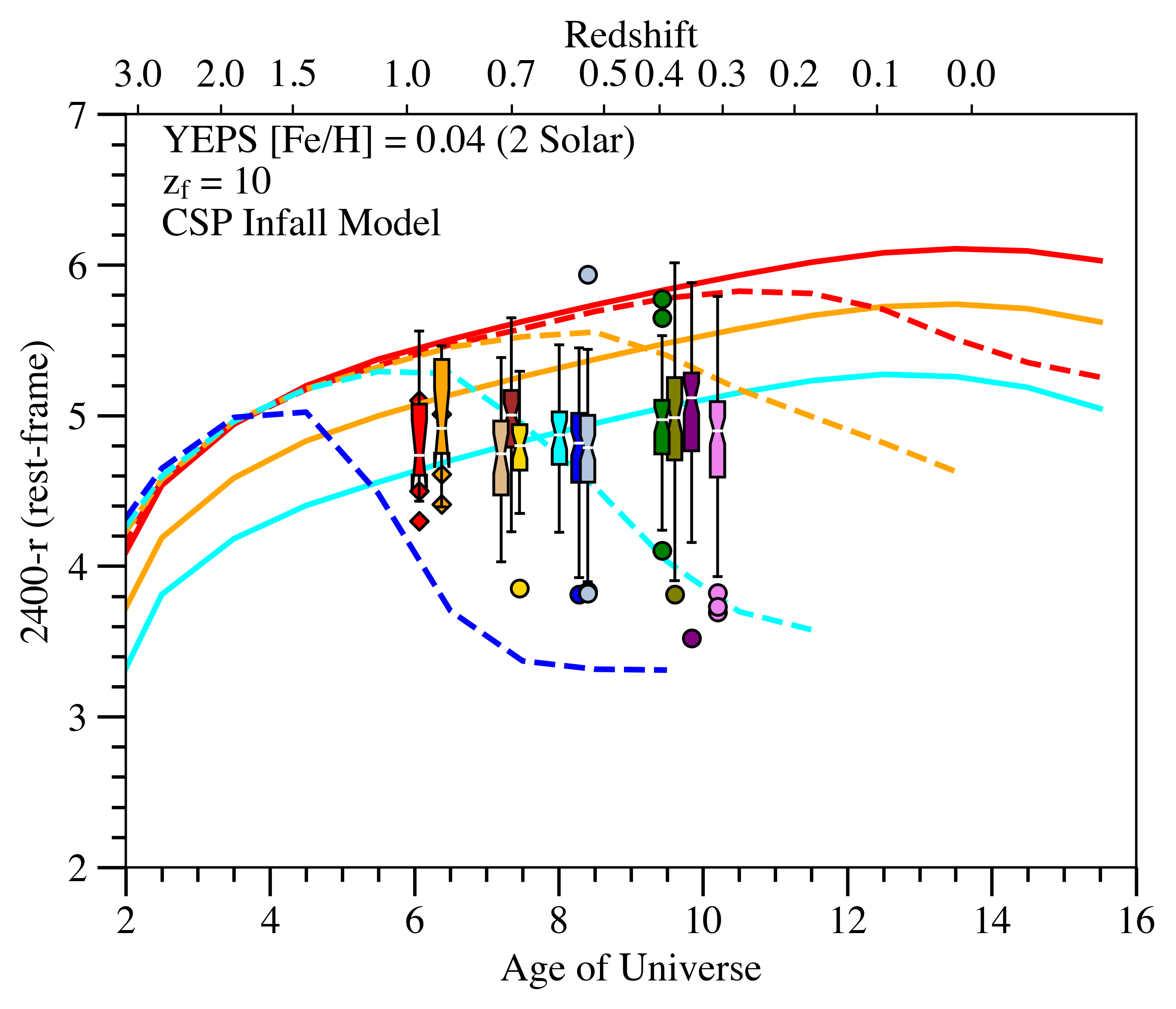}
\caption{YEPS infall (CSP) models showing the evolution of the rest-frame $2400-r$ (observed $F475W-F125W$ at z=0.96) color over redshift/lookback time for a range of initial helium abundances - $Y_{ini}=0.28,0.33,0.38,0.43$ (dashed lines), formation redshifts ($z_f$) and metallicities as detailed in the figure legends. Also included in every sub-plot is the evolution of the same color for infall models with $Y_{ini}=0.23$ (i.e. no upturn) for $Z$=\(Z_\odot\), 0.5\(Z_\odot\) \& 2\(Z_\odot\) (solid lines) at varying formation redshifts. Plotted on top for comparison are box plots which show the rest-frame $2400-r$ colors of all clusters between $z=0.3-1$. Photometric uncertainties in color are $<0.2$ magnitudes.}
\label{fig:yeps}
\end{figure*}

In order to analyse the evolution of the upturn over cosmic time, in Fig. \ref{fig:yeps} we compare our observed colors to those from YEPS spectrophotometric models of \cite{chung2017}. The YEPS models used in this figure are CSPs (with an infall chemical history) of given $Z$ and $z_f$, but also with varying initial helium abundances - $Y_{ini}$. This He-enhancement parameter represents the degree of UV upturn in stellar populations, with a larger $Y_{ini}$ giving rise to hotter HB stars at earlier cosmic times. There have been many observations of multiple stellar populations producing hot and extended HBs in Milky Way globular clusters (e.g. \citealt{lee2005b}; \citealt{piotto2005,piotto2007}) leading to a strong vacuum UV flux (e.g., see \citealt{dalessandro2012}), with direct spectroscopic evidence of enhanced helium in stars within such systems (e.g. \citealt{marino2014} and references therein).

In Fig. \ref{fig:yeps} we plot the time evolution of the rest-frame $\sim2400-r$ over redshift/lookback time. First, in every sub-plot we show the evolution of the YEPS models with solar-like $Y_{ini}=0.23$, for $Z$=\(Z_\odot\), 0.5\(Z_\odot\), 2\(Z_\odot\) (represented as solid lines in the plot). These form an important set of baseline models that \textit{do not} exhibit a UV upturn component, with the colors being driven largely by age and metallicity effects. Hence they act as a key point of comparison to both the observational data and the other models of increasing $Y_{ini}$, to determine how enhanced helium affects the observed colors.

In each sub-plot we then show the evolution of this near-UV color as given by the YEPS model for 4 more increasing initial helium abundances: $Y_{ini}=0.28, 0.33, 0.38, 0.43$ (represented as dashed lines in the plot). In the individual sub-plots we then vary the metallicity and the redshift of formation of these models. In each row, the $z_f$ is varied between three values of 2.5, 4 and 10 (from top to bottom). In each column, the $Z$ of the YEPS models are also varied between three values of 0.5\(Z_\odot\), \(Z_\odot\) and 2\(Z_\odot\) (from left to right). This creates a grid of models of varying $Z$ and $z_f$, allowing us to examine the combined effect of these parameters as well as $Y$ in determining the evolution of ETGs in the UV. 

Alongside the models, we plot the observed colors (rest-frame $2400-r$) of ETGs in our 12 clusters from all redshift bins between $0.3<z<1$. The colors of Abell 2744, Abell S1063, Abell 370, MACSJ0416, MACSJ1149, MACSJ0717, MACSJ2129, SDSS1004, MACSJ0744, RCS2327 and CL1226 (as seen in Fig. \ref{fig:3}) were k-corrected using the method described in section 3.1.1 to that of CL2011's rest-frame $2400-r$ ($F475W-F125W$ at $z=0.96$) . As detailed in section \ref{sec:kcorr}, given that all clusters were observed at wavelengths close to rest-frame $2400-r$, the results could be reliably compared after applying a relatively small k-correction that took into account the difference in rest-frame wavelengths and the shapes of the filters used to observe the clusters in the UV (as seen in Fig. \ref{fig:sed}). The colors are plotted in the form of box plots for an intuitive look at the statistics of each cluster. The box plots show the median of the color distribution of each cluster, the 25\% and 75\% quartiles and the entire range of colors in each cluster as indicated by the whiskers. Galaxies that have colors 1.5 times the interquartile range above and below the quartiles are plotted as individual points and are considered outliers. The notches around the median in each box plot represents the 95\% confidence interval of the median value for each cluster. Due to the small number of directly detected galaxies in CL2011 and CL1226, the notches around the median are large, representing the uncertainty in the median value. For these clusters we also plot the stacks separately from the box plots as these represent amalgam values of all galaxies in several magnitude bins (as seen in Fig. \ref{fig:3}). The models and the cluster data combined allow us to probe the evolution of the upturn in the general cluster population out to $z=0.96$.

From this figure we can see that the UV upturn is clearly detected in all clusters at $0.3<z<0.6$ and the $2400-r$ colors are relatively consistent from cluster-to-cluster, showing no significant signs of evolution. A subset of galaxies can be reproduced using models of standard $z_f$ and $Z$ combinations, but even in the most extreme cases that produce the bluest YEPS $Y_{ini}=0.23$ (no upturn) colors (i.e. $z_f$=2.5 and Z=0.5\(Z_\odot\) - see top left sub-plot), there still exists a large subset of bluer galaxies outside these models that can only be explained using models with enhanced $Y$. For more moderate and reasonable YEPS models with $z_f=4$ or higher (see middle and bottom row sub-plots), the argument for an enhanced $Y$ is even stronger in order to account for the larger proportion of blue galaxies outside the bounds of the non-upturn models. Increasing the $Y$ gradually also brings on the onset of the UV upturn at earlier redshifts/lookback times. This is affected by both $z_f$ and $Z$ at a given $Y$, with an earlier evolution to bluer colors for higher $z_f$ and lower $Z$ and vice versa. This is important as galaxies with different $z_f$ and/or $Z$ within a cluster can thus have different strengths and times for the onset of the upturn.

There is also a detection of the upturn for galaxies in the $z\sim0.7$ clusters, but at a slightly lower degree of  confidence, as the `blue tail' of galaxies appear to be receding, and as seen in figure \ref{fig:scatter}, have generally smaller observed scatters in their near-UV colors compared to their lower redshift counterparts. These factors may indicate that we are starting to see some signs of evolution in the strength and incidence of the upturn, which is backed up by the findings in \cite{ali2018c}, where SDSS1004 galaxies even when stacked together showed only a weak $3\sigma$ detection of 7.2 mags in rest-frame $1650-g$ (about one mag. redder than similar mass galaxies at $z=0.55$), in a color much more sensitive to the upturn than $NUV-r$. The fact that we still observe an upturn, albeit weakening, in $2400-r$ in the same cluster (and those of similar redshift) may suggest that even the most extreme He-rich stars at this age are no longer able to achieve the high surface temperatures needed to produce flux in the $FUV$ as their stellar envelopes are now too massive, and are thus contributing relatively more to the near-UV flux. As such we may be seeing the first signs of the upturn sub-population starting to disappear.

This is further confirmed by the data in CL1226 and CL2011 where we see that the majority of galaxies out to 3 mags below $M^*$ can be encapsulated by a standard non-upturn $Y_{ini}=0.23$ YEPS model of $Z$ roughly between above 2\(Z_\odot\) and 0.5\(Z_\odot\), assuming $z_f=2.5\sim4$ (top middle and middle sub-plots in Fig. \ref{fig:yeps}), without the need to invoke any significant He-enhancement above primordial levels. It is likely that by these redshifts, even the most extreme He-rich stars are no longer old enough (haven't had enough time to reach the HB) to achieve the surface temperatures required to output significant flux even in the near-UV. This gradual fading of the upturn is an excellent agreement with the results of \cite{lecras2016}, who also found the strength and incidence of the upturn decrease from $z=0.6$ to $z=1$ from the analysis of a number of near-UV indices of a large sample of LRGs. While non-upturn YEPS models do appear to fit the data relatively well at higher redshifts, it is important to note that even at these redshifts not all galaxies will be without an upturn. A combination of earlier $z_f$ and lower metallicity can lead to an earlier onset of the upturn, which may be applicable to some galaxies in a given cluster and explain the recent finding of a massive ETG exhibiting upturn at $z\sim1.4$ (\citealt{lonoce2020} -- although this is based on a single line at the $2\sigma$ level only, with other indicators disagreeing).

The overall color spread of the cluster galaxies is very similar across all clusters and in each individual redshift bin, indicating that all clusters exhibit similar upturns in their member galaxies irrespective of size or environment, as demonstrated in previous studies (\citealt{ali2019}; \citealt{phillipps2020}). Even between the clusters in the different redshift bins, we find the observed colors to be reasonably similar out to $z\sim0.6$, suggesting that the upturn has not evolved significantly, although there is a slightly smaller fraction of very blue galaxies at higher redshifts. These results are mostly consistent with our previous studies, where we detected strong $FUV$ emission from MACSJ1149/MACSJ0717 galaxies both directly and using stacking analysis to several magnitudes below $M^{*}$, which were also agreeing with colors measured in $z\sim0$ clusters such as Coma, Fornax and Perseus (\citealt{ali2018a,ali2018b,ali2018c}). This finding led to the conclusion that the upturn color has remained broadly similar in cluster galaxies out to $z\sim0.55$, which we reaffirm here.

Given a scenario in which the upturn color remains approximately constant out to $z\sim0.6$, then shows signs of weakening at $z\sim0.7$ and has mostly disappeared by $z\sim0.9$, the YEPS models (assuming a moderate \(Z_\odot\) and $z_f=4$ - middle sub-plot in Fig. \ref{fig:yeps}) suggest a minimum $Y_{ini}$ of 0.42 to explain the observations. This equates to a Y of 0.46 using the formula from \cite{chung2017} cited earlier ($Y=Y_{ini}+0.04$ for \(Z_\odot\)). An earlier formation epoch can reduce the requirement of high helium enhancement somewhat (by allowing more time for evolution), but this age can only be reasonably pushed back by 1 Gyr ($z=10$ corresponds to 0.5 Gyr after the Big Bang, which is the most likely redshift for the formation of the first galaxy of Milky Way mass -- \citealt{Naoz2006} -- whereas under the assumption of simple spherical infall the core regions of galaxies cannot form until $z=20$, yielding only an extra 200 Myr in time) at most and even with the most extreme case of $z_f=10$ (bottom middle sub-plot in Fig. \ref{fig:yeps}), the required $Y_{ini}$ only decreases to about 0.40. 

\begin{figure*}
\includegraphics[width=\textwidth]{Legend.png}
\includegraphics[width=0.325\textwidth]{Evolution_zf=4_feh=0_02_inf.png}
\includegraphics[width=0.325\textwidth]{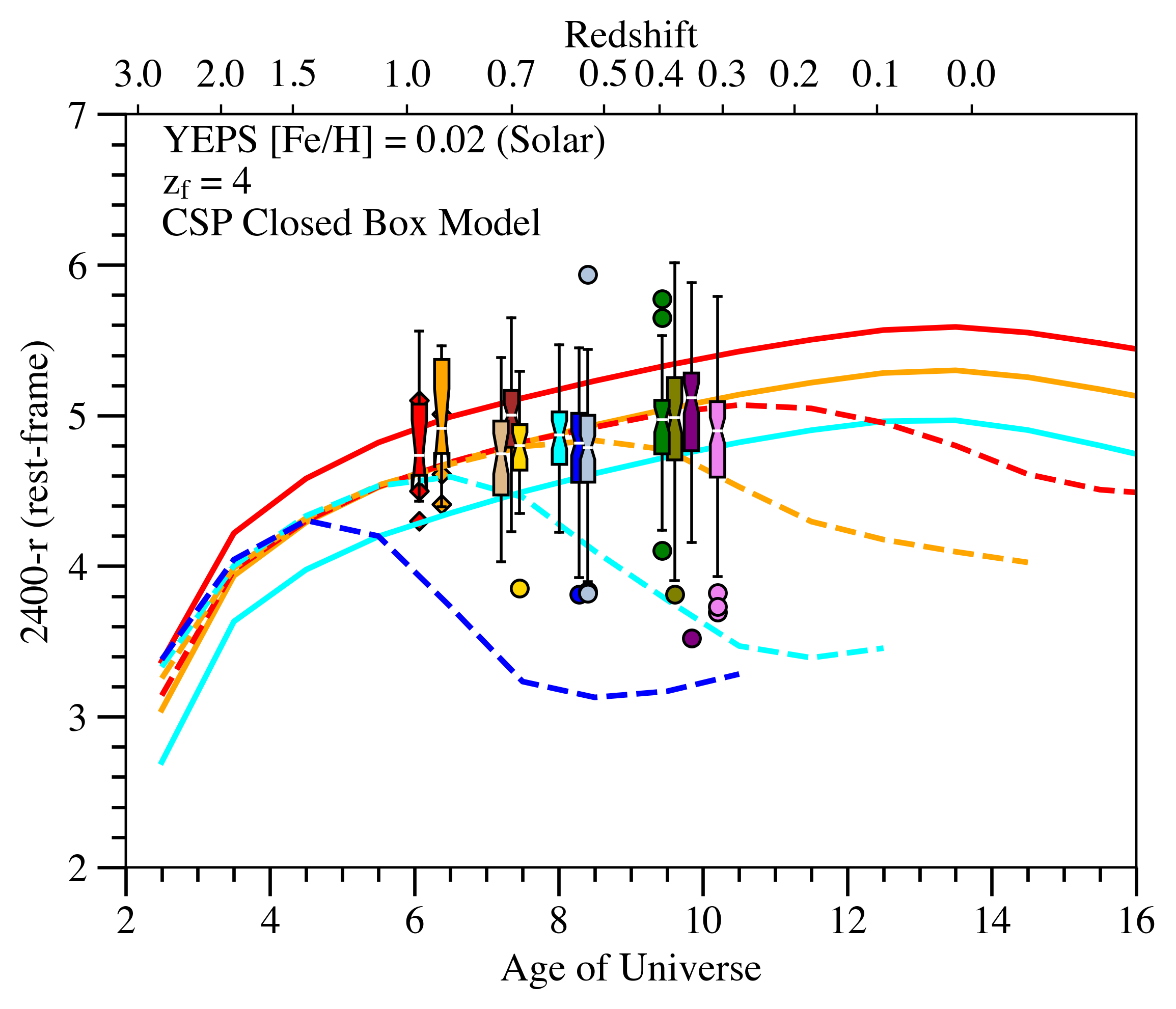}
\includegraphics[width=0.325\textwidth]{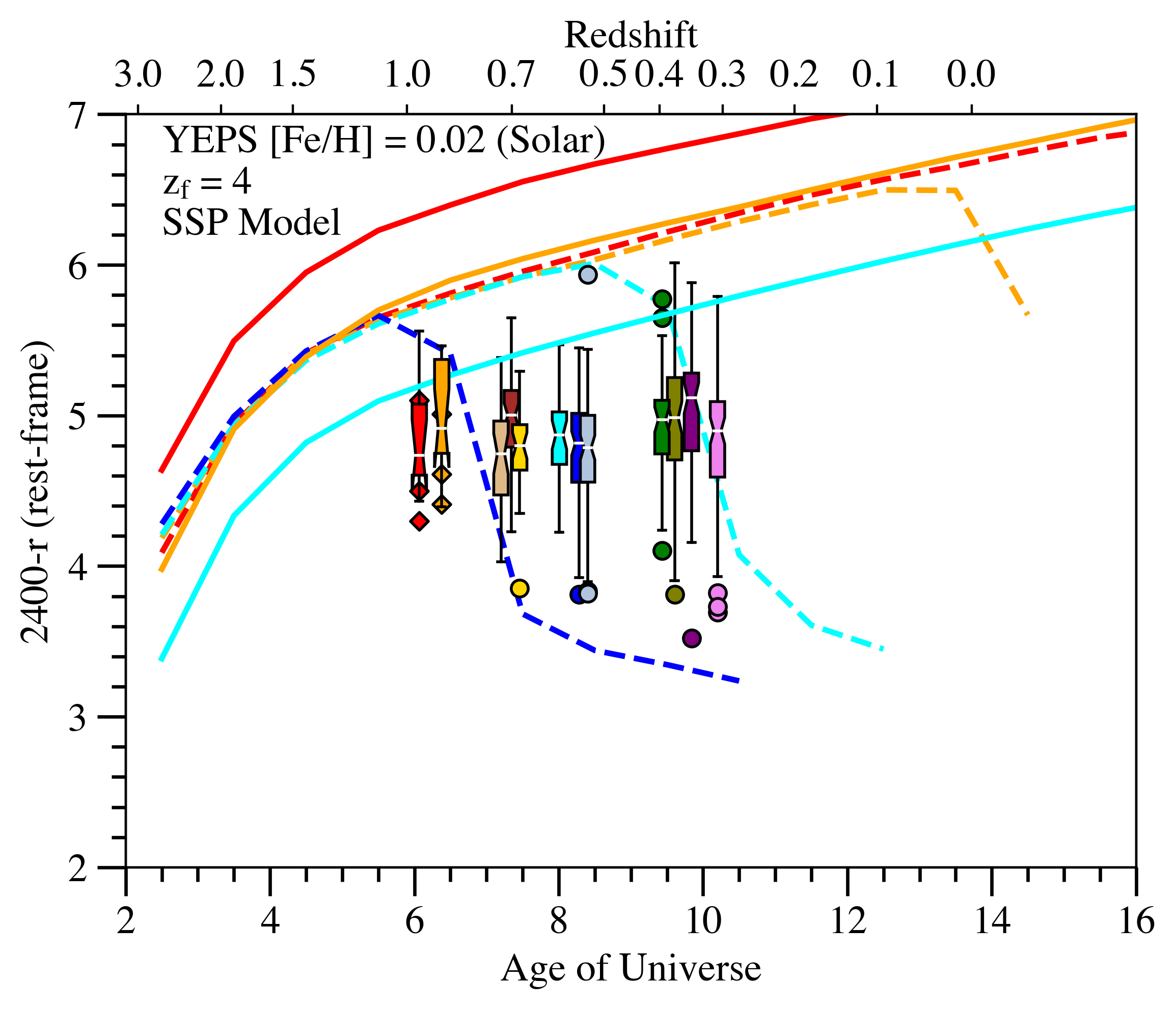}
\caption{YEPS models of different chemical formulations showing the evolution of the rest-frame $2400-r$ (observed $F475W-F125W$ at z=0.96) color over redshift/lookback time for a range of initial helium abundances - $Y_{ini}=0.28,0.33,0.38,0.43$ (dashed lines) with solar metallicity and formation redshift of 4. Also included in every sub-plot is the evolution of the same color for models with $Y_{ini}=0.23$ (i.e. no upturn) for $Z$=\(Z_\odot\), 0.5\(Z_\odot\) \& 2\(Z_\odot\) (solid lines). \textit{Left column:} YEPS composite stellar population infall model. \textit{Middle column:} YEPS composite stellar population simple closed box model. \textit{Right column:} YEPS simple stellar population model. Plotted on top for comparison are box plots that show the rest-frame $2400-r$ colors of all clusters between $z=0.3-1$. Photometric uncertainties in color are $<0.2$ magnitudes.}
\label{fig:chem}
\end{figure*}

\subsubsection{Comparison with different chemical evolutionary models}

In Fig.~\ref{fig:chem} we show the look-back time evolution of YEPS models with three different assumptions on the chemical evolution histories. The models are a simple stellar population and two variations of a composite stellar population - simple (closed box) and infall assumptions as described in \cite{kodama1997}. Using the metallicity distribution function from \cite{kodama1997}, simple stellar populations have been added up to mimic the composite stellar population of ETGs. For all of the models we shift the resulting SEDs to the redshift of the clusters and measure the integrated color at observed passbands. We overplot the observed colors of our clusters on these models as before.

The SSP models exhibit the reddest $UV-optical$ colors, while the closed box models being the bluest, and the infall models having colors in between. For a given $Y_{ini}$, SSPs appear to have a rapid onset of the upturn, while in CSPs the colors get bluer more gradually over time. This suggests that the upturn sub-population in CSPs take longer to fully populate the HB and become UV-bright compared to SSPs. Our observations appear to support the somewhat gradual mode of evolution of CSPs, given that we observe a strong upturn in all galaxies up to $z=0.6$, which then weakens (but does not fully disappear) at $z=0.7$ and has mostly ceased by $z=0.96$. Beyond our observations, CSPs generally provide a more accurate representation of ETGs, which will contain stellar populations with a range of ages and a metallicity distribution, unlike an SSP.

Amidst the CSP models, the closed box model being the bluest would require the reddest galaxies in our sample to have unusually high metallicities, much greater than 2\(Z_\odot\). Conversely, the highest metallicity required by the infall model to fit our reddest galaxies is $\sim$2\(Z_\odot\), which provides a very realistic upper limit for the metallicity of large cluster ETGs from their optical spectra, as confirmed in previous studies - e.g. in Coma (\citealt{price2011}). The infall models thus provide overall the most realistic and economical fit to our data for all three paratemeters of $Y_{ini}$, $z_f$ and $Z$ simultaneously. Infall models are also generally heralded to better replicate the evolution of observed galaxies, particularly with their ability to solve the `G-dwarf' problem of closed box models.

\subsection{Comparison with other UV upturn models}

We can briefly consider the implications of these results for other models of the origin of the UV upturn.

Low metallicity HB stars or high metallicity HB stars (apart from the issues we have mentioned earlier) would start to appear only at $z\sim 0.3$ \citep{yi1997}, well below the redshift where the presence of the UV upturn is already securely detected (at least $z=0.55$). Therefore our results confirm that these objects (with normal $Y$ abundance) cannot provide the sources of the UV upturn and account for its evolution with time.

The observed evolution, here and in previous studies \citep{lecras2016,ali2018c} is also not consistent with the binary model of \cite{Han2007} and \cite{Hernandez2014}, where the UV upturn color in these models does not change significantly until $z \sim 6$ and there is clear evidence of evolution in our data. Similarly, PAGB stars would only stop contributing to the UV light about 10 Gyrs prior to the present epoch \citep{Lee1999}, corresponding to $z \sim 3$, again far higher than observed here and in previous works.

\cite{Vazdekis2016} and \cite{Rusinol2019},
among others, suggest that residual star formation within ETGs
may contribute to the UV upturn. However, the results presented here (and the observed decline of the UV upturn at $z>0.6$ in \citealt{lecras2016} and \citealt{ali2018c}) cannot be explained in this fashion, as the residual star formation rate would have to decrease as a function of increasing redshift, opposite to all observed behavior for star-forming galaxies in the field and clusters \citep{Finn2005}. The galaxies in our sample, additionally, lie within rich clusters, where quenching of star formation is believed to be more efficient than in the field (but we note that the evolution of the UV upturn color appears to be similar between field and cluster environments -- \citealt{Atlee2009}, De Propris et al., in preparation).

\subsection{Implications for galaxy evolution}

While there exists several other models for the upturn, the observed evolution of the near-UV colors is best fitted by He-rich models. Since He-rich stars have been directly observed in local globular clusters and are directly linked to a stronger UV output in hot HB stars, this mechanism is the only one thus far proposed that does not require significant modifications to cosmology or theories of stellar evolution, and where local counterparts are observed to exist.

In this scenario ETGs form, at least from the standpoint of chemical evolution, in the same manner as globular clusters, albeit at generally much higher metallicities (but He-rich metal rich clusters are observed in M87 -- \citealt{peacock2017}). This would imply that most ETGs form their stellar populations and assemble their mass at very early times. Indeed, the stellar mass needed to provide the observed flux in the UV corresponds to about 10\% of the total stellar mass, and this must have been formed at $z > 2.5$ and probably closer to $z=4$ as discussed above. However, these are second generation stars, whose helium content must have been enriched by previous processing, likely in fast rotating massive stars \citep[e.g.,][]{decressin2007} or in massive AGB stars \citep[e.g.,][]{ventura2001} during the third dredge-up. Given what is known about the likely yields of element production and the resulting timescales, and within a closed box model, the vast majority of the stellar mass must already have been present within ETGs at these redshifts, especially if we also need to account for the observed radial gradients in the UV upturn populations within galaxies \citep{carter2011,jeong2012}. This therefore suggests a very early period of mass assembly for ETGs, as indicated by the discovery of several massive fully formed galaxies at redshifts approaching 4 \citep{guarnieri2019}.

\subsection{Caveats}
Finally, we note here the key caveats in our data and analysis (see body of paper for a more thorough discussion):
\begin{itemize}
    \item The upturn has been conventionally classified using $FUV-optical$ colors (centred at $\sim1500$\AA), the wavelength regime where the upturn is at its strongest, while we have used a $NUV-optical$ color for our analysis. Although this wavelength regime is not the most optimal and is partly affected by the tail end of the blackbody emission of the main sequence population in ETGs, this study (e.g. Fig. \ref{fig:sed}) and others (\citealt{schombert2016}; \citealt{phillipps2020}) have shown that it is still mostly dominated by the emission from the upturn sub-population. Hence the $NUV-optical$ color should also trace the evolution of the upturn with redshift, as is predicted by models and seen in our results.
    \item At high redshift, for CL1226 and CL2011 the data is not deep enough to individually detect galaxies below the cluster $M^*$, unlike the lower redshift clusters. Clusters at higher redshift are also generally not as rich as their lower redshift counterparts, which when combined with the aforementioned detection limits can lead to small number statistics particularly towards the faint end of the luminosity function. To overcome this issue we have stacked all galaxies below $M^*$ and were able to clearly detect the average flux per magnitude bin in this region, reaching down to a similar point in the luminosity function as the lower redshift clusters and probing the more general ETG population. While stacking does not allow for us to explore the scatter in the color caused by the upturn, it does allow us to probe the general evolution of the color with redshift.
    \item We have selected our clusters using the standard red sequence selection method as there are no spectroscopic or accurate photometric redshifts particularly for the high redshift cluster galaxies. However, to reduce contamination from star-forming and background/foreground objects, we have selected the red sequence simultaneously in both an optical and U-band colors. This method of cluster ETG selection is found to be rather successful from spectroscopic studies (\citealt{rozo2015}).
    \item While the He-enhanced models can definitely explain the evolution of the upturn with redshift most consistently and have local observed analogues such as in Milky Way globular and open clusters, there is still no theoretical construct by which one can get such anomalously He-enriched HB stars in old systems, potentially formed at very high redshifts ($z\sim3-4$) to allow for the required timeframes to evolve on to the HB. Some suggestions have been made, such as the disintegration of GCs providing He-enhanced HB stars in ETGs (\citealt{Goudfrooij2018}), but there is no observational evidence thus far to support such hypotheses. Further theoretical and observational efforts will be required in both the galaxy evolution and globular cluster fronts to discover any potential links between the systems (such as metal-rich GCs being observed in M87 - \citealt{peacock2017}) and to further constrain the source of the He-enhacement in a subset of the stellar population in these systems.
\end{itemize}

\section{Conclusions}
We have measured an approximate $NUV-r$ color for ETGs within clusters at $0.3 \lesssim z \lesssim 1$. At $z < 0.6$ we observe the classical UV upturn showing little evolution and homogeneous UV colors with large spread. Above this redshift we find evidence of a decline in the strength of the upturn, that largely disappears by $z=0.96$ (our most distant target). This is most consistent with composite stellar population models with an infall chemical history where the UV upturn is produced by a population of blue HB stars with $Z$=\(Z_\odot\) formed at $z=2.5\sim4$, in turn originating from a sub-component stellar population with high He content ($\sim 45-47\%$) similar to the second generation population in globular clusters. These models are able to best replicate the evolution and scatter of the UV colors across the entire redshift range of $0.3<z<1$ analysed in this study. Given the evolutionary timescales for these stars, the results imply a surprising degree of chemical evolution occurring within the first 2 Gyrs of the history of the Universe.

\acknowledgments

Based on observations made with the NASA/ESA Hubble Space Telescope, and obtained from the Hubble Legacy Archive, which is a collaboration between the Space Telescope Science Institute (STScI/NASA), the Space Telescope European Coordinating Facility (ST-ECF/ESA) and the Canadian Astronomy Data Centre (CADC/NRC/CSA). C. C. acknowledges the support provided by the National Research Foundation of Korea (2017R1A2B3002919).

\bibliography{references}{}
\bibliographystyle{aasjournal}

\end{document}